\documentclass[galaxies,review,submit,moreauthors,pdftex]{Definitions/mdpi}

\usepackage{natbib}
\usepackage{color}
\usepackage{mathtools}

\usepackage{soul}
\setitemize{parsep=6pt,itemsep=0pt,leftmargin=*,labelsep=5.5mm}
\setenumerate{parsep=6pt,itemsep=0pt,leftmargin=*,labelsep=5.5mm} 
\setlist[description]{itemsep=0mm}   
   
\usepackage{amsmath}
\usepackage{amssymb}
\usepackage{bm}
\usepackage{ulem}
\usepackage{upgreek}
\usepackage{enumerate}
\usepackage{enumitem}
\usepackage{subcaption}
\setlist{nosep}	
\usepackage{textcomp}

\def	\cm		{\,{\rm {cm}}}
\def	\K		{\,{\rm {K}}}
\def	\g		{\,{\rm {g}}}
\def	\s		{\,{\rm {s}}}

\def	\rad	{\,{\rm {rad}}}

\def \bea {\begin{eqnarray}}
\def \ena {\end{eqnarray}}                  

\def    \ba     {\bf  a}

\def	\ba			{{\bf a}}
\def	\be			{{\bf e}}

\def    \gas     	{{\rm gas}}

\font\mib=cmmib10

\def\bomega{\hbox{\mib\char"21}}

\def\bmu{\hbox{\mib\char"16}}
    
\graphicspath{{./figures/}}    

\firstpage{1} 
\pubvolume{xx}
\issuenum{1}
\articlenumber{5}
\pubyear{2020}
\copyrightyear{2020}
\history{Received: 14 April 2020; Accepted: 29 June 2020; Published: date}

   \pdfoutput=1

\def	\cm		{\,{\rm {cm}}}
\def	\K		{\,{\rm K}}
\def	\g		{\,{\rm {g}}}

\def	\max	{{\rm max}}

\def	\H	{{\rm H}}

\def	\km	{\,{\rm km}}

\def	\erg	{\,{\rm erg}}
\def	\gas	{\,{\rm gas}}
\def	\eff	{{\rm eff}}

\def	\D	{\,{\rm D}}
\def	\yr	{\,{\rm yr}}
\def	\s	{\,{\rm s}}
\def	\AU	{\,{\rm AU}}

\def \bea {\begin{eqnarray}}
\def \ena {\end{eqnarray}}

\def    \bv     	{\bf  v}

\def	\xhat		{\hat{\bf x}}
\def	\yhat		{\hat{\bf y}}

\Title{Rotational Disruption of Astrophysical Dust and Ice: Theory and Applications}

\Author{{Thiem Hoang} $^{1,2}$\orcidA{}}  

\AuthorNames{Thiem Hoang}

\address{%
$^{1}$ \quad Korea Astronomy and Space Science Institute, Daejeon 34055, Republic of Korea\\
$^{2}$ \quad University of Science and Technology, Korea, (UST), 217 Gajeong-ro Yuseong-gu, Daejeon 34113, Republic of Korea}

\corres{Correspondence: thiemhoang@kasi.re.kr}

\abstract{Dust is an essential component of the interstellar medium (ISM) and plays an important role in  many different astrophysical processes and phenomena. Traditionally, dust grains are known to be destroyed by thermal sublimation, Coulomb explosions, sputtering, and shattering. The first two mechanisms arise from the interaction of dust with intense radiation fields and high-energy photons (extreme UV), which work in a limited astrophysical environment. The present review is focused on a new destruction mechanism present in the {\it dust-radiation interaction} that is effective in a wide range of radiation fields and has ubiquitous applications in astrophysics. We first describe this new mechanism of grain destruction, namely rotational disruption induced by Radiative Torques (RATs) or RAdiative Torque Disruption (RATD). We then discuss rotational disruption of nanoparticles by mechanical torques due to supersonic motion of grains relative to the ambient gas, which is termed MEchanical Torque Disruption (METD). These two new mechanisms modify properties of dust and ice (e.g., size distribution and mass), which affects observational properties, including dust extinction, thermal and nonthermal emission, and polarization. We present various applications of the RATD and METD mechanisms for different environments, including the ISM, star-forming regions, astrophysical transients, and surface astrochemistry.}
\keyword{ISM: dust; galaxies: evolution; galaxies: ISM}
\begin{document}

\section{Introduction}
Interstellar dust is an essential component of the interstellar medium (ISM). It plays important roles in astrophysics, including gas heating, star and planet formation (see \citet{2003ARA&A..41..241D} for a review), and~grain-surface chemistry (\citet{Herbst:2009go,Hama:2013ja}). The~interaction of dust with radiation from starlight causes extinction, emission, and~polarization of light, which~are the basis of modern astrophysics. Indeed, thermal dust emission in far-infrared/submm is a powerful window to observe dense star-forming regions and to understand how stars and planets form. Dust~polarization induced by the alignment of dust grains with the magnetic field allows us to measure magnetic fields in various astrophysical environments \mbox{(see \citet{Andersson:2015bq,LAH15} for} recent reviews). Polarized dust emission is a critical foreground component that needs to be accurately removed for the detection of the Cosmic Microwave Background (CMB) B-mode (\citet{Kamionkowski:2016bb}). Furthermore, ice mantles on dust grains play a central role in astrochemistry because water and complex organic molecules (COMs) are thought to first form on the ice mantles and subsequently are released into the gas phase due to star formation activity (\citet{Herbst:2009go}). 

Formation of dust is believed to arise from two leading sources, including Asymptotic Giant Branch (AGB) stars (e.g., \citet{2013A&A...555A..99Z}) and core-collapse supernovae\mbox{ (CSNe; \citet{Sarangi:2018fg})}. Newly formed grains are then injected into the ISM and constantly reprocessed due to collisions with gas and dust (\citet{2016ApJ...831..147Z}). The~subsequent evolution of dust in the ISM consists of accretion of gas species (atoms and molecules) on the grain surface and grain~coagulation.

Destruction of dust is widely studied in the literature (see \citet{2004ASPC..309..347J}). Four~well-known mechanisms of dust destruction include thermal sublimation, Coulomb~explosions, sputtering (thermal and non-thermal), and~shattering by grain-grain collisions\mbox{ (see \citet{1979ApJ...231..438D,1994ApJ...431..321T,1994ApJ...433..797J}} for details). The~first two mechanisms are induced by {\it dust-radiation} interaction. While thermal sublimation is efficient in intense radiation fields where dust grains can be heated to sublimation temperatures of \mbox{$T_{\rm sub}\sim$ 1500--1800~K}  for silicate and graphite grains, Coulomb explosions require high-energy photons (extreme UV or X-ray) to be efficient (\citet{2006ApJ...645.1188W,2015ApJ...806..255H}). Note that the large majority of volume of astrophysical environments is filled with radiation fields where grains are only heated to a temperature of $T_{d}<T_{\rm sub}$, which corresponds to the local energy density spanning a wide range, from~a radiation strength of $U=1$ ({in units of the average interstellar radiation field (ISRF); \mbox{\citet{1983A&A...128..212M})}~to $U\sim 10^{11}$}, assuming $U\sim (T_{d}/16.4)^{1/6}$ for silicates (\citet{2011piim.book.....D}). Thus, the~current paradigm of dust evolution implies that dust properties do not change under such radiation~fields.  

The above paradigm of dust evolution is challenged by the early-time observations (within~weeks since the first light) toward type Ia supernovae (SNe Ia). Indeed, numerous observations reveal anomalous properties of dust, namely the predominance of small grains in the local interstellar medium of SNe Ia (\citet{2013ApJ...779...38P}), which is characterized by the unusually low value of total-to-selective extinction ratio, $R_{V}=A_{V}/E_{B-V}<2$ \mbox{(\citet{2014ApJ...789...32B,2015MNRAS.453.3300A,2016ApJ...819..152C})}. This~is inconsistent with the standard model of interstellar dust with the standard value of \mbox{$R_{V}=3.1$} \mbox{(\citet{2013ApJ...779...38P}) and requires} enhanced abundance of small grains \citet{2016P&SS..133...36N}. A~similar problem is demonstrated through mid-infrared emission excess (1--5 $\upmu$m) in the spectrum of massive young stellar clusters, which requires an enhanced abundance of small grains relative to large ones. Extinction~curves with a steep far-ultraviolet (UV) toward Gamma-ray bursts (GRBs) or quasars also reveal anomalous dust properties with the predominance of small grains. These puzzles cannot be explained with previously known mechanisms of dust destruction (\citet{2017ApJ...836...13H}). 

In addition to grain heating and charging, \mbox{dust-radiation interaction is well-known to induce} radiation force and radiative torques (RATs) on dust grains~\mbox{(\citet{1976Ap&SS..43..291D})}. Radiation~force accelerates grains to high speeds (\citet{Spitzer:1949bv}), resulting in the drift of grains through~the gas (\citet{2015ApJ...806..255H,2017ApJ...847...77H}), while RATs can spin-up grains to suprathermal rotation, that is,~rotation at velocities above their thermal angular value (\citet{1996ApJ...470..551D,2004ApJ...614..781A,2007MNRAS.378..910L,Hoang:2008gb,2009ApJ...695.1457H,Herranen:2019kj}). Note~that dust grains are widely known to be rotating suprathermally, as~required to reproduce starlight polarization and far-IR/submm polarized dust emission (see \citet{Andersson:2015bq,LAH15} for reviews). 

Recently, \citet{Hoang:2019da} discovered that irregular grains exposed to intense radiation fields (e.g., from~massive stars and supernovae) could be spun-up by RATs to extremely fast rotation such that the centrifugal stress exceeds the maximum tensile strength of the grain material, breaking the original grain into small grains (see also \citet{2019ApJ...876...13H}). This is called RAdiative Torque Disruption (RATD) mechanism. {It is noted that \citet{1979ApJ...231..404P} first mentioned rotational disruption as a potential consequence of grain suprathermal rotation induced by pinwheel torques from hydrogen formation. Rotational disruption by RATs for fluffy grains in the solar system is also noticed in \citet{2016ApJ...818..133S}}.

The RATD mechanism requires lower radiation energy to be effective than thermal sublimation. The~reason is that RATD tends to break the constituents of the grain that are loosely bound by Van der Waals force of binding energy $E_{b}\sim 0.01$ eV (e.g., for~composite grains), whereas thermal sublimation needs high energy to break chemical bonds of energy $\sim$0.1 eV (\citet{2019ApJ...876...13H}). Thus, this RATD mechanism could be effective for the majority of astrophysical environments (cf. to sublimation or Coulomb explosions), from~the average ISRF, star-forming regions, circumstellar regions, to~environments around cosmic transients. The~RATD mechanism is a ubiquitous process because radiation is ubiquitous in the~Universe.

Moreover, dust grains drifting through the ambient gas experience mechanical torques due to stochastic collisions with gas species (\citet{1952MNRAS.112..215G}). A~grain of irregular shape experiences {\it regular} mechanical torques, which are stronger than the stochastic \mbox{torques (\citet{2007ApJ...669L..77L,2018ApJ...852..129H})}. The fast relative motion of dust to gas can arise from radiation pressure or interstellar shocks (\citet{1980ApJ...241.1021D}). As~a result, grains can also be spun-up to suprathermal rotation by mechanical torques such that centrifugal stress is sufficient to break a small grain into smaller fragments. This~mechanism is termed MEchanical Torque Disruption (METD) and most effective for very small grains of size $a<10$ nm (\citet{2019ApJ...877...36H,2019ApJ...886...44T,Hoang:2020kh}). 

In particular, centrifugal forces on rapidly spinning grains have inevitable effects on the formation and desorption of molecules on/in the ice mantles of dust grains. The~reason is that water and complex molecules in the ice mantles sublimate at temperatures of $T_{d}>100\K$ (\citet{Herbst:2009go}), which~corresponds to \mbox{$U> 10^{5}$}, that is,~much stronger radiation fields than the ISRF for which RATD is important. As~shown in \citet{2020ApJ...891...38H} and~\citet{2019ApJ...885..125H}, the~grain suprathermal rotation also affects the chemical composition and metallicity of the gas because ice mantles on the grain surface are easily disrupted by centrifugal~stress. 

Since RATs are strongest for the grains of size comparable to the photon wavelength, that is,~$a\sim \lambda$ (\citet{2007MNRAS.378..910L}), the~RATD mechanism is most efficient for large grains of $a>0.1$ $\upmu$m for starlight of $\lambda\gtrsim$ 0.1 $\upmu$m. On~the other hand, the~METD mechanism is most efficient for very small grains (VSGs) or nanoparticles (size $a<10$ nm) because these tiny grains with small inertia moment rotate faster (\citet{2019ApJ...877...36H,Hoang:2020kh}). Thus, the~RATD and METD mechanisms determine the grain size distribution, including the upper cutoff and abundance of nanoparticles, which affects dust extinction, emission from microwave to infrared wavelengths, and~polarization. 

This review focuses on the new grain destruction mechanisms (RATD and METD) and their astrophysical applications. The~structure of the review is as follows. We first summarize the current knowledge of grain destruction mechanisms in Section~\ref{sec:des}. In~Section~\ref{sec:theory}, we describe the theory of a new mechanism of dust destruction (RATD) and provides an example of grain disruption by a point source. In~Section~\ref{sec:ice}, we describe rotational disruption of ice mantles and desorption of molecules from the icy grain mantle. In~Section~\ref{sec:nano}, we present the description of rotational disruption of nanoparticles by mechanical torques (METD). In~Section~\ref{sec:model}, we present modeling methods of dust extinction, emission, and~polarization that takes into account the rotational disruption effects. In~Section~\ref{sec:appl}, we discuss the applications of the RATD and METD mechanisms for different astrophysical environments, including~the diffuse ISM, star-forming regions, cosmic explosions, and~high-z galaxies. An~extended discussion on the implications of our new mechanisms for time-domain astrophysics and astrochemistry is presented in Section~\ref{sec:discuss}. Conclusions and future prospects are summarized in Section~\ref{sec:summ}.
 
\section{Destruction Mechanisms of Astrophysical~Dust}\label{sec:des}
Here we first review four well-known destruction mechanisms of astrophysical dust, including~thermal sublimation, Coulomb explosions, sputtering, and~shattering. The~first two mechanisms are associated with intense radiation fields, while the last two are related to relative motion between dust and~gas.

\subsection{Thermal~Sublimation}
Grains in a strong radiation field are heated to high temperatures and sublimate rapidly when their temperatures are above the sublimation threshold. The sublimation rate for a grain of radius $a$ and temperature $T_{d}$ is given by:
\bea
\frac{da}{dt}=-\frac{\nu_{0}}{n_{d}^{1/3}}\exp\left(\frac{-B}{k T_d}\right),\label{eq:dasdt}
\ena
where $n_{d}\sim 10^{22}$--$10^{23}\cm^{-3}$ is the atomic number density of dust, $B$ is the sublimation energy per atom, $\nu_0 = 2\times 10^{15} \s^{-1}$ and $B/k=68100 -20000N^{-1/3}\K $ for silicate grains,  $\nu_0 = 2\times 10^{14} \s^{-1}$ and $B/k=81200-20000N^{-1/3} \K $ for carbonaceous grains with $N$ being the total number of atoms of the grain (\citet{1989ApJ...345..230G,2000ApJ...537..796W}).

The sublimation time of a dust grain of size $a$ is defined as
\bea
t_{\rm sub}(T_d)=-\frac{a}{da/dt}
= an_{d}^{1/3}\nu_{0}^{-1}\exp\left(\frac{B}{k T_d}\right),\label{eq:tausub}
\ena
where $da/dt$ from Equation~(\ref{eq:dasdt}) has been~used.

Plugging the numerical parameters into the above equation, we obtain
\bea
t_{\rm sub}(T_d)=6.36\times 10^{3}a_{-5}\exp\left[68100\K\left(\frac{1}{T_d}-\frac{1}{1800\K}\right)\right] \s\label{eq:tausub_sil}~~~~~
\ena
for silicate grains, and~\bea
t_{\rm sub}(T_d)=1.36a_{-5}\exp\left[81200\K\left(\frac{1}{T_d}-\frac{1}{3000\K}  \right)\right]\s\label{eq:tausub_gra}~~~~~
\ena
for graphite grains, where $a_{-5}=a/(10^{-5}\cm)$

At $T_{d}\sim$ 1800 K, one has $t_{\rm sub}\sim 1000$ s for silicates, which is rather short for astronomical~timescales.

For a point source of radiation, one can estimate the sublimation distance of dust grains, $r_{\rm sub}$,~from~the central source as
\begin{eqnarray}
R_{\rm sub}\simeq 0.015\left(\frac{L_{\rm UV}}{10^{9}L_{\odot}}\right)^{1/2}\left(\frac{T_{\rm sub}}{180
0\K}\right)^{-5.6/2} {\rm pc},\label{eq:rsub}
\end{eqnarray}
where $L_{\rm UV}$ is the luminosity in the optical and UV, which is roughly one half of the bolometric luminosity, and~$T_{\rm sub}$ is the dust sublimation temperature between 1500--1800 K for silicate and graphite material (\citet{1995ApJ...451..510S,1989ApJ...345..230G}). This relation is obtained using $U\sim 2.5\times 10^{6}(L_{UV}/10^{9}L_{\odot}) (R/1pc)^{-2}$, which corresponds to $T_{d}\sim 22.3 U^{1/5.6}\K$ for graphite grains and $L_{bol}=2L_{UV}$. By~setting $T_{d}=1800\K$, one obtains 
\bea
R_{\rm sub}=0.007\left(\frac{L_{\rm UV}}{10^{9}L_{\odot}}\right)^{1/2}\left(\frac{T_{\rm sub}}{180
0\rm K}\right)^{-5.6/2} {\rm pc}.
\ena

Equation~(\ref{eq:rsub}) implies the increase of the sublimation radius with the source luminosity. For~active galactic nuclei (AGN) of $L\sim 10^{13}$ $L_{\odot}$, the~sublimation region is $R_{\rm sub}\sim 1$ pc.

\subsection{Coulomb Explosions and Ion Field~Emission}
Grains subject to an extreme UV radiation field are positively charged due to photoemission of electrons. Photoelectric emission can rapidly increase the grain charge, $Z$, electric surface potential, $\phi=eZ/a$, and~electric field on the surface, $E=eZ/a^{2}=(\phi/a)$.

Tensile stress experienced by a surface element $\delta A$ of charge $\delta Z=\sigma_{Z} \delta A$ is given by
\bea
\mathcal{S}=\frac{\delta F}{\delta A}=\frac{e\delta Z E}{\delta A}=\frac{(\phi/a)^2}{4\pi},\label{eq:SZ} 
\ena
where $\sigma_{Z}=\delta Z/\delta A=Z/(4\pi a^{2})$ has been~used.

When the tensile stress exceeds the maximum tensile strength that the material can support, $\mathcal{S}_{\max}$,~the~grain will be disrupted by Coulomb explosions. Setting $\mathcal{S}=\mathcal{S}_{\max}$, we derive the maximum surface potential and charge that the grain still survives:
\bea
\phi_{\max}\simeq 1.06\times 10^{3}\left(\frac{\mathcal{S}_{\max}}{10^{10} \erg \cm^{-3}}\right)^{1/2}a_{-5} {\rm V},\label{eq:phimax}\\
Z_{\max}\simeq 7.4\times 10^4 \left(\frac{\mathcal{S}_{\max}}{10^{10}\erg\cm^{-3}}\right)^{1/2}a_{-5}^{2}.\label{eq:Zmax}
\ena

When a grain is positively charged to a sufficiently strong electric field, the~emission of individual ions (i.e., ion field emission) from the grain surface can occur. Experiments show that with an electric field of $\phi/a \sim 3\times 10^{8}V\cm^{-1}$, ion field emission already occurs for some metals (see Table~1 in \citet{1970PSSAR...1..513T}). Thus, grains may gradually be destroyed by ion field emission without Coulomb explosions in the case of ideal material with $\mathcal{S}_{\max}\sim 10^{11} \erg\cm^{-3}$ (i.e., $\phi_{\max}/a \sim 3\times 10^{8}{\rm V}\cm^{-1}$).

Coulomb explosions and ion field emission are efficient in intense extreme UV or X-ray radiation fields only (see \citet{2006ApJ...645.1188W}) or for grains moving with relativistic speeds through the ambient radiation field (\citet{2015ApJ...806..255H}).

\subsection{Thermal and Nonthermal~Sputtering}
When grains move rapidly relative to the gas, they are gradually eroded by sputtering induced by the bombardment of energetic gaseous atoms/ions. The~physics is as follows. Upon~bombardment, energetic ions penetrate the dust grain and interact with the target atoms, transferring part of their kinetic energy to the target atoms via Coulomb nuclear and electronic interactions. If~the target atoms receive kinetic energy larger than their binding energy, they can escape from the grain surface, and~the grain loses its~mass.

In hot gas, sputtering is induced by thermal protons and referred to as thermal sputtering (\citet{1979ApJ...231..438D}). Subject to a supersonic gas flow, sputtering is induced by the bombardment of atoms, which is called nonthermal sputtering (\citet{1994ApJ...433..797J}). 

Let $Y_{\rm sp}$ be the average sputtering yield per impinging atom (i.e., H and He) with speed $v$. The~number of target atoms sputtered by the H bombardment per second is given by 
\bea
\frac{dN_{\rm sp}}{dt}=n_{\rm H}v\pi a^{2}Y_{\rm sp}.
\ena

{The rate of mass loss due to thermal sputtering is given by
\bea
\frac{4\pi \rho a^{2}da}{dt} = -\frac{m_{\rm H}\bar{A}_{sp}dN_{\rm sp}}{dt}=-\bar{A}_{sp}m_{\rm H}n_{\rm H}\langle v\rangle\pi a^{2}Y_{\rm sp},
\ena
yielding
\bea
\frac{da}{dt}=\frac{\bar{A}_{sp}m_{\H}n_{\H}\langle v\rangle Y_{sp}}{4\rho},
\ena
where $\rho$ is the grain mass density, $\bar{A}_{\rm sp}$ is the average atomic mass number of sputtered atoms, and~$v$ is replaced by the mean thermal speed $\langle v\rangle =(8kT_{gas}/\pi m_{\H})^{1/2}$. The~thermal sputtering time is equal to 
\bea
\tau_{\rm sp}=\frac{a}{da/dt} 
\simeq 1.3\times 10^{4}\hat{\rho}\left(\frac{12a_{-5}}{\bar{A}_{sp}n_{1}}\right)\left(\frac{10^{6}~\rm K}{T_{\rm gas}}\right)^{1/2}\left(\frac{0.1}{Y_{\rm sp}}\right) \rm yr,
\ena
where $\hat{\rho}=\rho/(3\g\cm^{-3})$.
}

The decrease in the grain radius per time unit due to nonthermal sputtering by grain drifting at speed $v_{d}$ through the gas is given by (see e.g.,~\citet{2015ApJ...806..255H})
\bea
\frac{da}{dt}&=&\frac{n_{\H}m_{\H}v_{d}Y_{\rm sp}\bar{A}_{\rm sp}}{4\rho},
\ena
which implies a characteristic timescale of nonthermal sputtering,
\bea
\tau_{\rm sp}&=&\frac{a}{da/dt}=\frac{4\rho a}{n_{\H}m_{\H}v_{d}Y_{\rm sp}\bar{A}_{\rm sp}}\nonumber\\&\simeq& 1.9\times 10^{4}\hat{\rho}\left(\frac{12}{\bar{A}_{\rm sp}}\right)\left(\frac{a_{-5}}{n_{1}v_{2}}\right) \left(\frac{0.1}{Y_{\rm sp}}\right) \rm yr.\label{eq:tau_sp}
\ena

The sputtering yield, $Y_{\rm sp}$, depends on projectile energy and properties of grain material. Following~\citet{1994ApJ...431..321T}, the~sputtering yield is given by
\bea
Y_{\rm sp}(E)=4.2\times 10^{14}\frac{\alpha S_{n}(E)}{U_{0}}\left(\frac{R_{p}}{R}\right)\left[1-\left(\frac{E_{\rm th}}{E}\right)^{2/3}\right]\left(1-\frac{E_{\rm th}}{E}\right)^{2},
\ena
where $U_{0}$ is the binding energy of dust atoms, $\alpha\simeq 0.3(M_{2}/M_{1})^{2/3}$  for $0.5<M_{2}/M_{1}<10$, $\alpha\approx 0.1$ for $M_{2}/M_{1}<0.5$, and~$E_{\rm th}$ is the threshold energy for sputtering given by 
\bea
E_{\rm th}&=&\frac{U_{0}}{g(1-g)} {~\rm for ~} M_{1}/M_{2}\le 0.3,\\
E_{\rm th}&=&8U_{0}\left(\frac{M_{1}}{M_{2}}\right)^{1/3} {~\rm for ~} M_{1}/M_{2}> 0.3,
\ena
and $g=4M_{1}M_{2}/(M_{1}+M_{2})^{2}$ is the maximum energy transfer of a head-on elastic collision. The~factor $R_{p}/R$ is the ratio of the mean projected range to the mean penetrated path length, as~given by \citet{1984NIMPB...2..587B}
\bea
\frac{R_{p}}{R}=\left(K\frac{M_{2}}{M_{1}}+1 \right)^{-1},
\ena
where $K$ is a free parameter, and~$K=0.1$ and $0.65$ for silicate and graphite grains, respectively (see \citet{1994ApJ...431..321T}).

\subsection{Grain~Shattering}
A grain moving in the gas has a chance to hit another grain, resulting in grain coagulation or grain shattering, depending on their relative velocity. The~threshold velocity for grain shattering depends on the grain size as (\citet{Chokshi:1993p3420})
\bea
v_{\rm shat}\simeq 6\left(\frac{a}{10^{-5}\cm}\right)^{-5/6} \rm m\s^{-1}.
\ena

If the relative grain velocity $v_{\rm gg}<v_{\rm shat}$, the~grains collide and stick together. For~$v_{\rm gg}>v_{\rm shat}$, collisions at high velocity produce shock waves inside the grains and shatter them in smaller fragments. For~$v_{\rm gg}>20\rm$ km$\s^{-1}$, the~evaporation of dust grain occurs when a part of grains is heated to evaporation~temperatures.

Grain shattering is expected to be efficient in breaking large grains into nanoparticles in magnetized shocks (e.g., \citet{1996ApJ...469..740J}).

The destruction time by grain shattering can be estimated by the mean time between two successive collisions:
\begin{eqnarray}
\tau_{\rm gg}=\frac{1}{\pi a^{2} n_{\rm gr}v_{\rm drift}}=\frac{4\rho a M_{g/d}}{3n_{\rm H}m_{\rm H}v_{\rm drift}}\simeq 7.6\times 10^{4}\hat{\rho}a_{-5}n_{1}^{-1}v_{\rm drift,3}^{-1}{~\rm yr},
\end{eqnarray}
where $n_{\rm gr}$ is the number density of dust grains, $M_{g/d}\sim n_{\H}m_{\H}/n_{gr}m_{gr}=100$ with $m_{gr}=4\pi a^{3}\rho/3$ is the gas-to-dust mass ratio, and~we have assumed the single-grain size~distribution.

\section{Rotational Disruption of Dust Grains by Radiative~Torques}\label{sec:theory}
In this section, we review a new mechanism of dust destruction, so-called RAdiative Torque Disruption (RATD), which is associated to grain suprathermal rotation by~RATs. 

\subsection{Radiative Torques of Irregular~Grains}
Let $u_{\lambda}$ be the spectral energy density of radiation field at wavelength $\lambda$. The~energy density of the radiation field is then $u_{\rm rad}=\int u_{\lambda}d\lambda$. To~describe the strength of a radiation field, let define $U=u_{\rm rad}/u_{\rm ISRF}$ with 
$u_{\rm ISRF}=8.64\times 10^{-13}\erg\cm^{-3}$ being the energy density of the average ISRF in the solar neighborhoord (\citet{1983A&A...128..212M}). Thus, the~typical value for the ISRF is $U=1$. The~dust temperature can be approximately given by $T_{d}=T_{0}U^{4+\beta}$ with $\beta$ the dust opacity index ($\kappa_{d}\propto \lambda^{-\beta}$) and $T_{0}$ the grain temperature at $U=1$. Approximately, one has $T_{0}=16.4$ K and $\beta= 2$ for silicates, and~$T_{0}=23.5$ K and $\beta= 1.5$ for graphite (\citet{2011piim.book.....D}).

Let $a_{\rm eff}=(3V/4\pi)^{1/3}$ be the effective size of the dust grain of irregular shape with volume $V$. Such an irregular grain exposed to an anisotropic radiation field experiences radiative torque (RAT) due to differential absorption and scattering of left-handed and right-handed photons. The~magnitude of RATs is defined as
\bea
{\Gamma}_{\lambda}=\pi a_{\rm eff}^{2}
\gamma_{\rm rad} u_{\lambda} \left(\frac{\lambda}{2\pi}\right){Q}_{\Gamma},\label{eq:GammaRAT}
\ena
where $\gamma_{\rm rad}$ is the anisotropy degree of the radiation field, and~${Q}_{\Gamma}$ is the RAT efficiency (\citet{1996ApJ...470..551D}). {$\gamma_{\rm rad}\approx 0.1$ for the ISRF (\citet{1996ApJ...470..551D}), $\gamma_{\rm rad}\sim$ 0.3--0.7 for molecular clouds (\citet{2007ApJ...663.1055B}), and~$\gamma_{\rm rad}=1$ for unidirectional radiation fields (e.g., from~a nearby star).} 

\begin{figure}[H]
\includegraphics[width=0.5\textwidth]{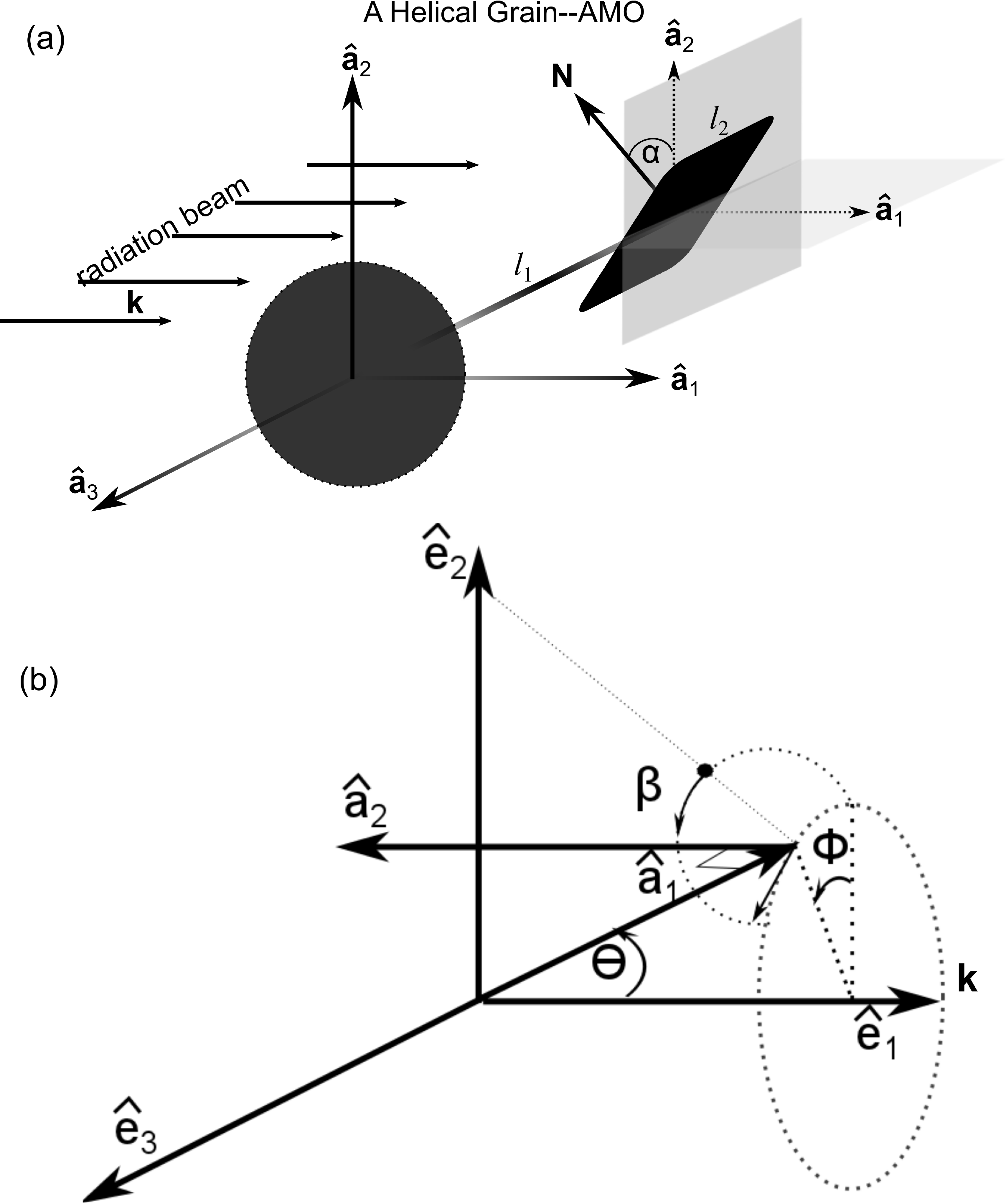}
\includegraphics[width=0.5\textwidth]{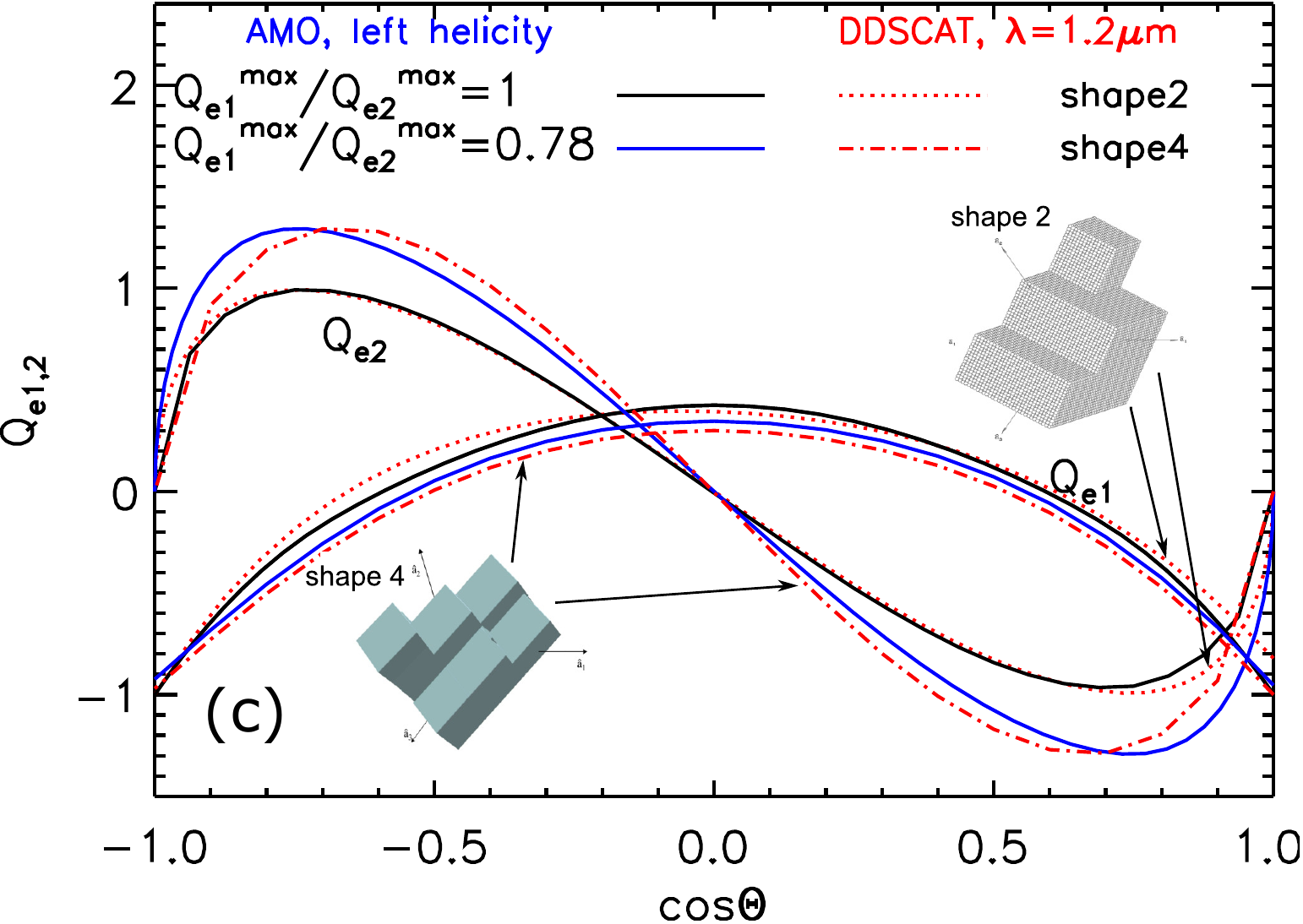}
\caption{Panel (\textbf{a}): A helical grain model described by three principal axes, $\hat{\ba}_{1}\hat{\ba}_{3}\hat{\ba}_{3}$, proposed by \citet{2007MNRAS.378..910L} to calculate analytically Radiative Torques (RATs), comprising an oblate spheroid and a massless mirror attached to the spheroid. The~normal vector of the mirror is titled by an angle $\alpha$ with respect to the principal axis $\hat{\ba}_{2}$. Panel (\textbf{b}): {Scattering coordinate frame of reference used for calculations of RATs}, described by three axes, $\hat{\be}_{1}\hat{\be}_{2}\hat{\be}_{3}$ where $\hat{\be}_{1}$ is defined along ${\bf k}$, and~$\hat{\be}_{2}\hat{\be}_{3}$ form a plane perpendicular to $\hat{\be}_{1}$. $\Theta$ is the angle between the axis of maximum moment of inertia, $\hat{\ba}_{1}$, with~the radiation direction ${\bf k}$, $\Phi$ is the precession angle of $\hat{\ba}_{1}$ around ${\bf k}$, and~$\beta$ is the angle that describes the rotation of the grain around $\hat{\ba}_{1}$. Panel (\textbf{c}): comparison of the functional form of RATs from AMO 
and numerical computation using DDSCAT. 
From~\citet{2007MNRAS.378..910L}.}
\label{fig:AMO}
\end{figure}

{A helical grain model suggested by \citet{2007MNRAS.378..910L} to obtain analytical formulae of RATs is shown in Figure~\ref{fig:AMO} (panel (a)). The~components of RATs as functions of the angle $\Theta$ (panel~(b)) obtained from AMO are in good agreement with numerical results using DDSCAT (panel (c)).}

\begin{figure}[H]
\includegraphics[width=0.5\textwidth]{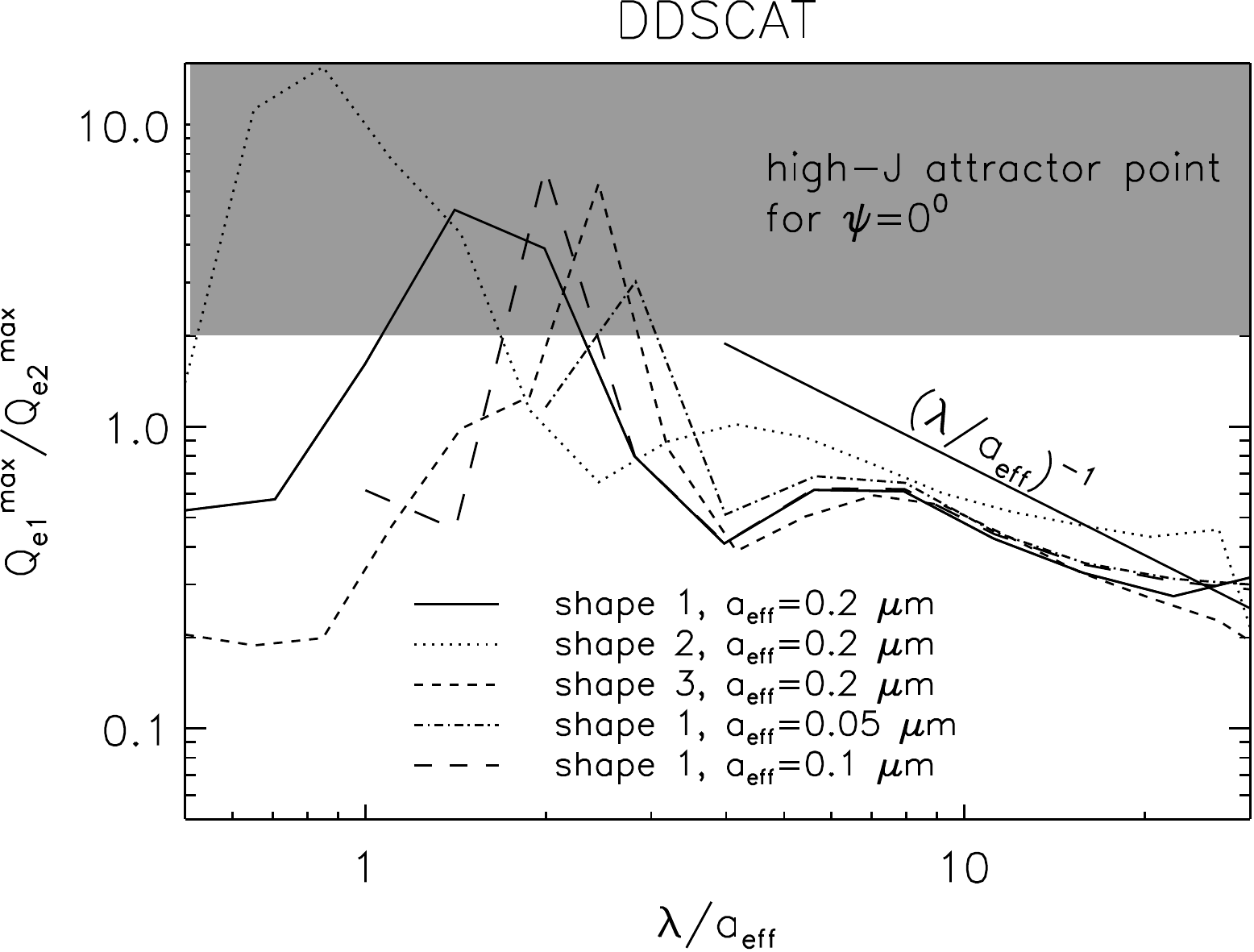}
\includegraphics[width=0.5\textwidth]{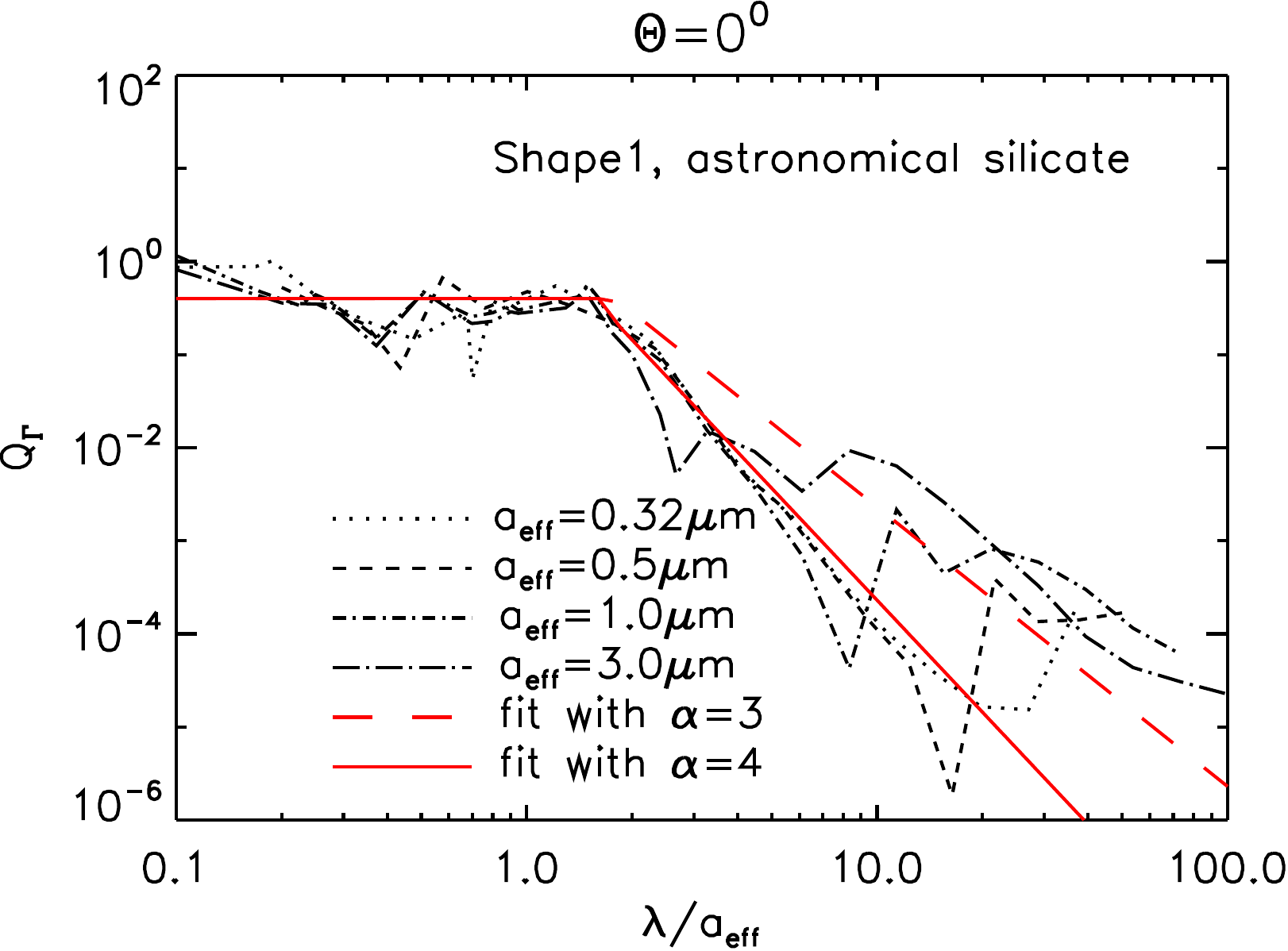}
\caption{Radiative torques computed for the different grain sizes and shapes using DDSCAT from \citet{2007MNRAS.378..910L}. Left and right panels show the ratio of maximum torque components and RAT magnitude as function of $\lambda/a_{\rm eff}$. Astronomical silicate is considered. Power-law fits to the computed results are shown in red~lines.}
\label{fig:RAT_DDA}
\end{figure}

Numerical calculations of RATs for several shapes and different optical constants using the DDSCAT code (\citet{Draine:2004p6718}) by \citet{2007MNRAS.378..910L} find slight differences in RATs among the realization (see Figure~\ref{fig:RAT_DDA}). The~magnitude of RAT efficiency, $Q_{\Gamma}$ can be approximated by a power-law (\citet{2007MNRAS.378..910L}):
\bea
Q_{\Gamma}\sim 0.4\left(\frac{{\lambda}}{1.8a_{\rm eff}}\right)^{\eta},\label{eq:QAMO}
\ena
where $\eta=0$ for $\lambda \lesssim 1.8a_{\rm eff}$  and $\eta=-3$ for $\lambda > 1.8a_{\rm eff}$. 

\begin{figure}[H]
\includegraphics[width=1.0\textwidth]{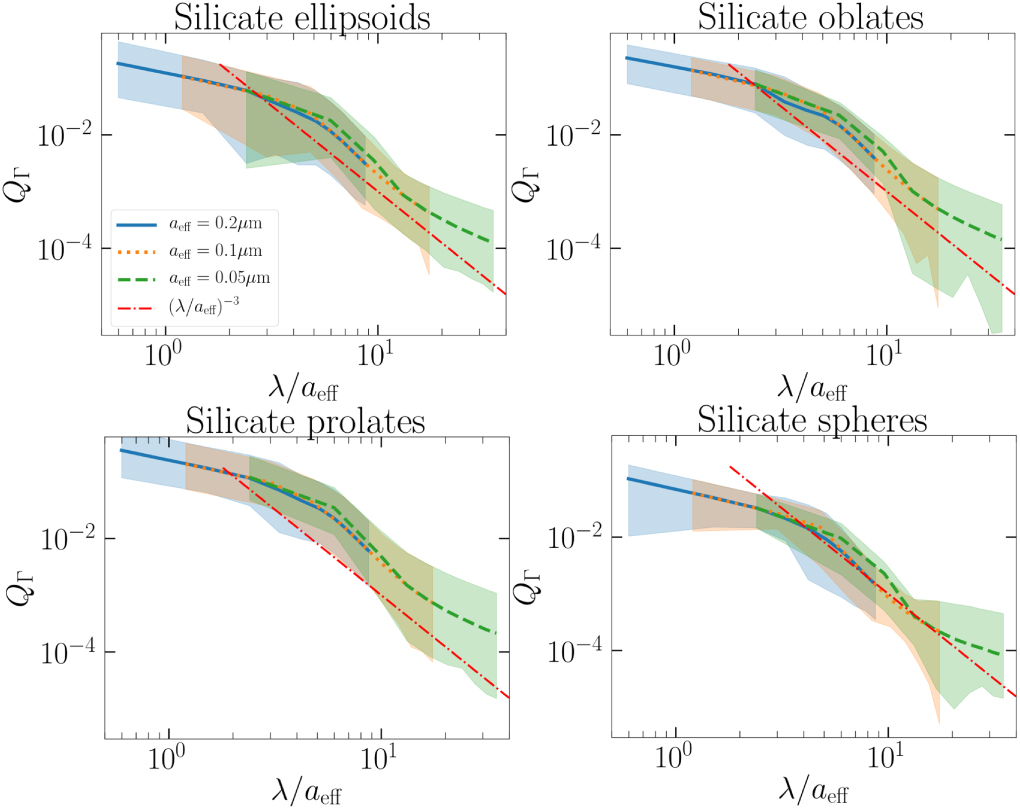}
\caption{RATs computed by T-matrix method for the numerous grain shapes of silicate material from \citet{Herranen:2019kj}. A~power-law fit to the numerical results of slope $\eta=-3$ is given for $\lambda/a_{\eff}\gtrsim 1$.}
\label{fig:RAT_Tmatrix}
\end{figure}

Recently, \citet{Herranen:2019kj} calculated RATs for an extensive sample of grain shapes using the T-matrix method, as~shown in Figure~\ref{fig:RAT_Tmatrix}. We can see that the analytical fit (Equation (\ref{eq:QAMO})) is in good agreement with their numerical calculations. Therefore, one can use Equation~(\ref{eq:QAMO}) for the different grain compositions and grain shapes, and~the difference is an order of~unity

\unskip

\unskip

Let $\overline{\lambda}=\int \lambda u_{\lambda}d\lambda/u_{\rm rad}$ be the mean wavelength of the radiation spectrum. For~the ISRF, \mbox{$\overline{\lambda}=1.2\upmu$m}. The~average radiative torque efficiency over the radiation spectrum is defined as
\bea
\overline{Q}_{\Gamma} = \frac{\int \lambda Q_{\Gamma}u_{\lambda} d\lambda}{\int \lambda u_{\lambda} d\lambda}.
\ena

For interstellar grains with $a_{\rm eff}\lesssim \overline{\lambda}/1.8$, $\overline{Q}_{\Gamma}$ can be approximated to (\citet{2014MNRAS.438..680H})
\bea
\overline{Q}_{\Gamma}\simeq 2\left(\frac{\overline{\lambda}}{a_{\rm eff}}\right)^{-2.7}\simeq 2.6\times 10^{-2}\left(\frac{\overline{\lambda}}{0.5\upmu \text{m}}\right)^{-2.7}a_{-5}^{2.7},\label{eq:QRAT_avg}
\ena
and $\overline{Q_{\Gamma}}\sim 0.4$ for $a_{\rm eff}>\overline{\lambda}/1.8$. 

For convenience, let $a_{\rm trans}=\bar{\lambda}/1.8$ be the transition size of grains from a flat to the power-law stage of RATs. Plugging $\overline{Q}_{\Gamma}$ into Equation~(\ref{eq:GammaRAT}) yields the radiative torque averaged over the radiation~spectrum,
\bea
\Gamma_{\rm RAT}=\pi a_{\rm eff}^{2}
\gamma_{\rm rad} u_{\rm rad} \left(\frac{\overline{\lambda}}{2\pi}\right)\overline{Q}_{\Gamma}
\simeq  5.8\times 10^{-29}a_{-5}^{4.7}\gamma_{\rm rad}U\overline{\lambda}_{0.5}^{-1.7}\erg,\label{eq:Gamma_avg1}
\ena
for $a_{\rm eff}\lesssim a_{\rm trans}$, and~\bea
\Gamma_{\rm RAT}\simeq & 8.6\times 10^{-28}a_{-5}^{2}\gamma_{\rm rad}U\overline{\lambda}_{0.5}\erg,\label{eq:Gamma_avg2}
\ena
for $a_{\rm eff}> a_{\rm trans}$, where $\overline{\lambda}_{0.5}=\overline{\lambda}/(0.5\upmu \text{m})$.

\subsection{Suprathermal Rotation of Dust Grains Induced by~RATs}
\unskip
\subsubsection{Rotational~Damping}
The well-known damping process for a rotating grain is sticking collisions with gas species, followed by their thermal evaporation. Thus, for~a gas with $90\%$ of H and $10\%$ of He in abundance, the~characteristic damping time is given by
\bea
\tau_{\gas}&=&\frac{3}{4\sqrt{\pi}}\frac{I_{1}}{1.2n_{\rm H}m_{\rm H}
v_{\rm th}a_{\rm eff}^{4}}\nonumber\\
&\simeq& 8.74\times 10^{4}\alpha_{1}a_{-5}\hat{\rho}\left(\frac{30\cm^{-3}}{n_{\H}}\right)\left(\frac{100\K}{T_{\gas}}\right)^{1/2}~{\rm yr},~~
\ena
where $v_{\rm th}=\left(2k_{\rm B}T_{\rm gas}/m_{\rm H}\right)^{1/2}$ is the thermal velocity of a gas atom of mass $m_{\rm H}$ in a plasma with temperature $T_{\gas}$ and density $n_{\H}$, and~$I_{1}=8\pi \rho \alpha_{1} a_{\rm eff}^{5}/15$ is the grain inertia moment with $\alpha_{1}$ being a geometrical factor of order unity (\citet{2009ApJ...695.1457H,1996ApJ...470..551D}). In~the following, we assume $\alpha_{1}=1$ for simplicity, and~set $I=I_{1}$, $a=a_{\rm eff}$.

IR photons emitted by the grain carry away part of the grain's angular momentum, resulting in the damping of the grain rotation. The~characteristic time of the IR damping is $\tau_{\rm IR}=\tau_{\gas}F_{\rm IR}$ where $F_{\rm IR}$ is a dimensionless coefficient . For~strong radiation fields or not very small sizes, grains can achieve an equilibrium temperature, and~$F_{\rm IR}$ is given by (see \citet{1998ApJ...508..157D}),
\bea
F_{\rm IR}\simeq \left(\frac{0.4U^{2/3}}{a_{-5}}\right)
\left(\frac{30 \cm^{-3}}{n_{\H}}\right)\left(\frac{100 \K}{T_{\gas}}\right)^{1/2}.\label{eq:FIR}
\ena 

Other rotational damping processes include plasma drag, ion collisions, and~electric dipole emission. These processes are mostly important for polycyclic aromatic hydrocarbons (PAHs) and very small grains (\citet{1998ApJ...508..157D,Hoang:2010jy,2011ApJ...741...87H}). Thus, the~total rotational damping rate of grains by gas collisions and IR emission can be written as
\bea
\tau_{\rm damp}^{-1}=\tau_{\gas}^{-1}+\tau_{\rm IR}^{-1}=\tau_{\gas}^{-1}(1+ F_{\rm IR}).\label{eq:taudamp}
\ena

For intense radiation fields with $U\gg 1$ and not very dense gas, one has $F_{\rm IR}\gg 1$. Therefore, $\tau_{\rm damp}\sim \tau_{\gas}/F_{\rm IR}\sim a_{-5}^{2}U^{2/3}$, which does not depend on the gas properties. In~this case, the~only damping process is the IR emission. For~dense environments and weak radiation fields, $F_{\rm IR}\ll 1$, and~gas damping~dominates.

\subsubsection{Maximum Grain Angular Velocity Spun-Up by~RATs}
Assuming a perfect internal alignment of the grain axis with angular momentum, the~equation of motion is given by
\bea
\frac{I_{1}d\omega}{dt}=\Gamma_{\rm RAT}-\frac{I_{1}\omega}{\tau_{\rm damp}},
\ena
where $\tau_{\rm damp}$ is given by Equation~(\ref{eq:taudamp}). The~assumption is valid for suprathermally rotating grains due to internal relaxation (\citet{1979ApJ...231..404P,1999ApJ...520L..67L,2009ApJ...697.1316H,2009ApJ...695.1457H,2014MNRAS.438..680H}).

For the radiation source with stable luminosity, the~radiative torque $\Gamma_{\rm RAT}$ is constant, and~the grain velocity is steadily increased over time. The~maximum angular velocity of grains spun-up by RATs is given by
\bea
\omega_{\rm RAT}=\frac{\Gamma_{\rm RAT}\tau_{\rm damp}}{I_{1}}.~~~~~\label{eq:omega_RAT0}
\ena.

For strong radiation fields with $U\gg 1$, such as $F_{\rm IR}\gg 1$, plugging $\Gamma_{\rm RAT}$ (Equations (\ref{eq:Gamma_avg1}) and (\ref{eq:Gamma_avg2})) and $\tau_{\rm damp}$ (Equation (\ref{eq:taudamp})) into the above equation, one obtains
\bea
\omega_{\rm RAT}&\simeq &7.1\times 10^{7}\gamma_{\rm rad,-1} a_{-5}^{1.7}U^{1/3}\bar{\lambda}_{0.5}^{-1.7}\rad\s^{-1},~~~\label{eq:omega_RAT}
\ena
for grains with $a\lesssim a_{\rm trans}$, and~\bea
\omega_{\rm RAT}&\simeq &1.1\times 10^{8}\frac{\gamma_{\rm rad,-1}}{a_{-5}}U^{1/3}\bar{\lambda}_{0.5}\rad\s^{-1},~~~
\ena
for grains with $a>a_{\rm trans}$.

For a general radiation field, the~maximum rotation rate induced by RATs is given by
\bea
\omega_{\rm RAT}\simeq 3.2\times 10^{7}\gamma_{\rm rad,-1} a_{-5}^{0.7}\bar{\lambda}_{0.5}^{-1.7}
\left(\frac{U}{n_{1}T_{2}^{1/2}}\right)\left(\frac{1}{1+F_{\rm IR}}\right)\rad\s^{-1},~~~\label{eq:omega_RAT}
\ena
for grains with $a\lesssim a_{\rm trans}$, and~\bea
\omega_{\rm RAT}\simeq 1.6\times 10^{8}\gamma_{\rm rad,-1}a_{-5}^{-2}\bar{\lambda}_{0.5}
\left(\frac{U}{n_{1}T_{2}^{1/2}}\right)\left(\frac{1}{1+F_{\rm IR}}\right)\rad\s^{-1},~~~
\ena
for grains with $a> a_{\rm trans}$. Here $\gamma_{\rm rad,-1}=\gamma_{\rm rad}/0.1$ is the anisotropy of radiation field relative to the typical anisotropy of the diffuse interstellar radiation~field.

\subsection{Centrifugal Stress Due to Grain~Rotation}

We assume that the grain is rotating around the axis of maximum inertia moment, denoted by z-axis, with~angular velocity $\omega$. This assumption is valid for suprathermal rotating grains in which internal relaxation can rapidly induce the perfect alignment of the axis of the major inertia with the angular momentum which corresponds to the minimum rotational energy state \citet{1979ApJ...231..404P}. Let us consider a slab $dx$ at distance $x$ from the center of mass. The~average tensile stress due to centrifugal force $dF_{c}$ acting on a plane located at distance $x_{0}$ is equal to
\bea
dS=\frac{\omega^{2}x dm}{\pi (a^{2}-x_{0}^{2})}= \frac{\rho \omega^{2}(a^{2}-x^{2})xdx}{a^{2}-x_{0}^{2}},
\ena
where the mass of the slab $dm=\rho dAdx$ with $dA=\pi(a^{2}-x^{2})$ the area of the circular~slab.  

The surface average tensile stress is then given by
\bea
S_{x}&=&\int_{x_{0}}^{a} dS_{x}=\frac{\rho\omega^{2}a^{2}}{2}\int_{x_{0}/a}^{1}\frac{(1-u)du}{1-u_{0}}\nonumber\\
&=& \frac{\rho \omega^{2}a^{2}}{4}\left(\frac{(1-u_{0})^{2}}{1-u_{0}}\right)=\frac{\rho \omega^{2}a^{2}}{4}\left[1-\left(\frac{x_{0}}{a}\right)^{2}\right],~~~\label{eq:Smax}
\ena
where $u=(x/a)^{2}$. 

Equation~(\ref{eq:Smax}) reveals that the tensile stress is maximum at the grain center and decreases with decreasing the mantle thickness $(a-x_{0})$. 

By plugging the numerical numbers into Equation~(\ref{eq:Smax}), one obtains
\bea
S_{x}\simeq 7.5\times 10^{9}\hat{\rho}\omega_{10}^{2}a_{-5}^{2} \left[1-\left(\frac{x_{0}}{a}\right)^{2} \right] \erg \cm^{-3},\label{eq:Sx}
\ena
where $\omega_{10}=\omega/(10^{10}\rm rad\s^{-1})$. The~equation reveals that the centrifugal stress is maximum at the grain center ($x_{0}=0$) and decreases with increasing $x_0$.

\subsection{Tensile Strength of~Dust}
Mechanical properties of dust grains is described by its tensile strength, that is,~the maximum strength that the grain material still withstands against an applied tension, denoted by $S_{\max}$. The~tensile strength depends on the internal structure of dust as well as its composition, which is poorly known for astrophysical~dust. 

In general, compact grains are expected to have higher $S_{\rm max}$ than composite/fluffy grains. For~instance, polycrystalline bulk solid has $S_{\max} \sim 10^{9}$--$10^{10} \erg \cm^{-3}$ (\citet{1974ApJ...190....1B,1979ApJ...231..438D}), while ideal materials, that is,~diamond, have $S_{\max} \sim 10^{11} \erg \cm^{-3}$ (see \citet{Hoang:2019da}). large grains of $a\gtrsim 0.1\upmu$m) likely have a composite structure, such that the tensile strength is lower, with~$S_{\rm max} \sim 10^{6}$--$10^{8}\erg\cm^{-3}$, depending on the radius of monomers (\citet{2019ApJ...876...13H}). Nanoparticles or VSGs are likely to have compact structures, thus, their tensile strengths are expected to be large of $S_{\max}\gtrsim 10^{9}\erg\cm^{-3}$.

We now consider a composite grain model as proposed by \citet{1989ApJ...341..808M}. This~composite model relies on the fact that upon entering the ISM, original silicate and carbonaceous grains are shattered (e.g., by~shocks) into small fragments. The~subsequent collisions of these fragments reform interstellar composite grains. Following \citet{1989ApJ...341..808M}, individual particles are assumed to be compact and spherical of radius $a_{p}$. Particles can be of silicate or carbonaceous materials. 
Let $P$ be the porosity which is defined such that the mass density of the porous grain is $\rho=\rho_{0}(1-P)$ with $\rho_{0}$ being the mass density of fully compact grain. The~value $P=0.2$ indicates an empty volume of $20\%$. 

Let $\bar{E}$ be the mean intermolecular interaction energy at the contact surface between two particles and $h$ be the mean intermolecular distance at the contact surface. Let $\beta$ be the mean number of contact points per particle between 1--10. The~volume of interaction region is $V_{\rm int}=(2ha_{p}^{2})$. \mbox{Following \citet{1995A&A...295L..35G},} one can estimate the tensile strength as given by the volume density of interaction energy
\bea
S_{\rm max}=3\beta(1-P)\frac{\bar{E}}{2ha_{p}^{2}}.\label{eq:Smax_comp1}
\ena

We can write $\bar{E}=\alpha 10^{-3}$ eV where $\alpha$ is the coefficient of order unity \mbox{when the interaction between} contact particles is only van der Waals forces. The~contribution of chemical bonds  between ice~molecules can increase the value of $\alpha$. The~tensile strength can be rewritten as (see \citet{1997A&A...323..566L}):
\bea
S_{\max}&\simeq& 1.6\times 10^{6}(1-P)\left(\frac{\beta}{5}\right)\left(\frac{\bar{E}}{10^{-3}\rm eV}\right)\left(\frac{\alpha}{1}\right)\nonumber\\
&&\times \left(\frac{a}{5\rm nm}\right)^{-2}\left(\frac{0.3\rm nm}{\H}\right) ~\erg\cm^{-3}.\label{eq:Smax_comp}
\ena

The tensile strength decreases rapidly with increasing the particle radius, as~$a_{p}^{-2}$, and~decreases with increasing the porosity $P$. In~the following, we fix the porosity $P=0.2$, as~previously assumed for {\it Planck} data modeling (\citet{Guillet:2017hg}), and~adopt the typical value of $\alpha=1$. 

Numerical simulations for porous grain aggregates from \citet{2019ApJ...874..159T} find that the tensile strength decreases with increasing the monomer radius as
\bea
S_{\max}\sim 0.12\frac{F_{c}}{r_{0}^{2}}\phi_{ini}^{1.8}
\sim 6\times 10^{6} \left(\frac{\gamma}{100 mJ m^{-2}}\right)\left(\frac{r_{0}}{0.1\upmu \text{m}}\right)^{-1}\phi_{ini}^{1.8} \erg\cm^{-3},
\ena
where $F_{c}$ is the maximum force needed to separate two sticking monomers,
$\gamma$ is the surface energy per unit area of the material, and~$r_{0}$ is the monomer radius, and~$\phi_{ini}$ is the initial volume filling~factor.
 
\subsection{Grain Disruption Size and Disruption~Time}\label{sec:disr}

When the grain rotation rate is sufficiently high such as the tensile stress, $S_{x}$ (Equation (\ref{eq:Smax})), can~exceed the tensile strength, $S_{\rm max}$, the~grain is instantaneously disrupted into fragments. The~critical angular velocity is obtained by setting $S_{x}=S_{\rm max}$:
\bea
\omega_{\rm disr}=\frac{2}{a}\left(\frac{S_{\max}}{\rho} \right)^{1/2}
\simeq \frac{3.6\times 10^{8}}{a_{-5}}S_{\max,7}^{1/2}\hat{\rho}^{-1/2}~\rad\s^{-1},\label{eq:omega_cri}
\ena
where $S_{\max,7}=S_{\max}/(10^{7}\erg\cm^{-3})$, and~we assumed $x_{0}=0$, that is,~the disruption occurs along the plane going through the grain center (\citet{Hoang:2018es}).

For strong radiation fields such that $F_{\rm IR}\gg 1$, from~Equations~(\ref{eq:omega_RAT}) and (\ref{eq:omega_cri}), one can obtain the disruption grain size:
\bea
\left(\frac{a_{\rm disr}}{0.1\upmu \text{m}}\right)^{2.7}&\simeq&5.1\gamma_{\rad,-1}^{-1}U^{-1/3}\bar{\lambda}_
{0.5}^{1.7}S_{\max,7}^{1/2},~~~~~\label{eq:adisr_comp1}
\ena
for $a_{\rm disr}\le a_{\rm trans}$. 

For an arbitrary radiation field and $a\le a_{\rm trans}$, one obtains
\bea
\left(\frac{a_{\rm disr}}{0.1\upmu \text{m}}\right)^{1.7} \simeq 3.8\gamma_{\rm rad,-1}^{-1}\bar{\lambda}_
{0.5}^{1.7}S_{\max,7}^{1/2}
 (1+F_{\rm IR})\left(\frac{n_{1}T_{2}^{1/2}}{U}\right),~~~\label{eq:adisr_comp2}
\ena
which depends on the local gas density and temperature due to gas rotational~damping.

\begin{figure}[H]
\centering
\includegraphics[width=0.8\textwidth]{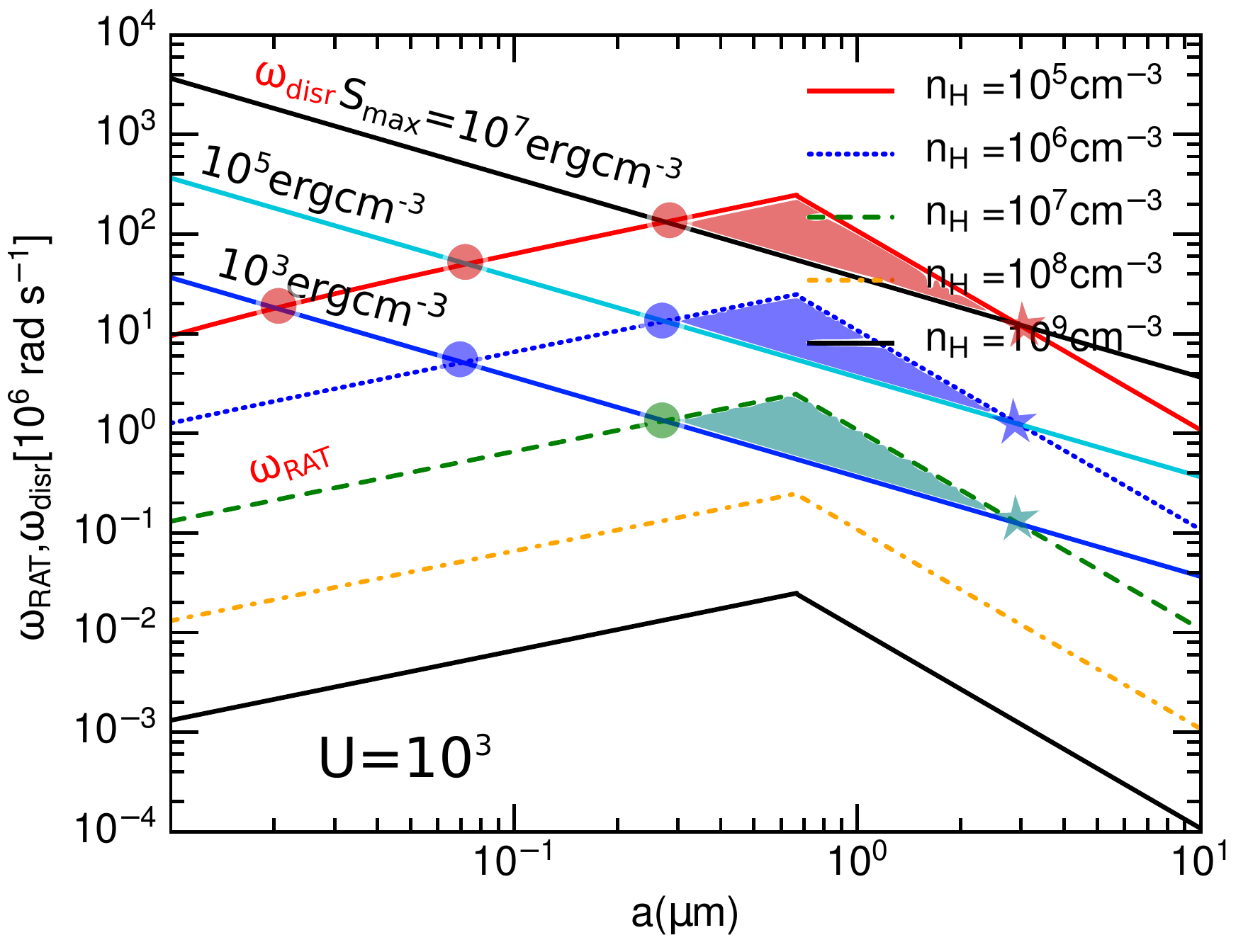}
\caption{The grain rotation rate spun-up by RATs ($\omega_{\rm RAT}$) and disruption rate ($\omega_{\rm disr}$) as functions of the grain size for different gas densities. The~radiation strength $U=10^{3}$ and three values of tensile strengths $S_{\max}=10^{3}, 10^{5}, 10^{7}\erg\cm^{-3}$ are considered. The~peak of $\omega_{\rm RAT}$ occurs at $a=a_{\rm trans}$. The~intersection of $\omega_{\rm RAT}$ and $\omega_{\rm disr}$ can occur at a lower grain size (marked by a circle) and an upper size (marked by a star), and~the shaded area denotes the range of grain sizes in which grains are disrupted by~RATs.}
\label{fig:omegaRAT_disr}
\end{figure}

Due to the decrease of the rotation rate for $a>a_{trans}$ (see Figure~\ref{fig:omegaRAT_disr}), there exist a maximum size of grains that can still be disrupted by centrifugal stress (\citet{2020ApJ...891...38H}):
\bea
a_{\rm disr,max}\simeq 5.0\gamma\bar{\lambda}_{0.5}\left(\frac{U}{n_{1}T_{2}^{1/2}}\right)\left(\frac{1}{1+F_{\rm IR}}\right)\hat{\rho}^{1/2}S_{\max,7}^{-1/2}~\upmu \text{m}.~~~~~\label{eq:adisr_up}
\ena 

For the standard parameters of the diffuse ISM of $U=1$, one gets $a_{\rm disr,max}\sim$ 5  $\upmu$m for the typical physical parameters in Equation~(\ref{eq:adisr_up}). This is much larger than the maximum grain size of $a_{\max}\sim$ 0.25--0.3 $\upmu$m obtained from modeling of observational data (\citet{Mathis:1977p3072,1995ApJ...444..293K,2009ApJ...696....1D}). So, all available grains of a $\gtrsim$ $a_{\rm disr}$ are disrupted. In~dense~ regions, grains~are expected to grow to large sizes due to coagulation and accretion \mbox{(e.g., \citet{Chokshi:1993p3420,Ossenkopf:1993p4016})}. Therefore, not all grains of $a\gtrsim a_{\rm disr}$ can be disrupted, and~we will find both $a_{\rm disr}$ and $a_{\rm disr,max}$ for grains in star-forming~regions.

The characteristic timescale for rotational desorption can be estimated as (\citet{Hoang:2019da}):
\bea
t_{\rm disr,0}=\frac{I\omega_{\rm disr}}{dJ/dt}=\frac{I\omega_{\rm disr}}{\Gamma_{\rm RAT}} \simeq 10^{5} (\gamma U)^{-1}\bar{\lambda}_{0.5}^{1.7}\hat{\rho}^{1/2}S_{\max,7}^{1/2}a_{-5}^{-0.7}{~ \yr}\label{eq:tdisr1}
\ena
for $a_{\rm disr}<a \lesssim a_{\rm trans}$, and~\bea
t_{\rm disr,0}\simeq 7.4(\gamma U)^{-1}\bar{\lambda}_{0.5}^{-1}\hat{\rho}^{1/2}S_{\max,7}^{1/2}a_{-5}^{2}{~\yr}\label{eq:tdisr2}
\ena
for $a_{\rm trans}<a<a_{\rm disr,max}$.

In Table~\ref{tab:destr}, we compare the timescale of RATD obtained for a strong radiation field with the time from various destruction mechanisms. It is obvious that RATD is the most efficient mechanism to destroy large grains, while the METD mechanism is efficient for nanoparticles. The~efficiency of RATD over thermal sublimation is straightforward. While thermal sublimation requires high energy to break molecular bonds within the grain of 10 eV, the~RATD only requires low energy to break the Van der Waals force between the monomers/constituents which is only 0.01~eV. 

\begin{table}[H]
\begin{center}
\caption{Characteristic timescales of dust destruction by different~mechanisms.}\label{tab:destr}
\begin{tabular}{l l } \toprule
{\textbf{Mechanism}} & {\textbf{{Timescale} (yr)}}\cr
\midrule
Rotational disruption (RATD) & $1.0 a_{-5}^{-0.7}\bar{\lambda}_{0.5}^{1.7}U_{6}^{-1}S_{\rm max,9}^{1/2}$\cr
Thermal sputtering & $1.3\times 10^{4}\hat{\rho}(12/\bar{A}_{sp})a_{-5}n_{1}^{-1}T_{6}^{-1/2}(0.1/Y_{\rm sp})$ \cr
Non-thermal sputtering & $1.9\times 10^{3}\hat{\rho}(12/\bar{A}_{sp})a_{-5}n_{1}^{-1}v_{\rm drift,3}^{-1}(0.1/Y_{\rm sp}) $ \cr
Grain-grain collision & $7.6\times 10^{4}\hat{\rho}a_{-5}n_{1}^{-1}v_{\rm drift,3}^{-1}$\cr
Rotational disruption (METD) & $1.2 \hat{\rho}a_{-7}^{4}n_{4}^{-1}v_{\rm drift,1}^{-3}S_{\rm max,9}$\cr
\bottomrule
\end{tabular}\\
\begin{tabular}{@{}c@{}} 
\multicolumn{1}{p{\textwidth -.88in}}{\footnotesize {{\it Notes}:~$a_{-5}=a/(10^{-5}~\rm cm), U_{6}=U/10^{6}$, $\bar{A}_{sp}$ mean atomic mass number of sputtered atoms}; {$S_{\rm max,9}=S_{\rm max}/(10^{9}~\rm erg~cm^{-3})$}; {$n_{1}=n_{\rm H}/(10~\rm cm^{-3}),T_{6}=T_{\rm gas}/(10^{6}~\rm K) $}; {$v_{\rm drift,3}=v_{\rm drift}/(10^{3}~\rm km~s^{-1})$,~and $Y_{\rm sp}$ sputtering yield}.}
\end{tabular}
\end{center}
\end{table}
\unskip
\vspace{-6pt}
\subsection{Example of RATD by a Point Radiation~Source}
For a point radiation source with bolometric luminosity $L$, the~radiation energy density at distance $d_{\rm pc}$ in units of pc is given by
\begin{eqnarray}
u_{\rm rad}=\int u_{\lambda}d\lambda =\int  \frac{L_{\lambda}}{4\pi c d^{2}}d\lambda
\simeq1.06\times 10^{-6}\left(\frac{L_{9}}{d_{\rm pc}^{2}}\right) \rm erg~ cm^{-3},\label{eq:urad}
\end{eqnarray}
where  $L_{9}=L/(10^{9}L_{\odot})$, and~the radiation strength is equal to
\begin{eqnarray}
U\simeq 1.2\times 10^{6}\left(\frac{L_{9}}{d_{\rm pc}^{2}}\right).\label{eq:Urad}
\end{eqnarray}

For a massive star of $L\sim 10^{5}L_{\odot}$ at distance $d=1$ pc, Equation~(\ref{eq:Urad}) implies $U\sim 10^{2}$, and~for a supernova of $L\sim 10^{9}L_{\odot}$, $U\sim 10^{6}$, which are much stronger than the standard~ISRF.

For a given radiation field of constant bolometric luminosity $L$ and mean wavelength $\bar{\lambda}$, one can calculate $\omega_{\rm RAT}$ for a grid of grain sizes, assuming the gas density ($n_{\rm H}$) and temperature ($T_{\rm gas}$) for the local environment. We calculate $a_{\rm disr}$ and $t_{\rm disr}$ for the physical parameters of the standard ISM with $n_{\rm H}=30~\rm cm^{-3}$ and $T_{\rm gas}=100~\rm K$, and~an H\,{\sc ii} region with $n_{\rm H}=1.0~\rm cm^{-3}$ and $T_{\rm gas}= 10^{6}~\rm K$.

\begin{figure}[H]
\centering
\includegraphics[width=1.0\textwidth]{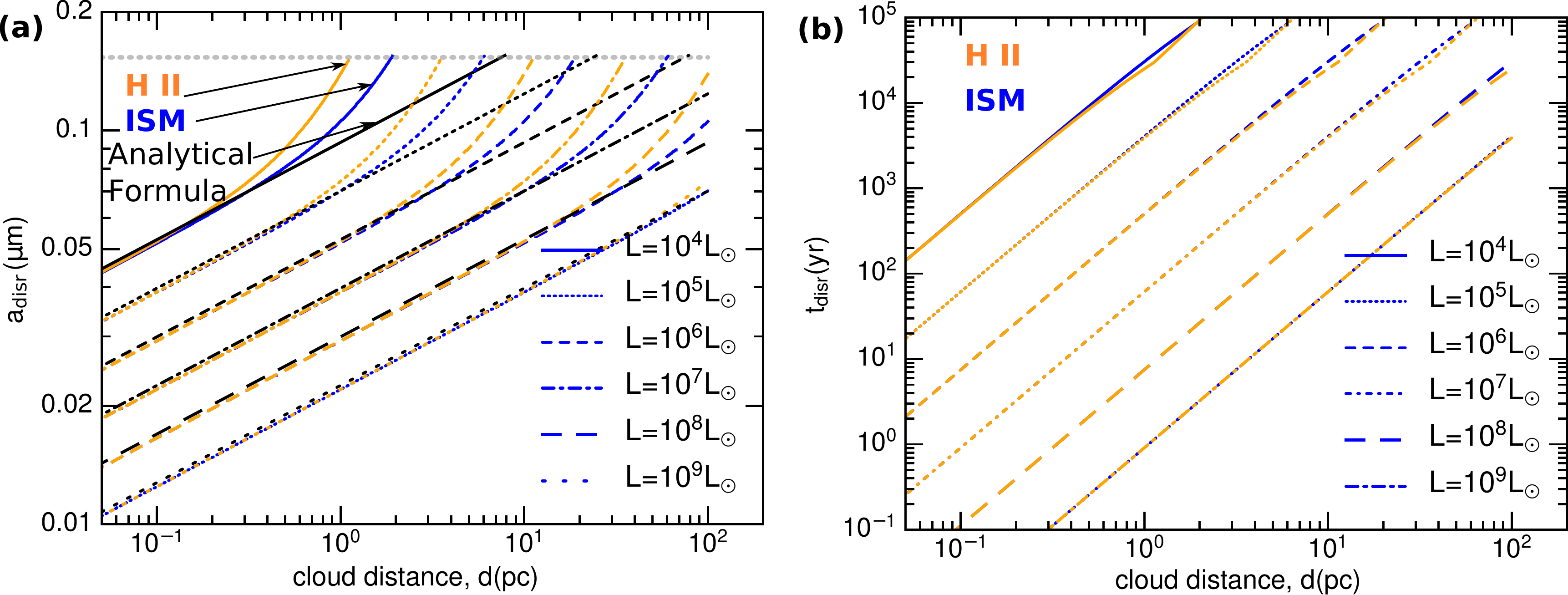}
\caption{{ Grain disruption size and disruption time vs. cloud distance from the central source for massive stars and young massive stellar clusters (YMSCs) of different luminosity, assuming~grain tensile strength $S_{\rm max}=10^{9}~\rm erg~cm^{-3}$}. Panel ({\bf a}): Grain disruption size vs. cloud distance computed for the interstellar medium (ISM) (blue lines) and H\,{\sc ii} regions (orange lines). Results obtained from an analytical formula in the absence of gas damping are shown in black lines. The~horizontal line in the top marks \mbox{$a_{\rm disr}=\bar{\lambda}/1.8$}. Panel ({\bf b}): Grain disruption time vs. cloud distance computed for the ISM and H\,{\sc ii} regions. The~disruption time is short, below~ $\sim$1 Myr for YMSCs of $L\sim 10^{6}-10^{9}L_{\odot}$. \mbox{From~\citet{Hoang:2019da}.}}
\label{fig:atdisr_dis}
\end{figure}

Figure~\ref{fig:atdisr_dis} (panel (a)) shows the grain disruption size as a function of the cloud distance for $L= 10^{4}$--$10^{9}L_{\odot}$ for the ISM (blue lines) and H\,{\sc ii} regions (orange lines). The~results obtained from an analytical formula where the grain rotational damping by gas collisions is disregarded is shown in black lines for comparison. The~disruption size $a_{\rm disr}$ increases rapidly with increasing cloud distance and reaches $a_{\rm disr}\sim \bar{\lambda}/1.8\sim 0.16~\upmu$m (marked by a horizontal line in the figure) at some distance. Beyond~this distance, grain disruption ceases to occur due to the decrease of radiation energy density (see Figure~\ref{fig:atdisr_dis}). For~$L= 10^{4}L_{\odot}$, which is typical for OB stars, we get $a_{\rm disr}\sim 0.1~\upmu$m for $d\sim 1$ $\rm pc$ for the ISM. For~more luminous stars of $L= 10^{6}L_{\odot}$, $a_{\rm disr}\sim 0.1~\upmu$m for $d\sim 10$ pc (see dashed line). For~a YMSC of $L=10^{9}L_{\odot}$, one obtains $a_{\rm disrp}\sim 0.05~\upmu$m for $d\sim 30$ pc, and~$a_{\rm disr}\sim 0.02~\upmu$m for $d\sim 1$ pc. 

For a given $L$, $a_{\rm disr}$ for the ISM and H\,{\sc ii} regions is similar at small distances. At~large distances from the source, $a_{\rm disr}$ for H\,{\sc ii} regions is larger than for the ISM and for the case without gas damping (black lines in Figure~\ref{fig:atdisr_dis}). The~reason is that at large distances, rotational damping by gas collisions becomes dominant over the rotational damping by infrared emission, resulting in the increase of $a_{\rm disr}$ with the gas damping rate which scales as $n_{\rm H}T_{\rm gas}^{1/2}$. This can also be seen through the increase in the critical radiation strength required to disrupt grains with the gas damping~rate. 

Figure~\ref{fig:atdisr_dis} (panel (b)) shows the disruption time $t_{\rm disr}$ of $a=a_{\rm disr}$ grains as a function of the cloud distance for the different values of $L$. The~disruption time increases rapidly with the cloud distance and decreases with increasing $L$. For~grains at 10 pc, one obtains $t_{\rm disr}\sim$ 50--30,000 yr for $L \sim 10^{9}$--$10^{6}L_{\odot}$.

\unskip

\section{Rotational Desorption of Ice Mantles by Radiative~Torques}\label{sec:ice}
Ice mantles are formed due to accretion of gas species on the grain surface in cold and dense regions of hydrogen density $n_{\H}=n(\H)+2n(\H_{2}) \sim 10^{3}$--$10^{5}\cm^{-3}$ or the visual extinction $A_{V}> 3$ (\citet{1983Natur.303..218W}). Irradiated by a nearby star or a young star located at the cloud center, icy grains are sublimated when heated to high temperatures of $T_{d}>100\K$. Here, we discuss the rotational desorption of ice mantles from the grain surface by centrifugal force due to~RATs.

\subsection{Rotational Desorption of Ice Mantles and Molecule~Desorption}

\begin{figure}[H]
\centering
\includegraphics[width=0.4\textwidth]{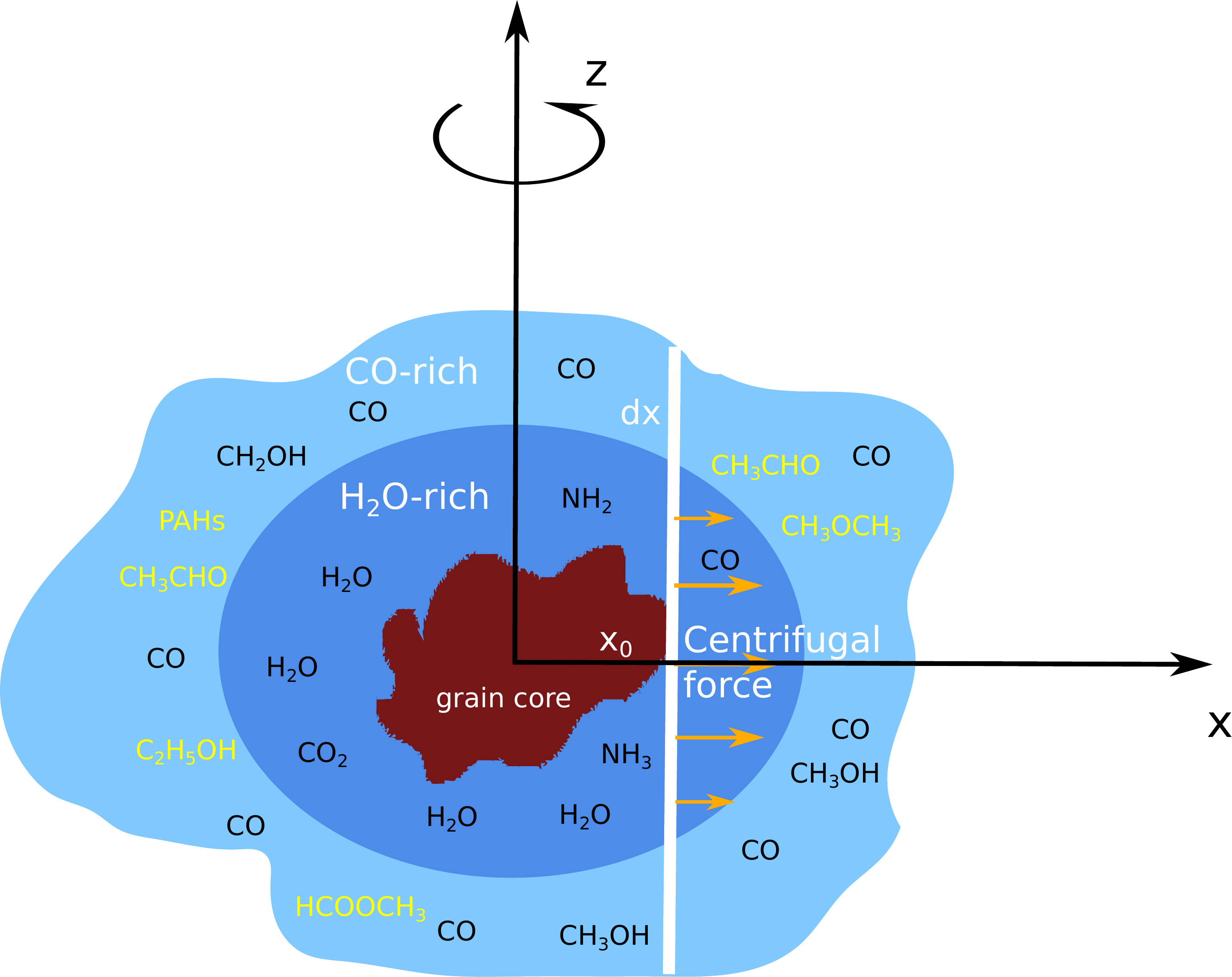}
\includegraphics[width=0.4\textwidth]{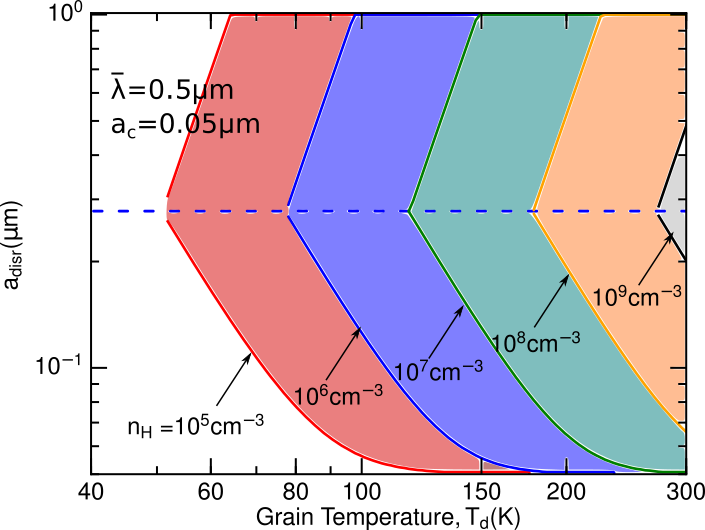}
\caption{Left panel: Illustration of a rapidly spinning core-mantle grain of irregular shape, comprising an icy water-rich (blue) and CO-rich (light blue) mantle layers. The~core is assumed to be compact silicate material, and~complex organic molecules are formed in the ice mantle of the core. Centrifugal~force field on a slab $dx$ is illustrated, which acts to pull off the ice mantle from the grain core at sufficiently fast rotation. Right panel: Range of desorption sizes of ice mantles, constrained by $a_{\rm disr}$ (lower boundary) and $a_{\rm disr, max}$ (upper boundary), as~a function of the grain temperature for the different gas densities for $\bar{\lambda}=0.5$ $\upmu \text{m}$, assuming a fixed core radius $a_{c}$ and the varying mantle thickness. The~horizontal dashed line denotes the transition grain size $a_{\rm trans}=\bar{\lambda}/1.8$. Shaded regions mark the range of grain sizes disrupted by RATD. From~\citet{2020ApJ...891...38H}.}
\label{fig:ice_disr}
\end{figure}

Here we consider a grain model consisting of an amorphous silicate core covered by a double-layer ice mantle (see Figure~\ref{fig:ice_disr}, left panel). Let $a_{c}$ be the radius of silicate core and $\Delta a_{m}$ be the average thickness of the mantle. The~exact shape of icy grains is unknown, but~we can assume that they have irregular shapes as required by strongly polarized H$_{2}$O and CO ice absorption features \mbox{(\citet{1996ApJ...465L..61C,2008ApJ...674..304W})}. Thus, one can define an effective radius of the grain, $a$, which is defined as the radius of the sphere with the same volume as the grain. The~effective grain size is $a \approx a_{c}+\Delta a_{m}$. The~silicate and carbonaceous cores are assumed to have a typical radius of 0.05  $\upmu \text{m}$ \citet{1989IAUS..135..345G}.

The tensile strength of the bulk ice is $S\sim 2\times 10^{7}\erg\cm^{-3}$ at low temperatures. As~the temperature increases to 200--300 K, the~tensile strength is reduced significantly to $5\times 10^{6}\erg\cm^{-3}$ \citet{Litwin:2012ii}. The~adhesive strength between the ice mantle and the solid surface has a wide range, depending on the surface properties (\citet{Itagaki:1983ud,Work:2018bu}). Here, we adopt a conservative value of $S_{\max}=10^{7}\erg\cm^{-3}$ for pure ice mantles for our numerical calculations. For~the grain core, a~higher value of $S_{\max}=10^{9}\erg\cm^{-3}$ is~adopted.

When the rotation rate is sufficiently high such as the tensile stress $S_{x}$ (Equation (\ref{eq:Smax})) can exceed the maximum limit of the ice mantle, $S_{\rm max}$, the~ice mantle is separated from the grain surface, which is termed {\it rotational desorption}. The~critical rotational velocity of the mantle desorption is determined by $S_{x}=S_{\rm max}$:
\bea
\omega_{\rm desr}=\frac{2}{a(1-x_{0}^{2}/a^{2})^{1/2}}\left(\frac{S_{\max}}{\rho_{\rm ice}} \right)^{1/2}\simeq \frac{6.3\times 10^{8}}{a_{-5}(1-x_{0}^{2}/a^{2})^{1/2}}\hat{\rho}_{\rm ice}^{-1/2}S_{\max,7}^{1/2}~\rad\s^{-1}.\label{eq:omega_disr}
\ena

Above, we assume that the grain is spinning along the principal axis of maximum inertia moment. This assumption is valid because internal relaxation within the rapidly spinning grain due to Barnett effect rapidly brings the grain axis to be aligned with its angular momentum (\citet{1979ApJ...231..404P,1999MNRAS.305..615R}).

The grain disruption size of ice mantles is given by
\bea
a_{\rm desr}\simeq 0.13\gamma^{-1/1.7}\bar{\lambda}_
{0.5}(S_{\max,7}/\hat{\rho}_{\rm ice})^{1/3.4} (1+F_{\rm IR})^{1/1.7}\left(\frac{n_{1}T_{2}^{1/2}}{U}\right)^{1/1.7}\upmu \text{m},~~~\label{eq:adisr_low}
\ena
for $a_{\rm disr}\lesssim a_{\rm trans}$ and $x_{0}\ll a$, which depends on the local gas density and temperature due to gas damping. The~equation indicates that all grains in the size range $a_{\rm trans}>a>a_{\rm disr}$ would be~disrupted. 

In the absence of rotational damping, the~characteristic timescale for rotational desorption of ice mantles can be estimated from Equations~(\ref{eq:tdisr1}) and (\ref{eq:tdisr2}):
\bea
t_{\rm desr,0}=\frac{I\omega_{\rm disr}}{\Gamma_{\rm RAT}}\simeq 6\times 10^{4}(\gamma U)^{-1}\bar{\lambda}_{0.5}^{1.7}\hat{\rho}_{\rm ice}^{1/2}S_{\max,7}^{1/2}a_{-5}^{-0.7}{~\rm yr}\label{eq:tdisr_ice}
\ena
for $a_{\rm disr}<a \lesssim a_{\rm trans}$, and~\bea
t_{\rm desr,0}\simeq& 4\times 10^{3}(\gamma U)^{-1}\bar{\lambda}_{0.5}^{-1}\hat{\rho}_{\rm ice}^{1/2}S_{\max,7}^{1/2}a_{-5}^{2}{~\rm yr}\label{eq:tdisr_ice2}
\ena
for $a_{\rm trans}<a<a_{\rm disr,max}$.

\begin{figure}[H]
\centering
\includegraphics[width=0.9\textwidth]{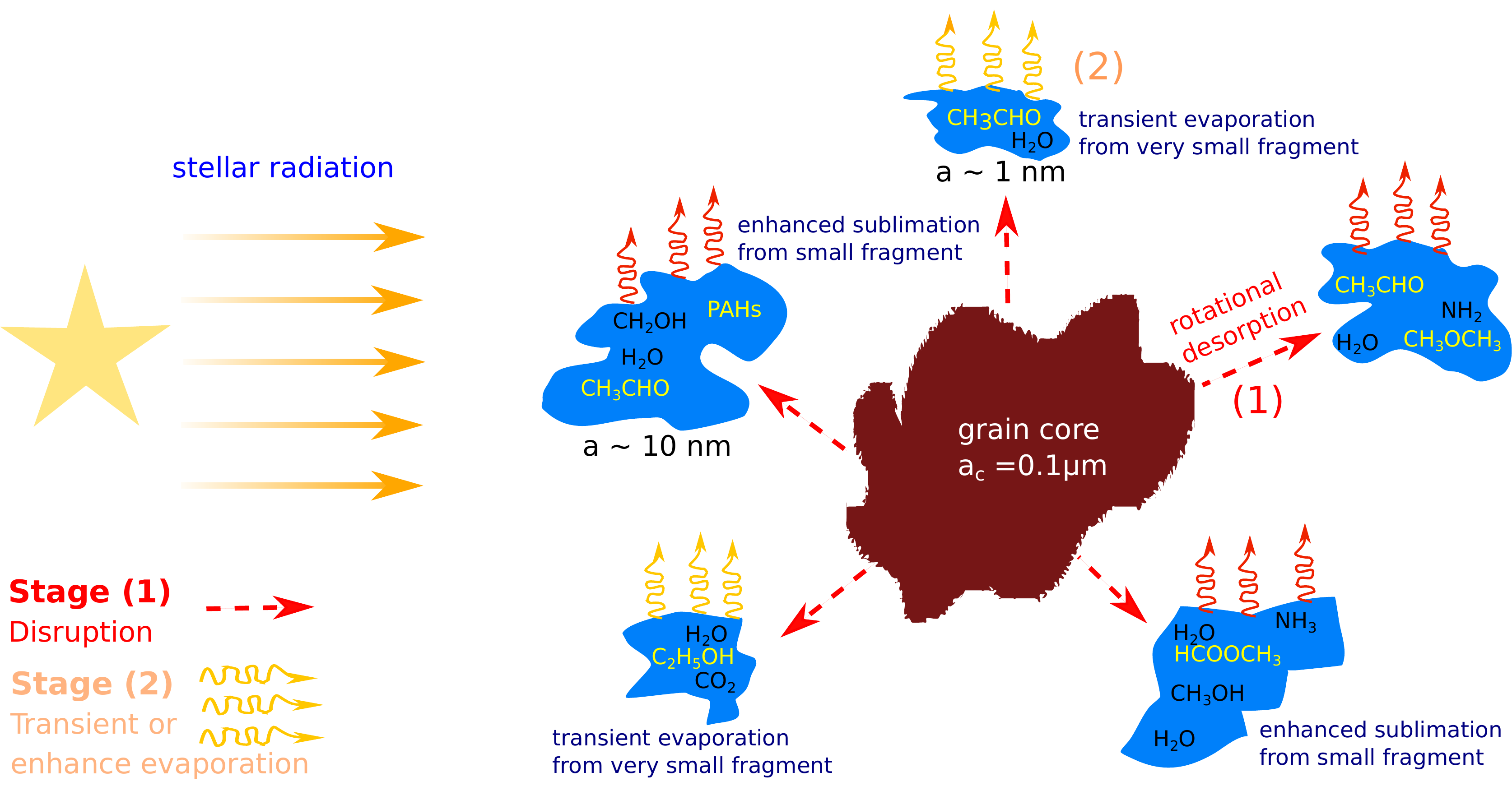}
\caption{Schematic illustration of the rotational desorption process of complex organic molecules (COMs) from icy grain mantles comprising two steps: (1) disruption of icy mantes into small fragments by Radiative Torque Disruption (RATD), and~(2) rapid evaporation of COMs due to thermal spikes for very small fragments or increased sublimation for larger fragments. From~\citet{2020ApJ...891...38H}.}
\label{fig:ice_ROTD}
\end{figure}

\begin{figure}[H]
\includegraphics[width=0.5\textwidth]{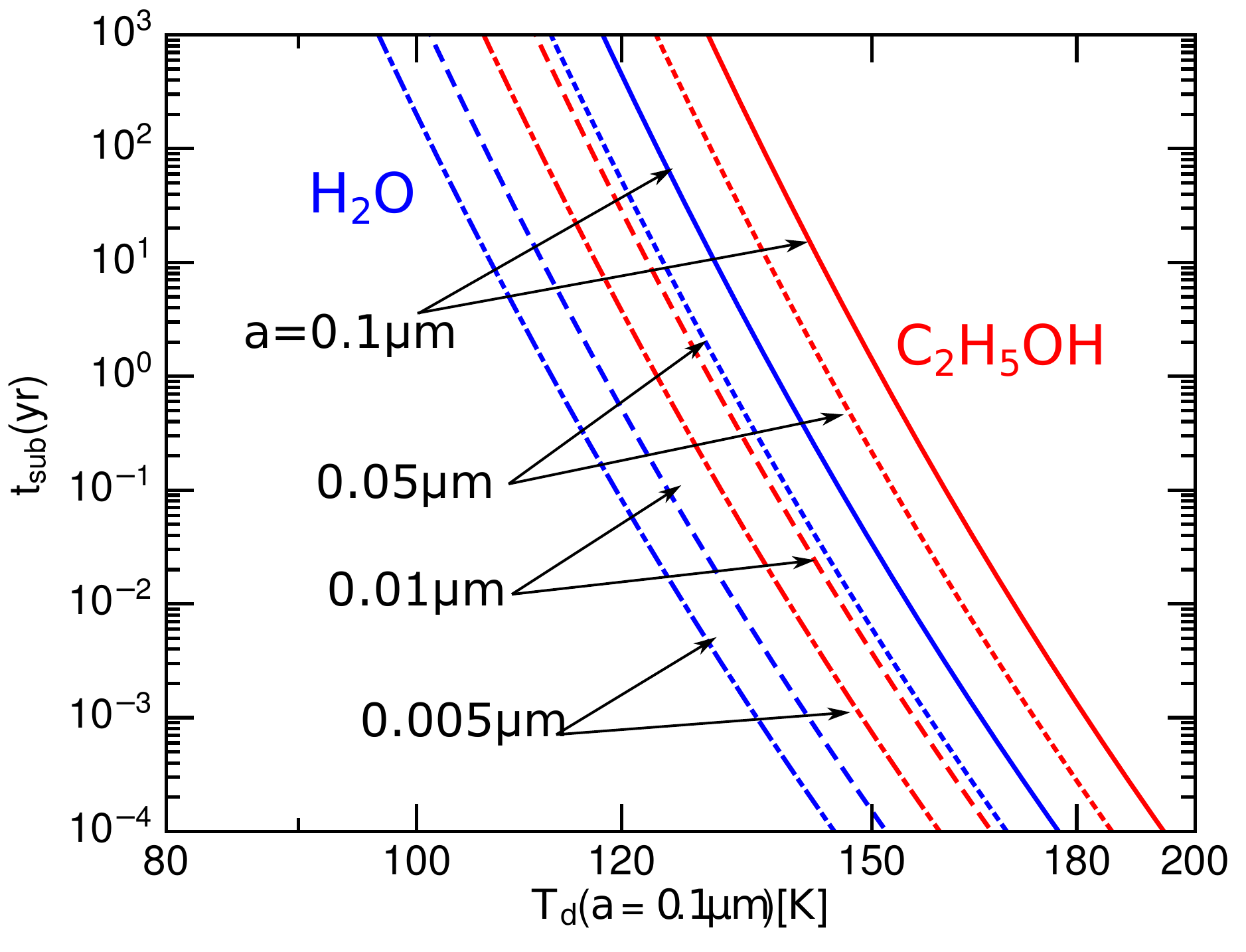}
\includegraphics[width=0.5\textwidth]{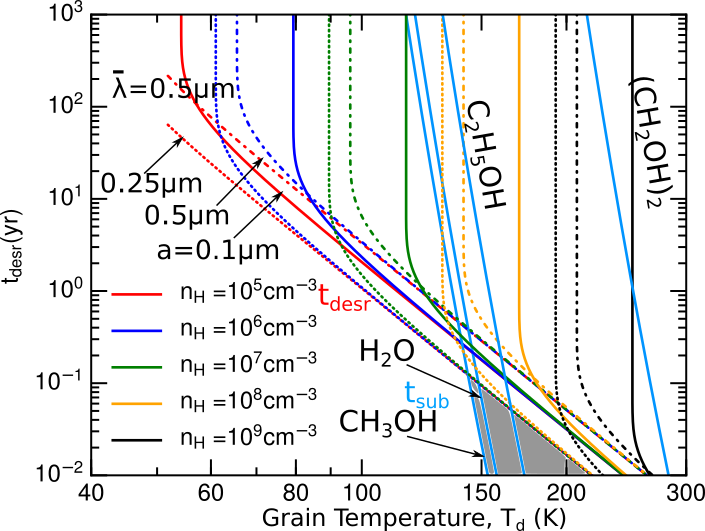}
\caption{Left panel: Decrease of sublimation time with decreasing the grain size. Right panel: Comparison of rotational desorption time and thermal sublimation time of various~molecules.}
\label{fig:ice_rothermal}
\end{figure}

{The subsequent effect of rotational disruption of ice mantles is the desorption of molecules from the icy fragments, such as water and complex organic molecules (COMs), at~temperatures below their sublimation threshold. Figure~\ref{fig:ice_ROTD} illustrates a two-step rotational desorption process of molecules from icy grain mantles induced by suprathermal rotation due to RATs. In~Figure~\ref{fig:ice_rothermal} (left panel), we show the thermal desorption time of icy grains of various sizes, which shows the decrease of $t_{\rm sub}$ with decreasing the grain size. In~the right panel, we compare the desorption time with the sublimation time for several molecules.} 

\unskip

\unskip

\subsection{Ro-Thermal Desorption of Molecules from Ice~Mantles}
The problem of thermal desorption from a non-rotating grain is well studied in the literature (\citet{1972ApJ...174..321W,1985A&A...144..147L}). The~underlying physics is that when the grain is heated to high temperatures, molecules on the grain surface acquire kinetic energy from thermal fluctuations within the grain lattice and escape from the~surface.

Let $\tau_{\rm des,0}$ be the desorption rate of molecules with binding energy $E_{b}$ from a grain at rest ($\omega=0$) which is heated to temperatures $T_{d}$. Following \citet{1972ApJ...174..321W}, one has
\bea
\tau_{\rm sub,0}^{-1}=\nu_{0}\exp\left(-\frac{E_{b}}{kT_{d}}\right),
\ena
where $\nu_{0}$ is the characteristic vibration frequency of molecules given by
\bea
\nu_{0} =\left(\frac{2N_{s}E_{b}}{\pi^{2}m}\right)^{1/2}
\ena
with $N_{s}$ being the surface density of binding sites (\citet{1987ppic.proc..333T}).

Table~\ref{tab:Ebind} lists the binding energy and sublimation temperature measured from experiments for popular~molecules.

\begin{figure}[H]
\includegraphics[width=0.5\textwidth]{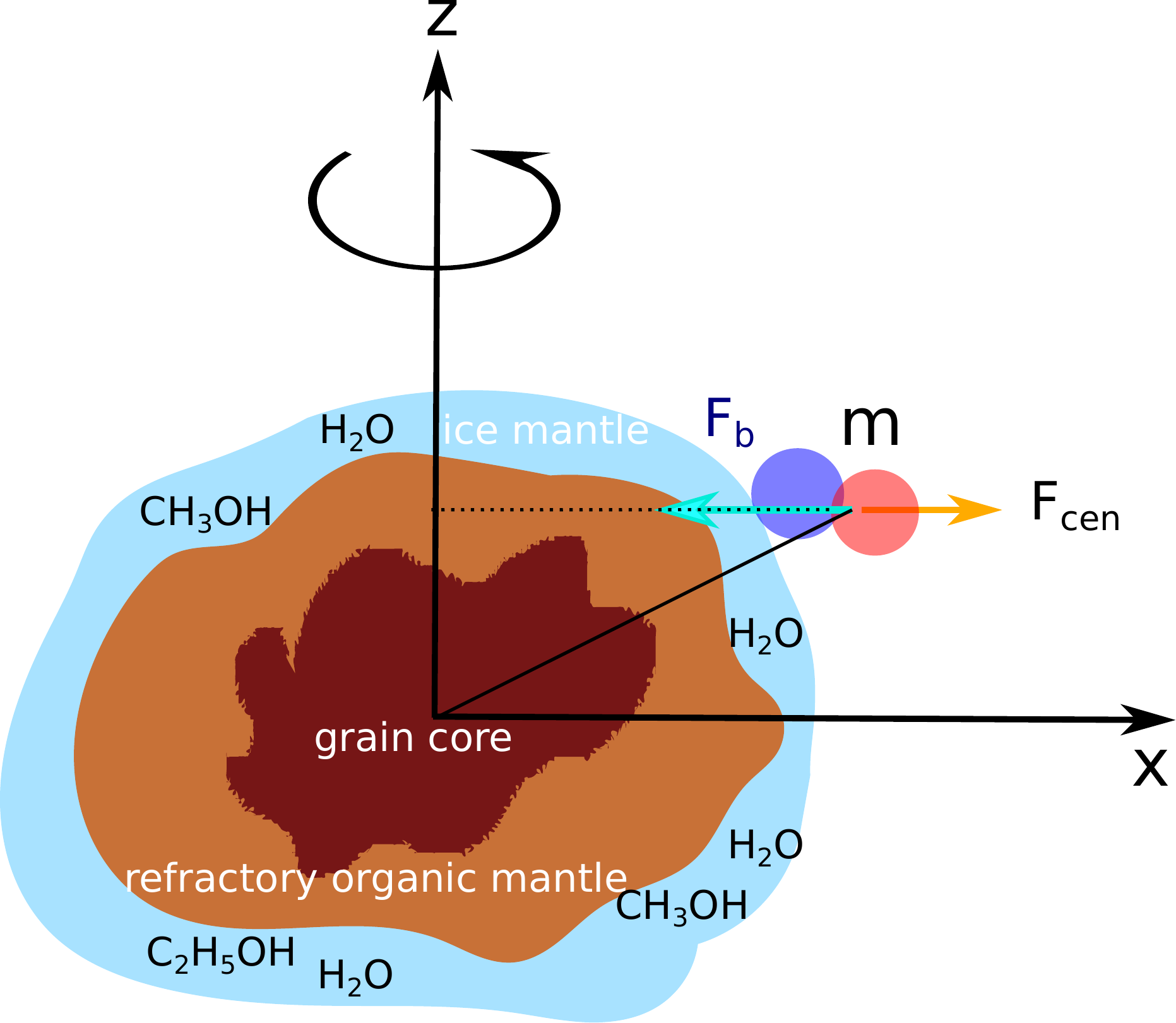}
\includegraphics[width=0.5\textwidth]{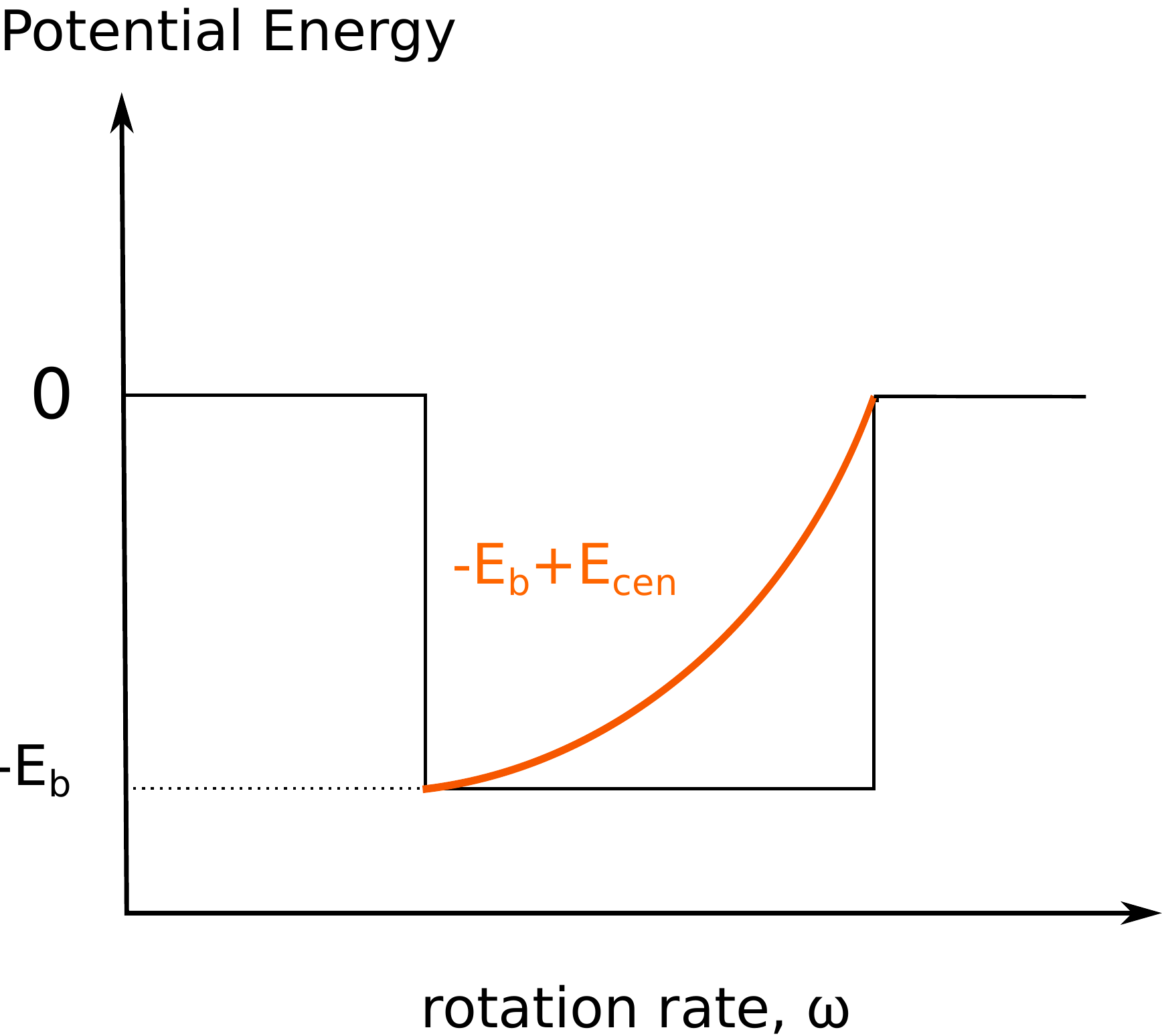}
\caption{Left panel: Illustration of a spinning dust grain with ice mantle. A~molecule on the surface experiences binding force and centrifugal force in opposite directions. Right panel: Illustration of the potential energy of a molecule on the rotating grain. The~potential barrier is reduced significantly as the angular velocity $\omega$ increases as a result of centrifugal potential ($E_{\rm cen}$). From~\citet{2019ApJ...885..125H}.}
\label{fig:ETD_mod}
\end{figure}

\begin{table}[H]
\begin{center}
\caption{Binding energies and sublimation temperatures for selected molecules on an ice~surface.}\label{tab:Ebind}
\begin{tabular}{l l l} \toprule
{\textbf{Molecules}} & {\boldmath{$E_{b}/k$ (K) $^a$}} & {\boldmath{$T_{\rm sub}$ (K)}}\cr
\midrule

$\rm H_{2}O$ & 5700 & 152 $^b$ \cr
$\rm CH_{3}OH$ & 5530 & 99 $^b$ \cr
$\rm HCOOH$ & 5570 & 155 $^c$ \cr
$\rm CH_{3}CHO$ & 2775 & 30 $^c$ \cr
$\rm C_{2}H_{5}OH$ & 6260 & 250 $^c$ \cr
$\rm (CH_{2}OH)_2$ & 10,200 & 350 $^c$ \cr
$\rm NH_{3}$ & 5530 & 78 $^b$\cr
$\rm CO_{2}$ & 2575 & 72 $^b$ \cr
$\rm H_{2}CO$ & 2050 & 64 $^b$ \cr
$\rm CH_{4}$ & 1300 & 31 $^b$ \cr
$\rm CO$ & 1150 & 25 $^b$ \cr
$\rm N_{2}$ & 1140 & 22 $^b$  \cr
\bottomrule
\end{tabular}\\
\begin{tabular}{@{}c@{}} 
\multicolumn{1}{p{\textwidth -.88in}}{\footnotesize {$^a$~See Table~4 in \citet{2013ApJ...765...60G}}; {$^b$~See Table~1 from \citet{1993prpl.conf.1177M}}; {$^c$~See \citet{2004MNRAS.354.1133C}}.}
\end{tabular}
\end{center}
\end{table}
\vspace{-6pt}
In the presence of grain suprathermal rotation, the~centrifugal force acting on a molecule of mass $m$ at distance $r\sin\theta$ from the spinning axis (see Figure~\ref{fig:ETD_mod}, left panel) is
\bea
{\bf F}_{\rm cen}=m{\bf a}_{\rm cen}=m\omega^{2}(x\xhat+y\yhat),\label{eq:Fcen}
\ena
where $\ba_{\rm cen}$ is the centrifugal acceleration, and~the unit vectors $\xhat,\yhat$ describes the plane perpendicular to the spinning~axis.

We can define centrifugal potential $\phi_{\rm cen}$ such as ${\bf a}_{\rm cen}=-\nabla\phi_{\rm cen}$. Then, the~corresponding potential~is 
\bea
\phi_{\rm cen}=\omega^{2}\left(\frac{x^{2}+y^{2}}{2}\right)=\frac{1}{2}\omega^{2}r^{2}\sin^{2}\theta.
\ena

Assuming that molecules are uniformly distributed over the grain surface of radius $a=r$, then, one can obtain the average centrifugal potential as follows:
\bea
\langle \phi_{\rm cen}\rangle=\frac{\omega^{2}a^{2}\langle \sin^{2}\theta\rangle}{2}=\frac{\omega^{2}a^{2}}{3},\label{eq:phi_cen}
\ena
{where $\langle \sin^{2}\theta\rangle=2\int_{0}^{\pi/2} \sin^{2}\theta \sin\theta d\theta=2/3$.}

As a result, the~{\it effective} binding energy of the molecule becomes
\bea
E_{b,rot}=E_{b}-m\langle \phi_{\rm cen}\rangle,\label{eq:Ebind_rot}
\ena
which means that molecules only need to overcome the reduced potential barrier of $E_{b}-E_{\rm cen}$ where $E_{\rm cen}=m\langle\phi_{\rm cen}\rangle$ to be ejected from the grain surface. The~rotation effect is more important for molecules with higher mass and low binding~energy.

Figure~\ref{fig:ETD_mod} (right panel) illustrates the potential barrier of molecules on the surface of a rotating grain as a function of $\omega$. For~slow rotation, the~potential barrier is determined by binding force. As~$\omega$ increases, the~potential barrier is decreased due to the contribution of centrifugal~potential.

The molecule is instantaneously ejected from the surface if the grain is spinning sufficiently fast such that $E_{b,rot}=0$. From~Equation~(\ref{eq:Ebind_rot}), one can obtain the critical angular velocity for the direct ejection as follows:
\bea
\omega_{\rm ej}=\left(\frac{3E_{b}}{ma^{2}}\right)^{1/2}\simeq \left(\frac{10^{10}}{a_{-5}}\right)\left(\frac{(E_{b}/k)}{1300\K}\frac{m_{\rm CO}}{m}\right)^{1/2}\rm rad\s^{-1},~~~\label{eq:omega_ej}
\ena
where $a_{-5}=a/(10^{-5}\cm)$.

The ejection angular velocity decreases with increasing grain size and molecule mass $m$, but~it increases with the binding energy $E_{b}$. 

The rate of ro-thermal desorption (sublimation) rate is given by
\bea
\tau_{\rm sub,rot}^{-1}=\nu_{0}\exp\left(-\frac{E_{b}-m\langle \phi_{\rm cen}\rangle}{kT_{d}}\right),\label{eq:tsub_rot}
\ena
where the subscript $\rm sub$ stands for sublimation, and~the second exponential term describes the probability of desorption induced by centrifugal~potential.

Equation~(\ref{eq:tsub_rot}) can be written as
\bea
\tau_{\rm sub,rot}^{-1}=\tau_{\rm sub,0}^{-1}RD(\omega),
\ena
where the function $RD(\omega)$ describes the effect of grain rotation on the thermal desorption as given by
\bea
RD(\omega)&=&\exp\left(\frac{m\langle \phi_{\rm cen}\rangle}{kT_{d}} \right)=\exp\left(\frac{m\omega^{2}a^{2}}{3kT_{d}} \right)\\
&&\simeq1.7\exp\left[a_{-5}^{2}\left(\frac{m}{m_{\rm CO}}\right)\left(\frac{\omega}{10^{9}\s^{-1}}\right)^{2}\left(\frac{20\K}{T_{d}}\right)\right]\nonumber,
\ena
which indicates the rapid increase of the ro-thermal desorption rate with the grain size $a$ and angular velocity $\omega$.

\begin{figure}[H]
\includegraphics[width=0.5\textwidth]{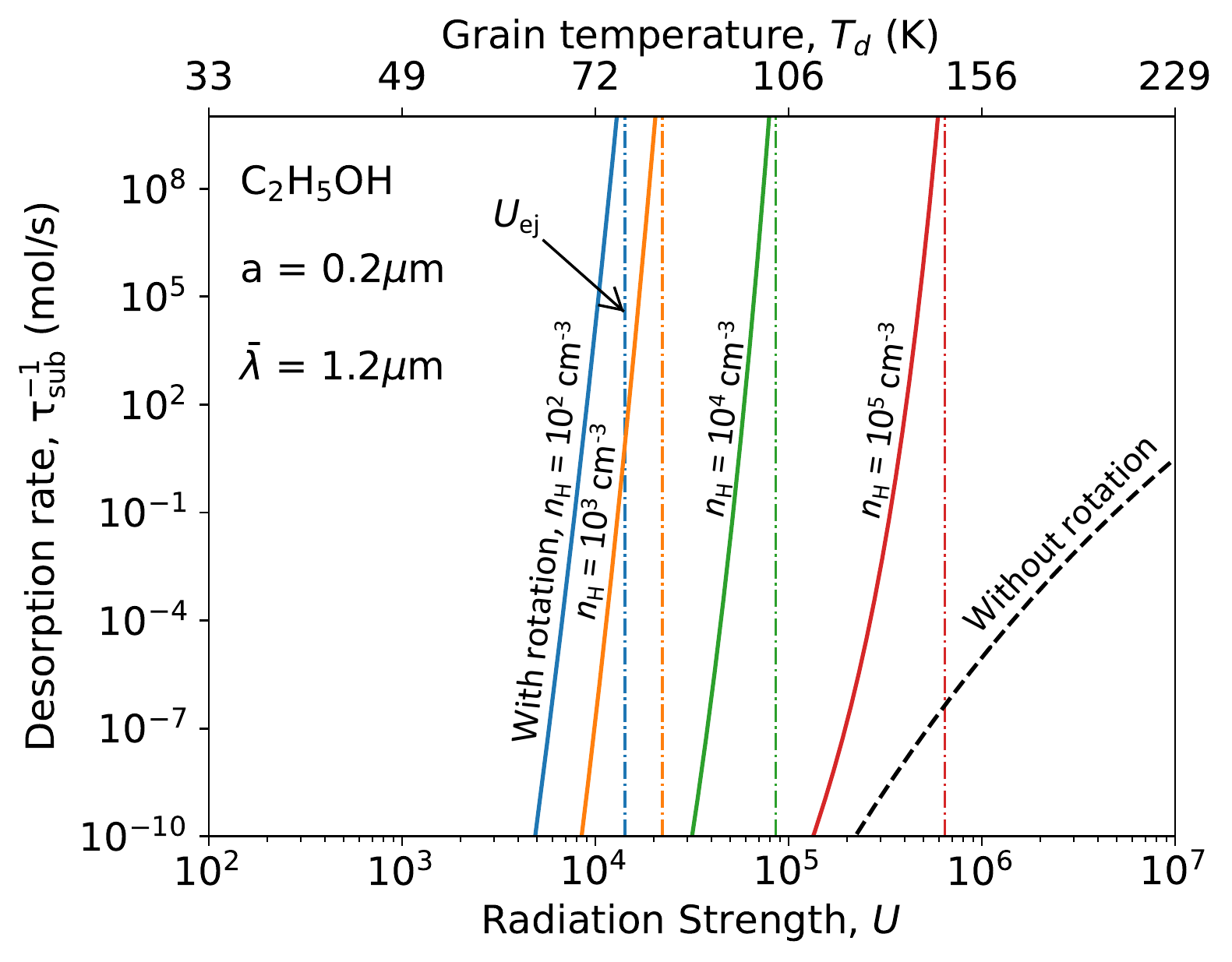}
\includegraphics[width=0.5\textwidth]{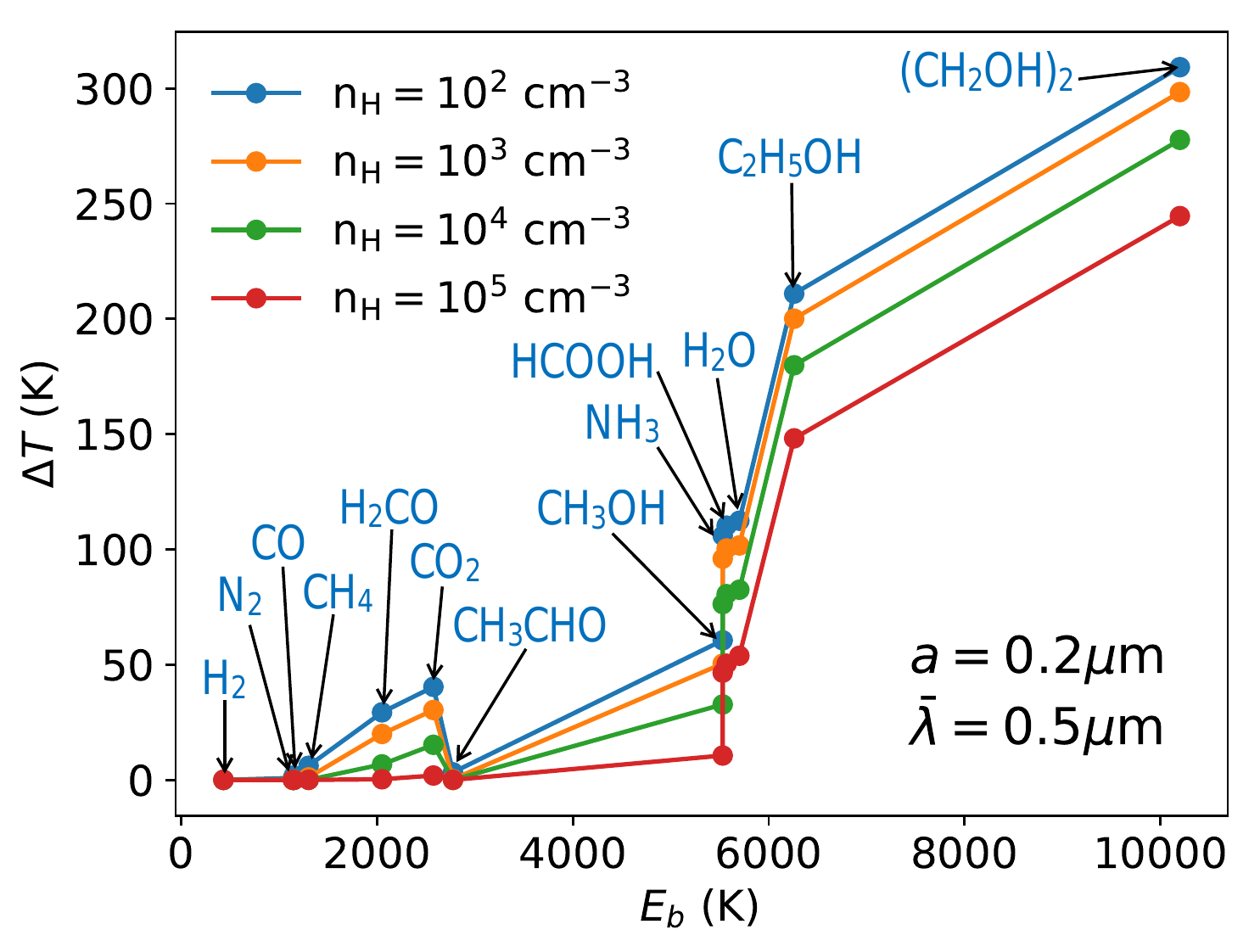}
\caption{Left panel: ro-thermal desorption rate of molecules from a thin ice mantle compared to the classical sublimation time. Right panel: decrease of the grain temperature required for desorption at the same rate as classical sublimation. From~\citet{2019ApJ...885..125H}.}
\label{fig:rothermal}
\end{figure}

Figure~\ref{fig:rothermal} (left) shows the comparison of the timescale of ro-thermal sublimation to classical thermal sublimation for various molecules, assuming the different gas density. The~right panel shows the decrease of grain temperature that results in the same desorption rate. The~effect of ro-thermal desorption is more efficient for molecules having high binding energy, such as water and ethanol. Figure~\ref{fig:rothermal} (right) shows the decrease of the grain temperature required to produce the same desorption rate as the classical mechanism. The~decrease is larger for molecules with higher binding energy (sublimation temperature).

\unskip
\section{Rotational Disruption of Nanoparticles by Stochastic Mechanical~Torques}\label{sec:nano}
Radiation pressure from strong radiation fields accelerate dust grains to high speeds, causing them to drift through ambient gas with velocity $v_{d}$ and experience mechanical torques \mbox{(\citet{2015ApJ...806..255H})}. Rapidly spinning nanoparticles can also be disrupted by centrifugal stress, {as first studied by \citet{2019ApJ...877...36H} for slow speeds in interstellar shocks and generalized for high speeds in \citet{Hoang:2020kh}.}

\subsection{Low-Energy~Regime}
Let us estimate the rotational excitation of grains due to sticking collisions of gas species. Each~atom colliding with the grain surface at radius ${\bf r}$ transfers its entire momentum $m_{\H}v$ to the grain, inducing an {\it impulsive torque} of $\delta {\bf J}= {\bf r}\times m_{\H}\bv$ (see e.g.,~\citet{1952MNRAS.112..215G}). The~increase of $(\delta J)^{2}$ from each impact is given by
\bea
(\delta J)^{2} = (a\cos\theta m_{\H}v_{d})^{2}= m_{\H}^{2}v_{d}^{2}a^{2}\cos^{2}\theta,\label{eq:dJsqr}
\ena
where $\theta$ is the polar angle of the radius vector ${\bf r}$, and~the projectiles are impinging along the horizontal~plane.

By averaging the above equation over the grain surface, one has $\langle \cos^{2}\theta\rangle= 1/2$. Thus, Equation~(\ref{eq:dJsqr}) becomes
\bea
\langle(\delta J)^{2}\rangle = \frac{1}{2}m_{\H}^{2}v_{d}^{2}a^{2}.\label{eq:dJ_sqr}
\ena

Using the random walk theory for stochastic collisions, one can derive the total increase of squared angular momentum per unit of time as follows:
\bea
\frac{\langle(\Delta J)^{2}\rangle}{\Delta t} = R_{\rm coll}(\delta J)^{2} = \frac{n_{\H}v_{d} \pi a^{2}m_{\H}^{2}v_{d}^{2}a^{2}}{2},\label{eq:dJ2}
\ena
where the collision rate $R_{\rm coll}= n_{\H}v_{d} \pi a^{2}$ has been~used.

After traversing a time interval $\Delta t$, the~total average increase of the squared angular momentum is equal to
\bea
\langle(\Delta J)^{2}\rangle = \frac{n_{\rm H}m_{\H}^{2}v_{d}^{3}\pi a^{4}}{2}\Delta t.\label{eq:deltaJ}
\ena

The {\it rms} angular velocity of grains can now be calculated using the total angular momentum $\Delta J$ from Equation~(\ref{eq:deltaJ}):
\bea
\omega_{\rm rms}^{2}=\langle \omega^{2}\rangle &=& \frac{\langle(\Delta J)^{2}\rangle}{I^{2}}=\frac{n_{\rm H}m_{\H}^{2}v^{3}\pi a^{4}}{2I^{2}}\Delta t.\label{eq:omegasqr}
\ena

A rotating grain experiences rotational damping due to sticking collision with gas atoms. Note~that sticking collisions do not damp grain rotation due to averaging effect, but~subsequent thermal evaporation of atoms that carry away part of the grain angular momentum results in grain rotational damping (see e.g.,~\citet{1998ApJ...508..157D}). Consider a grain rotating along the $z$-axis with angular velocity $\omega_{z}$. 

The mean decrease of grain angular momentum per unit of time is equal to
\bea
\bigg\langle\frac{\Delta J_{z}}{\Delta t}\bigg\rangle_{\H}&=& -R_{\rm coll}\langle \delta J_{z}\rangle\nonumber\\
&=&-\frac{2}{3}n_{\H}m_{\H}\pi a^{4}\omega_{z}\langle v\rangle=-\frac{I\omega_{z}}{\tau_{\rm H}}.
\ena

For the drift velocity with $v\gg v_{T}$, one has $\langle v\rangle =v_{d}$. Therefore, the~rotational damping time is
\bea
\tau_{\rm H}&=&\frac{3I}{2n_{\rm H}m_{\rm H}\pi a^{4}v_{d}}=\frac{4\rho a}{5n_{\rm H}m_{\rm H}v_{d}}\simeq 572\left(\frac{\hat{\rho} a_{-6}}{v_{2}n_{1}}\right)~\rm yr,~~~~~\label{eq:tauH}
\ena
where $n_{1}=n/(10\cm^{-3}), v_{2}=v_{d}/(100\km\s^{-1})$.

Rapidly spinning dust grains emit strong electric dipole radiation (\citet{1998ApJ...508..157D}), which also damps the grain rotation on a timescale of
\bea
\tau_{\rm ed}=\frac{3I^{2}c^{3}}{\mu^{2}kT_{\gas}}\simeq  2.25\times 10^{8} \left(\frac{a_{-6}^{7}}{3.8\hat{\beta}}\right)\left(\frac{100\K}{ T_{\gas}}\right)\rm yr,\label{eq:taued}
\ena
where $\mu$ is the grain dipole moment and $\hat{\beta}=\beta/(0.4\D)$ with $\beta$ being the dipole moment per structure due to polar bonds present in the dust grain (\citet{1998ApJ...508..157D,Hoang:2010jy,2016ApJ...824...18H}). 

Comparing $\tau_{\rm ed}$ with $\tau_{\H}$, one can see that, for~small grains of $a>1$ nm, the~electric dipole damping time is longer than the gas damping~time. 

Due to the rotational damping, the~grain looses angular momentum on a timescale of $\tau_{\rm H}$. Therefore, Equation~(\ref{eq:omegasqr}) yields
\bea
\omega_{\rm rms}^{2}\equiv \langle \omega^{2}\rangle &=& \frac{n_{\rm H}m_{\H}^{2}v^{3}\pi a^{4}}{2I^{2}}\tau_{\rm H},\label{eq:omegasqr2}
\ena
which can be rewritten as
\bea
\frac{\omega_{\rm rms}^{2}}{\omega_{T}^{2}}=\frac{s_{d}^{2}}{2},\label{eq:omega_sqr}
\ena
{where $s_{d}=v_{d}/v_{\rm th}$ with $v_{\rm th}=\left(2kT_{\rm gas}/m_{\rm H}\right)^{1/2}$ is the dimensionless parameter,} and the thermal angular velocity
\bea
\omega_{T}&=&\left(\frac{3kT_{\rm gas}}{I}\right)^{1/2}\nonumber\\
&\simeq& 9\times 10^{7}a_{-6}^{-5/2}T_{2}^{1/2}\hat{\rho}^{-1/2}~ \rad\s^{-1}.
\ena

Equation~(\ref{eq:omega_sqr}) reveals that nanoparticles can be spun-up to suprathermal rotation if $s_{d}>1.5$.

\subsubsection{Mechanical Torque Disruption (METD) Mechanism}
The basic idea of rotational disruption by stochastic mechanical torques (i.e., METD mechanism), is similar to that of RATD (see Section~\ref{sec:theory}). Using Equation~(\ref{eq:omega_cri}), the~critical angular velocity for the disruption is given by
\bea
\omega_{\rm cri}=\frac{2}{a}\left(\frac{S_{\rm max}}{\rho} \right)^{1/2}
\simeq 3.65\times 10^{10}\left(\frac{S_{\rm max,9}^{1/2}}{a_{-6}\hat{\rho}^{1/2}}\right)\rm rad\s^{-1},~~~~\label{eq:omega_cri2}
\ena
where $S_{\rm max,9}=S_{\rm max}/(10^{9}\erg\cm^{-3})$ is the tensile strength in units of $10^{9}\erg\cm^{-3}$ as expected for~nanoparticles. 

The time required to spin-up a grain of size $a$ to $\omega_{\rm cri}$, so-called rotational disruption time, is~evaluated as follows:
\bea
\tau_{\rm disr}&=&
\frac{J_{\rm cri}^{2}}{(\Delta J)^{2}/(\Delta t)}=\frac{2(I\omega_{\rm cri})^{2}}{n_{\H}m_{\H}^{2}v_{d}^{3}\pi a^{4}}= \frac{512\pi a^{4}\rho S_{\max}}{225n_{\H}m_{\H}^{2}v_{d}^{3}}\nonumber\\
&\simeq & 2.4\times 10^{4}\left(\frac{a_{-6}^{4}}{v_{2}^{3}}\right)\left(\frac{S_{\rm max,9}}{n_{1}\hat{\rho}}\right){\rm yr}~~~.\label{eq:tMET_disr}
\ena

The above equation implies that nanoparticles of $a\sim 1$ nm moving at $v_{d}\sim 100 \km\s^{-1}$ are disrupted in $t_{\rm disr}\sim 2$ yr, while the grain rotation is damped in $\tau_{\H}\sim 50$ yr by gas collisions or in $\tau_{\rm ed}\sim 20$ yr by electric dipole~emission.

We note that METD only occurs when the required time is shorter than the rotational damping time. Let $a_{\rm disr}$ be the grain disruption size as determined by $\tau_{\rm disr}=\tau_{\H}$. Thus, comparing Equations~(\ref{eq:tMET_disr}) and (\ref{eq:tauH}), one obtains:
\bea
a_{\rm disr}=\left(\frac{25m_{\H}v_{d}^{2}}{128\pi S_{\rm max}}\right)^{1/3}\simeq 5.5S_{\max,9}^{-1/3}\left(\frac{v_{d}}{300\rm km\s^{-1}}\right)^{2/3} ~~\rm nm,~~~\label{eq:aMET_disr}
\ena
which implies that very small grains ($a<5$ nm) moving at $v_{d}\sim 300\rm km\s^{-1}$ are disrupted by centrifugal stress, assuming strong grains of $S_{\max}\sim 10^{9}\erg\cm^{-3}$. The~rotation of larger grains (i.e., $a>a_{\rm disr}$) is damped by gas collisions before reaching the critical~threshold.

For a given grain size, the~critical speed required for rotational disruption is given by the condition of $\tau_{\rm disr}\lesssim \tau_{\H}$, which yields
\bea
v_{d}\gtrsim \left(\frac{128\pi a^{3}S_{\rm max}}{45m_{\rm H}}\right)^{1/2}\simeq 733a_{-6}^{3/2}S_{\max,9}^{1/2} \km\s^{-1}.~~~\label{eq:vdisr}
\ena
or the dimensionless parameter:
\bea
s_{d}\gtrsim \left(\frac{64\pi a^{3}S_{\rm max}}{45kT_{\rm gas}}\right)^{1/2}\simeq 565a_{-6}^{3/2}T_{2}^{-1/2}S_{\max,9}^{1/2}.\label{eq:sdisr}
\ena

The above equations indicate that the velocity required for METD decreases rapidly with decreasing grain size and with tensile strength. Smallest nanoparticles of sizes $a\sim 1$ nm only require $v_{d}\sim 23 \km\s^{-1}$ while small grains of $a\sim 0.01$ $\upmu \text{m}$ require much higher velocities for rotational disruption, assuming $S_{\rm max}\lesssim 10^{9}\erg\cm^{-3}$.

The rotation of nanoparticles experiences damping and excitation by various interaction processes, including ion collisions, plasma drag, and~infrared emission \mbox{(see \citet{1998ApJ...508..157D,Hoang:2010jy})}. A~detailed analysis of the different damping processes for grains in magnetized shocks is presented  {in \citet{2019ApJ...877...36H,2019ApJ...886...44T}.}

\subsubsection{Slowing-Down Time by Gas Drag~Force}
For hypersonic grains, the~main gas drag arises from direct collisions with gas atoms, and~the Coulomb drag force is subdominant (\citet{1979ApJ...231...77D}). Assuming the sticky collisions of atoms followed by their thermal evaporation, the~decrease in the grain momentum is equal to the momentum transferred to the grain in the opposite direction:
\bea
F_{\rm drag}\equiv \frac{dP}{dt} = m_{\H}v_{d}\times n_{\rm H}v_{d}\pi a^{2}.
\ena

The gas drag time is given by 
\bea
\tau_{\rm drag}= \frac{m_{\rm gr}v_{d}}{dP/dt}=\frac{4\rho a}{3n_{\rm H}m_{\rm H}v_{d}}\simeq 763\left(\frac{\hat{\rho}a_{-6}}{n_{1}v_{2}}\right)~\yr.\label{eq:tdrag}
\ena

Comparing Equations~(\ref{eq:tdrag}) with (\ref{eq:tMET_disr}) one can see that the disruption occurs much faster than the drag time for $v>100$ $\km\s^{-1}$ and small grains of $a<0.01$ $\upmu \text{m}$.

\subsection{High-Energy~Regime}
The penetration depth of impinging protons is approximately equal to (\citet{1979ApJ...231...77D}):
\bea
R_{\rm H}(E)\simeq \left(\frac{0.01}{\hat{\rho}}\right)\left(\frac{E}{1~\rm keV}\right) \upmu \text{m}\simeq 0.008 \left(\frac{v_{2}^{2}}{\hat{\rho}}\right)\upmu \text{m},~~~\label{eq:RH}
\ena
which reveals that for high-velocity collisions, impinging particles can pass through the grain because $R_{\rm H}> 2a$. As~a result, they only transfer part of their momentum to the grain. We will first find the fraction of ion momentum transferred to the grain and quantify the efficiency of~METD.

\textls[-5]{For interstellar grains with $a<1$ $\upmu \text{m}$ and energetic ions, we have $\Delta E \ll p^{2}/2m$, \citet{2017ApJ...847...77H}~derived}
\begin{eqnarray}
\delta p =  \frac{2mp\delta E}{2p^{2} - m\delta E} \approx  p\left(\frac{\delta E}{2E}\right)=p f_{p}(E,a),\label{eq:dp}
\end{eqnarray}
where $\Delta E$ is the energy loss passing the grain, and~$f_{p}(E,a)=\delta E/(2E)$ is the fraction of the ion energy transferred to the grain which is a function of $E$ and $a$.

Let $dE/dx=nS(E)$ where $S(E)$ be the stopping cross-section of the impinging ion of kinetic energy $E$ in the dust grain of atomic density $n$  (\citet{1981spb1.book....9S}). The~energy loss of the ion due to the passage of the grain is given by
\bea
\delta E= \frac{4a}{3}nS(E),~\label{eq:deltaE}
\ena
where the grain is approximated as slab of thickness $4a/3$. Thus,
\bea
f_{p}=\frac{2a}{3E}nS(E),\label{eq:fE}
\ena
and $f_{p}(E,a)=1$ for sticking~collisions.

Following Equation~(\ref{eq:dJ_sqr}), the~impulsive angular momentum from a collision is then given by
\bea
(\delta J)^{2}=\frac{a^{2}}{2}(\delta p)^{2}=\frac{a^{2}p^{2}}{2}f_{p}(E,a)^{2}.
\ena
which yields the average value
\bea
\langle (\delta J)^{2}\rangle = \frac{a^{2}p^{2}}{2}f_{p}(E,a)^{2}.\label{eq:dJ2_hi}
\ena

Following the similar procedure as in Section~\ref{sec:disr}, one obtain 
\bea
\langle\frac{(\Delta J)^{2}}{\Delta t}\rangle = \langle\frac{(\Delta J)^{2}}{\Delta t}\rangle_{S}f_{p}^{2},\label{eq:DJ2_hi}
\ena
where $S$ denotes sticking collisions considered in the previous subsection, and~$\langle\frac{(\Delta J)^{2}}{\Delta t}\rangle_{S}$ is given by Equation~(\ref{eq:dJ2}).

The increase of the grain angular velocity is given by 
\bea
\omega_{\rm rms}^{2}=\frac{(\Delta J)^{2}/\Delta t}{I^{2}}\times t=\left(\frac{n_{\H}m_{\H}^{2}v^{3}\pi a^{4}}{2I^{2}}\right)f_{p}^{2}\times t.~~~
\ena

If the incident ion passes through the grain, the~grain rotational damping by gas collisions is not important, and~the damping by electric dipole emission takes over. Since $\tau_{\rm ed}$ is rather long for nanoparticles of $a>1$ nm (see \citet{Hoang:2010jy}), the~grain angular velocity continues to increase to the critical limit, that is,~at $\omega_{\rm disr}$, that is,~the disruption occurs, in~disruption time equal to
\bea
\tau_{\rm disr}=\left(\frac{2I^{2}\omega_{\rm disr}^{2}}{n_{\H}m_{\H}^{2}v^{3}\pi a^{4}}\right)\frac{1}{f_{p}^{2}}=\tau_{\rm disr,S}\left(\frac{1}{f_{p}^{2}}\right),\label{eq:tdisr_high}
\ena
where $\tau_{\rm disr,S}$ is the disruption time for sticky collisions given by Equation~(\ref{eq:tMET_disr}). The~rate of rotational disruption is decreased rapidly with $E$ when $f_{p}<1$. In~the above analysis, we assumed a slab model to calculate the fraction of the momentum transfer (see \citet{Hoang:2020kh} for details).\footnote{Equations (30) and (31) in \citet{Hoang:2020kh} missed a factor $m_{\H}$ and $I/m_{\H}$, respectively, but~the final formulae are correct.}

For grain velocity below the Bohr velocity of $v_{0}=e^{2}/\hbar\approx c/137\approx 2189 \km\s^{-1}$, nuclear~interactions dominate, and~the stopping cross-section in units of erg cm$^{2}$ is given \mbox{by (\citet{1981spb1.book....9S})}
\bea
S_{n}(E) = 4.2\pi a Z_{1}Z_{2}e^{2}\frac{M_{1}}{(M_{1}+M_{2})}s_{n}(\epsilon_{12}),
\ena
where $M_{i}$ and $Z_{i}$ are the atomic masses and numbers charge of the projectile ($i=1$) and target ($i=2$) atom, and~$a$ is the screen length for the nuclei-nuclei interaction potential given by
\bea
a\simeq 0.885a_{0}(Z_{1}^{2/3}+Z_{2}^{2/3})^{-1/2},~~a_{0}=0.529\AA,
\ena
and 
\bea
\epsilon_{12}=\left(\frac{M_{2}E}{M_{1}+M_{2}}\right)\left(\frac{a}{Z_{1}Z_{2}e^{2}}\right).
\ena

We adopt the approximate function of $s_{n}$ as in \citet{1994ApJ...431..321T}:
\bea
s_{n}=\frac{3.441\sqrt{\epsilon_{12}}\ln(\epsilon_{12}+2.718)}{1+6.35\sqrt{\epsilon_{12}}+\epsilon_{12}(-1.708+6.882\sqrt{\epsilon_{12}})}.
\ena

For grain velocities above $v_{0}$, electronic interactions dominate, and~the stopping power can be approximated as
\begin{eqnarray}
nS_{e}(E) \approx \frac{2nS_{m}(E/E_{m})^{\eta}}{1+(E/E_{m})},\label{eq:dEdx_aprx}
\end{eqnarray}
where $\eta$ is the slope, $E_{m}=100$ keV and $S_{m}$ is the stopping power at $E=E_{m}$. For~graphite, we find that $\eta=0.2$ and $nS_{m}= 1.8\times 10^{6}$ keV/cm. For~quartz material, $\eta=0.25$ and $nS_{m}=1.3\times 10^{6}$ keV/cm.

Drag force in the high-velocity regime is given by (see also \citet{2017ApJ...847...77H})
\bea
F_{\rm drag}=R_{\rm coll}\delta p=n_{\H}\pi a^{2}m_{\H}v_{d}^{2}\left(\frac{1}{2 f_{p}}\right).
\ena
For high-energy regime, the~drag force is found to decrease with the velocity instead of increasing as in the classical low-energy regime (\citet{2017ApJ...847...77H}).

The drag time is equal to
\bea
\tau_{\rm drag}=\frac{m_{\rm gr}v_{d}}{F_{\rm drag}}=\tau_{\rm drag,S}\left(\frac{1}{2f_{p}}\right),\label{eq:tdrag_hi}
\ena
where $\tau_{\rm drag,S}$ is given by Equation~(\ref{eq:tdrag}).

\subsection{Grain Disruption Sizes Vs. Grain~Velocity}

\begin{figure}[H]
\includegraphics[width=0.5\textwidth]{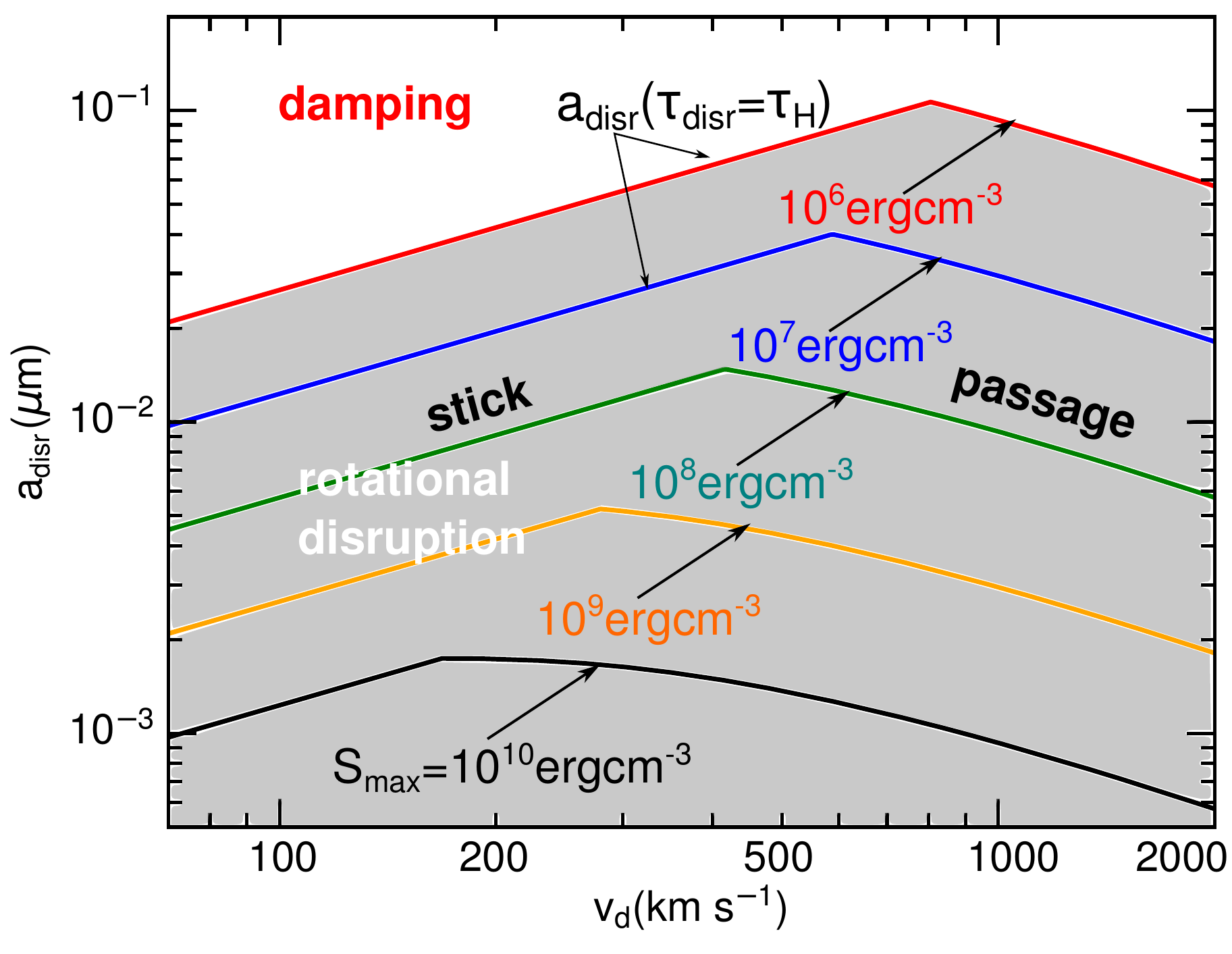}
\includegraphics[width=0.5\textwidth]{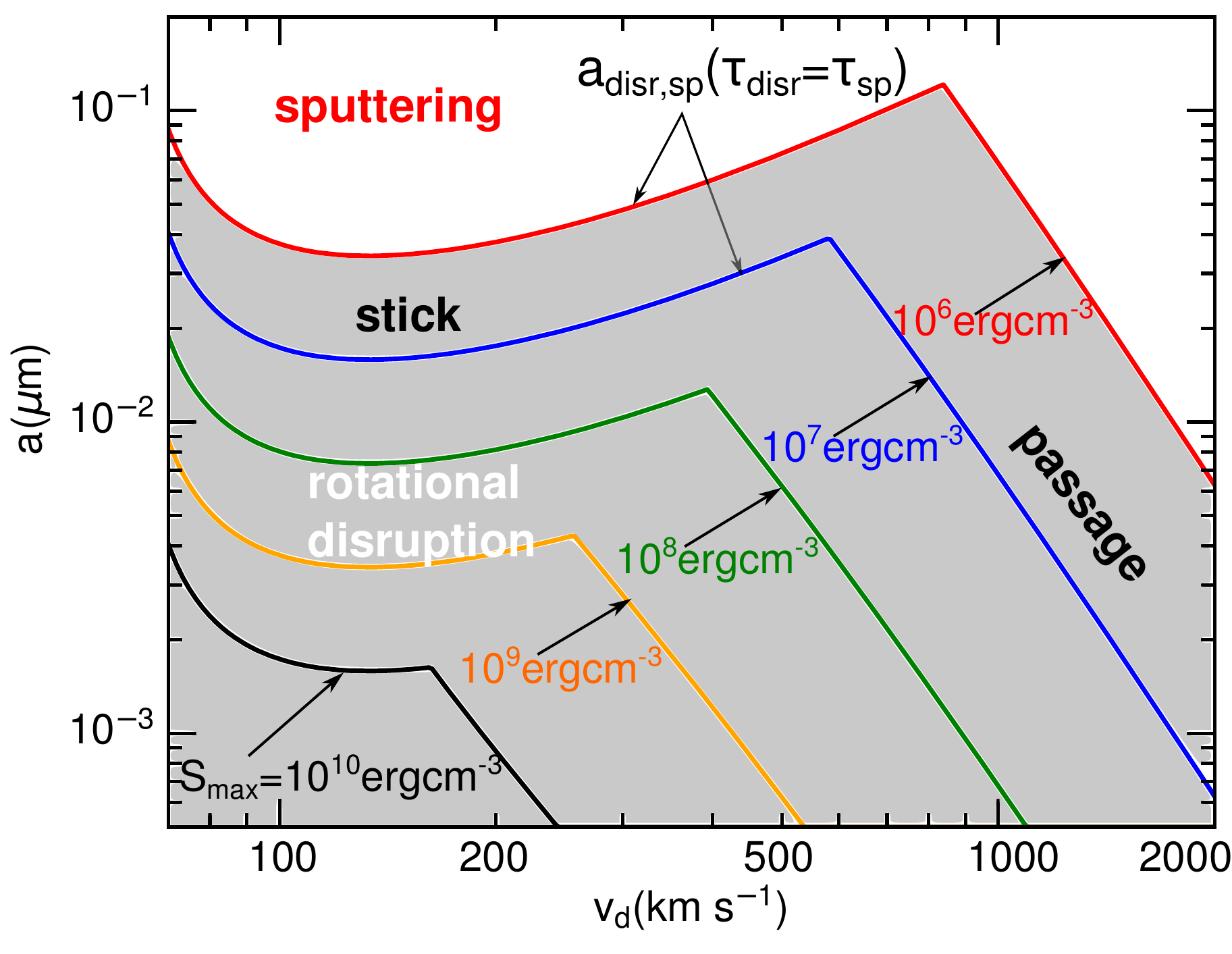}
\caption{Grain disruption size vs. grain velocity assuming the different tensile strength of grain material. The~solid lines mark the boundary between rotational disruption and damping ($\tau_{\rm disr}=\tau_{\H}$) and $\tau_{\rm disr}=\tau_{\rm sp}$ (right panel). Shaded areas mark the parameter space where rotational disruption (METD) is faster than rotational damping (left panel) and nonthermal sputtering (right panel). From~\citet{Hoang:2020kh}.}
\label{fig:aMETD_disr}
\end{figure}

To obtain grain disruption size $a_{\rm disr}$ for arbitrary velocities $v_{d}$, we first calculate $\tau_{\rm disr}$ for a range of grain sizes and compare with rotational damping time $\tau_{\H}$. Figure~\ref{fig:aMETD_disr} (left panel) shows the values of $a_{\rm disr}$ as a function of grain velocity for various tensile strength. The~disruption size $a_{\rm disr}$ increases with increasing $v_{d}$ and then decreases due to the decrease of ion momentum transfer to the grain ($f_{p}<1$).

Figure~\ref{fig:aMETD_disr} (right panel) shows the variation of $a_{\rm disr,sp}$ with $v_{d}$. Shaded areas mark the parameter space ($v_{d},a$) in which METD is faster than nonthermal sputtering. For~weak grains (e.g., of~fluffy structure) with $S_{\max}\sim 10^{7}\erg\cm^{-3}$, grains of $a\sim 0.02$ $\upmu \text{m}$ can be disrupted for $v_{d}<600\km\s^{-1}$. For~very strong grains of ideal structures with $S_{\max}\sim 10^{9}\erg\cm^{-3}$, very small grains of $a\sim 0.004$ $\upmu \text{m}$ can be rotational disrupted for $v_{d}<250\km\s^{-1}$. Nonthermal sputtering dominates the destruction of large grains (i.e., $a>0.1$ $\upmu \text{m}$) or at high velocities of $v_{d}>1000 \km\s^{-1}$.

\section{Effects of Rotational Disruption on Dust Extinction, Emission, and~Polarization}\label{sec:model}
The RATD and METD processes affects the grain size distribution that determines the observable properties of astrophysical dust, including extinction, thermal emission, and~polarization. Nanoparticles produced by RATD emit microwave emission via spinning dust mechanism. Below, we will present modeling of the dust extinction and polarization for grains in the ISM with various radiation strength $U$.

\subsection{Grain Size Distribution From~RATD}
Grain size distribution is a fundamental parameter of dust. For~the diffuse ISM, significant efforts have been made to infer the reliable size distribution \mbox{{(e.g., \citet{Mathis:1977p3072,2001ApJ...548..296W})}}.

The most popular dust model consists of two separate populations of amorphous silicate grains and carbonaceous (graphite) grains (see \citet{2001ApJ...548..296W,2007ApJ...657..810D}. For~both populations, the~grain size distribution is usually described by a power-law (\citet{Mathis:1977p3072}, hereafter~MRN):
\begin{equation}
\frac{1}{n_{\H}}\frac{dn_{j}}{da} = C_{j}a^{-\eta},
\label{eq:MRN}
\end{equation}
where $dn_{\rm j}$ is the number density of grains of material $j=sil$ or gra between $a, a+da$, $n_{\rm H}$ is the number density of hydrogen, and~$\eta=3.5$, and~the lower and upper cutoff are $a_{\rm min}$ = 10{\AA} and \mbox{$a_{\max} = 0.25$ $\upmu \text{m}$}. We take constant $C_j$ from \citet{2001ApJ...548..296W} for MRN size distribution as follows: $C_{\rm sil}=10^{-25.11}$cm$^{2.5}$ and $C_{\rm gra}=10^{-25.13}$cm$^{2.5}$. 

The MRN size distribution is widely used for describing dust in our galaxy.
We can plausibly assume that the standard MRN size distribution is suitable for the standard ISRF with radiation strength $U=1$. Toward a stronger radiation field ($U>1$), the~RATD effect changes the upper cutoff of the original size distribution and the abundance of small grains vs. large grains. The~new size distribution of grains should depend on the internal structure of~grains. 

To describe the size distribution in the presence of RATD, we can adopt the power law as in Equation~(\ref{eq:MRN}), but~with the different model parameters that are constrained by the constant dust-to-mass ratio. In~general, the~RATD effect disrupts large grains of size $a=[a_{\rm disr}-a_{\rm disr,max}]$ into smaller ones. In~the diffuse ISM, large grains above $1$ $\upmu$m are not expected, thus, the~RATD mechanism determines the upper limit of the grain size distribution of the ISM because \mbox{$a_{\rm disr,max}>1$ $\upmu$m}. Therefore, we set $a_{\max}=a_{\rm disr}$.

To account for the RATD effect, we fix the constant $C$ and change the slope $\eta$. Such a new slope $\eta$ is determined by the dust mass conservation as given by:
\bea
\int_{a_{\rm min}}^{a_{\rm disr}} a^{3} a^{\eta} da =\int_{a_{\rm min}}^{a_{\rm max,MRN}} a^{3} a^{-3.5} da,
\ena
which yields
\bea
\frac{a_{\rm disr}^{4+\eta} - a_{\rm min}^{4+\eta}}{4+\eta}=\frac{a_{\rm max,MRN}^{0.5} - a_{\rm min}^{0.5}}{0.5}.
\label{eq:slope}
\ena

We obtain $\eta$ by numerically solving the above equation (see \citet{2020ApJ...888...93G} for more details).


\subsection{Dust Extinction and Starlight~Polarization}\label{sec:ext}
The radiation intensity of starlight is reduced mainly due to the absorption and scattering \mbox{(i.e., extinction)} of dust along the line of sight. The~extinction efficiency of light by a dust grain is defined by
$Q_{\rm ext} = C_{\rm ext}/(\pi a^{2})$ where $C_{\rm ext}$ is the extinction cross-section. We use a mixed-dust model comprising silicate and graphite materials (\citet{2001ApJ...548..296W,2007ApJ...663..866D}) and take $C_{\rm ext}$ calculated for oblate spheroidal grains of axial ratio $a/b=2$ from \citet{2013ApJ...779..152H}.  

The extinction of stellar light at wavelength $\lambda$ in magnitude per H atom is given by (see e.g.,~\citet{2013ApJ...779..152H}):
\bea \label{eq:11}
\frac{A(\lambda)}{N_{\rm H}} = \sum_{j=sil,gra} 1.086 \displaystyle\int\limits_{a_{\rm min}}^{a_{\rm disr}}  C_{\rm ext}^{j}(a)\left(\frac{1}{n_{\rm H}}\frac{dn^{j}}{da}\right)da,
\ena
where $N_{\rm H}=\int n_{\rm H}dz=n_{\rm H}L$ with $L$ the path length is the column density, $dn^{j}/da$ is the grain size distribution of dust component $j$.

Starlight is polarized due to extinction by aligned dust grains (\citet{Hall:1949p5890,Hiltner:1949p5856}). We assume that only silicate grains can be aligned with the magnetic field, whereas graphite grains are not efficiently aligned (\citet{2006ApJ...651..268C}; see \citet{2016ApJ...831..159H} for a theoretical explanation).\footnote{Although carbonaceous grains are expected to be aligned via k-RAT mechanism (see \citet{2019ApJ...883..122L}), their~degree of alignment is not yet quantified, in~contrast to silicate grains that have the alignment degree quantified in \citet{2016ApJ...831..159H} using numerical simulations.} For~the magnetic field in the plane of the sky, the~degree of starlight polarization per H atom due to aligned grains in $\%$ is given by (\citet{2017ApJ...836...13H}):
\begin{align} \label{eq:16}
\frac{P_{\rm ext}(\lambda)}{N_{\rm H}} = 100 \displaystyle\int\limits_{a_{\rm align}}^{a_{\rm disr}} \frac{1}{2} C_{\rm pol}(a) f_{\rm align}(a) \left(\frac{1}{n_{\rm H}}\frac{dn}{da}\right)da,
\end{align}
where $C_{\rm pol}$ = $Q_{\rm pol} \pi a^{2}$ is the polarization cross-section with $Q_{\rm pol}$ the polarization efficiency, and~$f_{\rm align}(a)$ is the alignment function describing the grain-size dependence of the grain alignment degree. For~our modeling, we consider the oblate grain shape and take data of $Q_{\rm pol}$ computed by \citet{2013ApJ...779..152H}. 

The alignment function can be modeled by the following function:
\begin{align} \label{eq:17}
f_{\rm align}(a) = 1-\exp\left[-\left(\frac{0.5 a}{a_{\rm align}}\right)^{3}\right],
\end{align}
which yields the perfect alignment $f_{\rm align}(a)=1$ for large grains of $a\gg a_{\rm align}$ and adequately approximates the numerical results from \citet{2016ApJ...831..159H} as well as results from inverse modeling of starlight polarization (\citet{2014MNRAS.438..680H,2017ApJ...836...13H}).

\begin{figure}[H]
\includegraphics[width=0.5\textwidth]{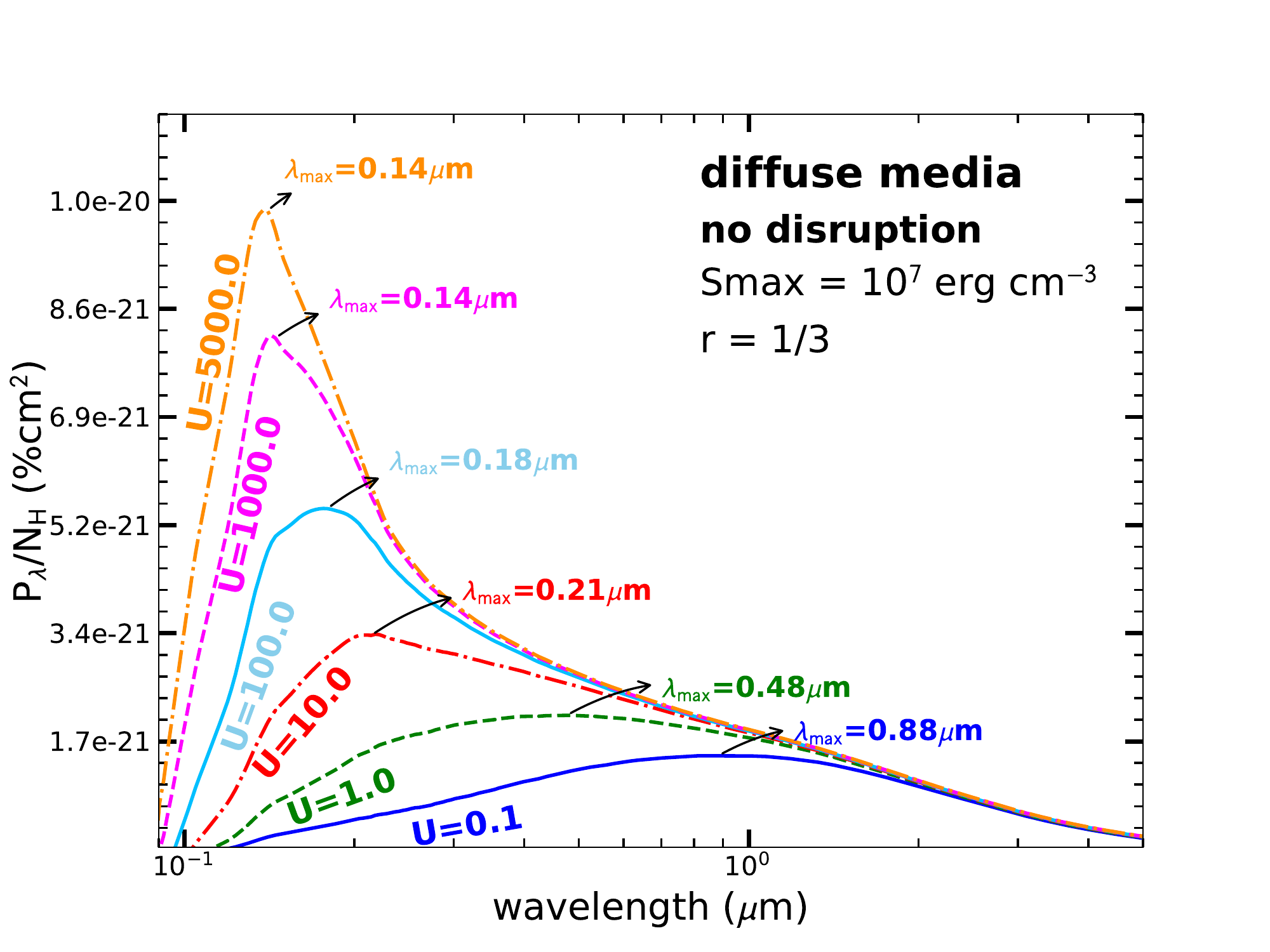}
\includegraphics[width=0.5\textwidth]{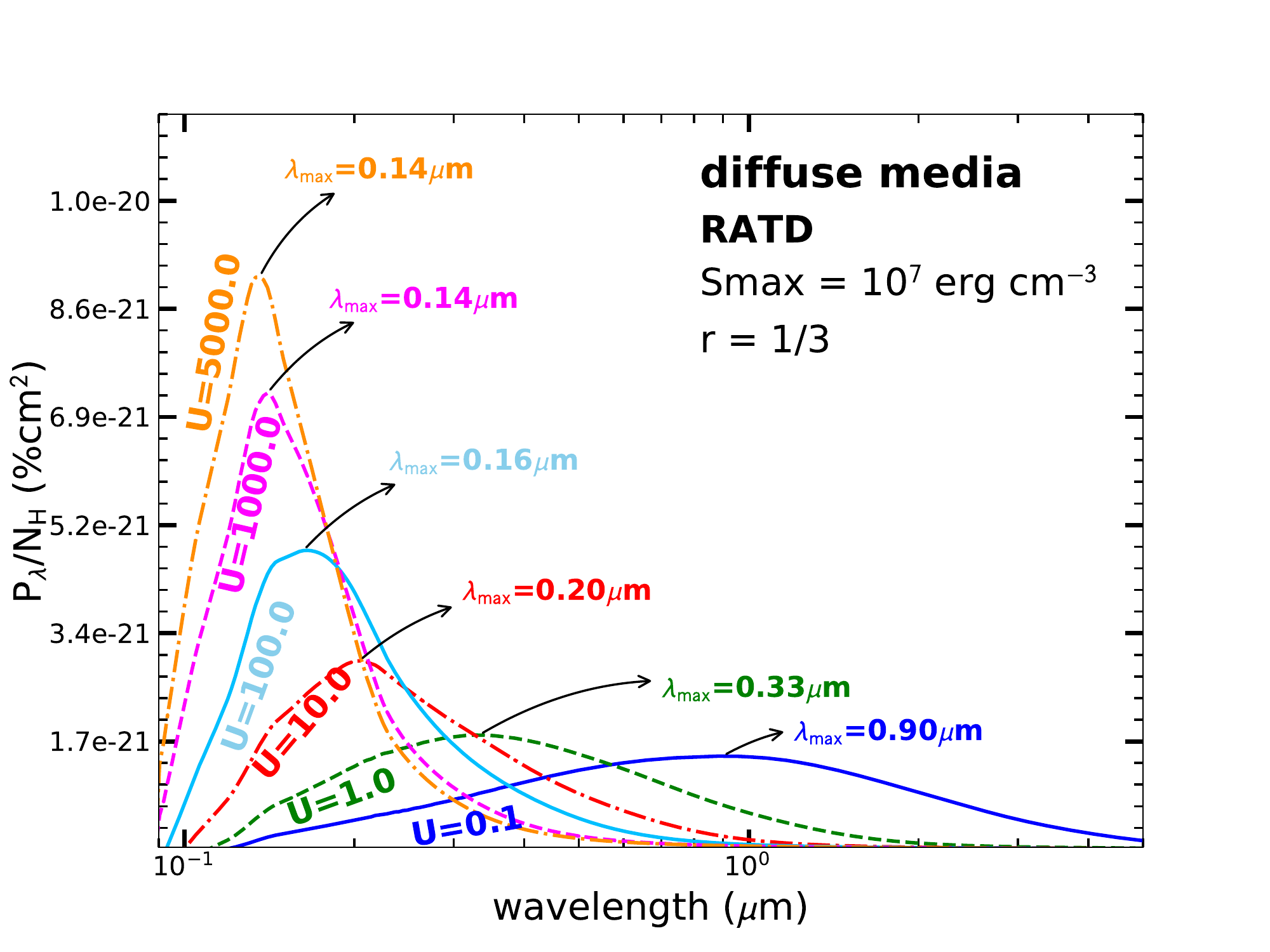}
\caption{Polarization spectrum due to extinction of starlight by dust grains aligned with axial ratio $r=1/3$ by RATs for the diffuse media with various radiation field strengths for two cases without RATD and with RATD. A~tensile strength $S_{\max}=10^{7}\erg\cm^{-3}$ is considered. From~\citet{2020ApJ...896...44L}.}
\label{fig:PabsDiff}
\end{figure}

Figure~\ref{fig:PabsDiff} (left panel) shows that the polarization spectrum with $r=1/3$ in the diffuse media peaks at $\lambda_{\max} \sim 0.48$ $\upmu$m when RATD is not taken into account. The~polarization at $U=1$ reflects the polarization spectrum from the typical interstellar radiation field. As~the radiation field strength increases, $\lambda_{\max}$ moves to shorter wavelengths because of the enhancement of small grains. Figure~\ref{fig:PabsDiff} (right panel) shows the results obtained when RATD is taken into account. The~optical-NIR polarization decreases but the UV polarization increases with increasing $U$ due to the conversion of largest grains into smaller ones. As~a result, the~width of the polarization spectrum becomes narrower as the radiation field strength~increases.

\unskip
\subsection{Thermal Emission and Polarized~Emission}\label{sec:em}
Dust grains heated by starlight re-emit thermal radiation in infrared. For~the optically thin regime, the~total emission intensity and polarized intensity are respectively given by (\citet{2009ApJ...696....1D}):
\begin{equation} 
\begin{split}
\frac{I_{\rm em} (\lambda)}{N_{\H}} &= \sum_{j=sil,car}\int ^{a_{\rm disr}}_{a_{\rm min}}Q_{\rm abs} \pi a^2 \int dT B_{\lambda}(T)\frac{dP(T)}{dT} \frac{1}{n_{\H}}\frac{dn_{j}}{da} da,\\
\frac{I_{\rm pol} (\lambda)}{N_{\H}} &= \int ^{a_{\rm disr}}_{a_{\rm min}}f_{\rm align}(a)Q_{\rm pol}\pi a^2 \int dT B_{\lambda}(T)\frac{dP(T)}{dT} \frac{1}{n_{\H}}\frac{dn_{sil}}{da} da,
\end{split}
\label{eq:Ipol_Iem}
\end{equation}
where $dP/dT$ is the temperature distribution function which depends on the grain size and radiation strength $U$, and~$B_{\lambda}(T)$ is the Planck function as given by
\begin{equation} 
B_{\lambda}(T) = \frac{2hc^2}{\lambda^5}\frac{1}{e^{hc/(k T\lambda)}-1}.
\label{eq:blackbody}
\end{equation}
Above, we disregard the minor effect of grain alignment on the thermal emission, which is considered in \citet{2009ApJ...696....1D}.

The polarization degree of thermal emission is then given by
\begin{equation} 
P_{\rm em} (\lambda)= 100\times \left(\frac{I_{\rm pol}}{I_{\rm em}}\right).
\label{eq:Pem_ratio}
\end{equation}

\begin{figure}[H]
\includegraphics[width=0.5\textwidth]{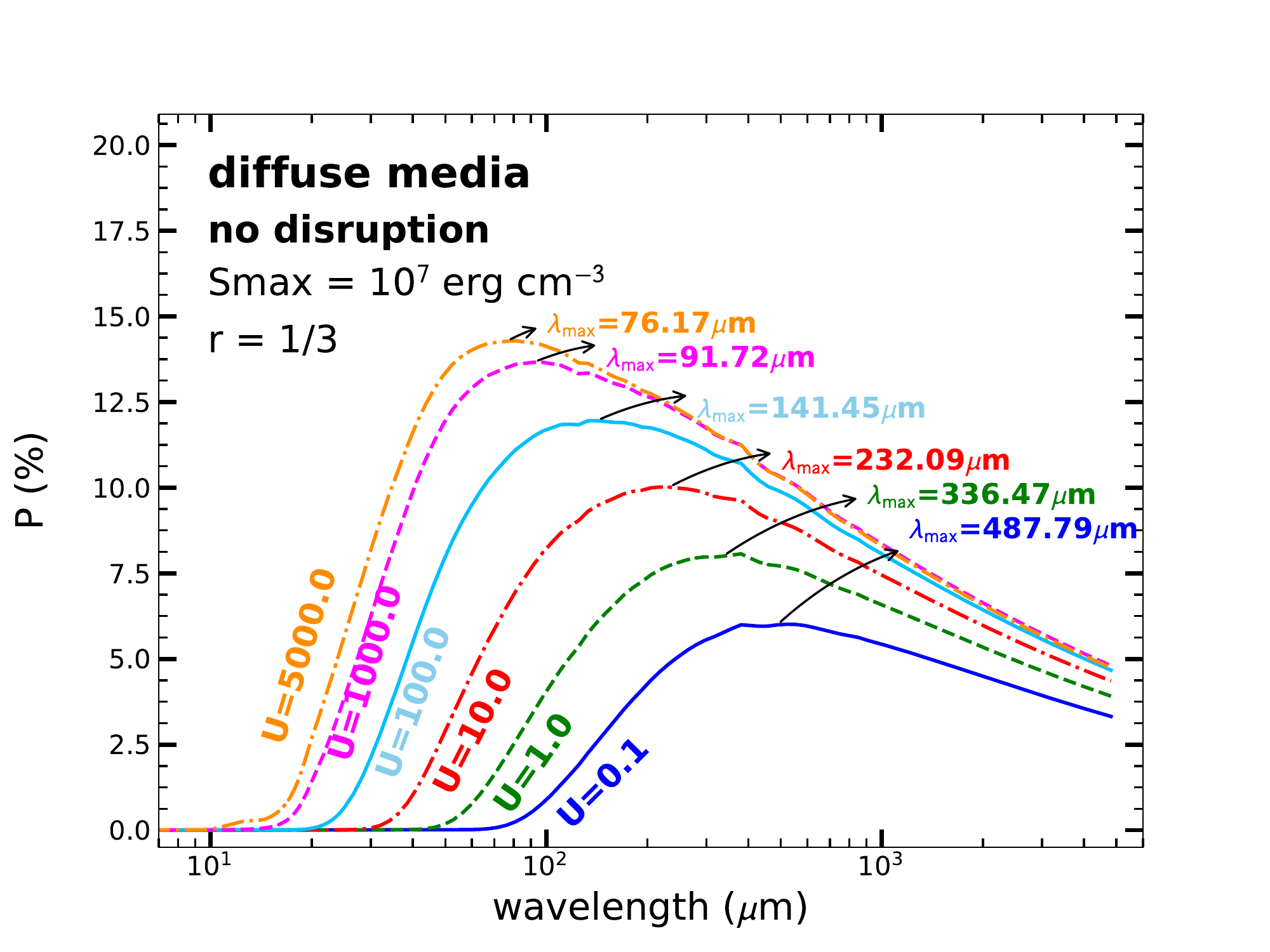}
\includegraphics[width=0.5\textwidth]{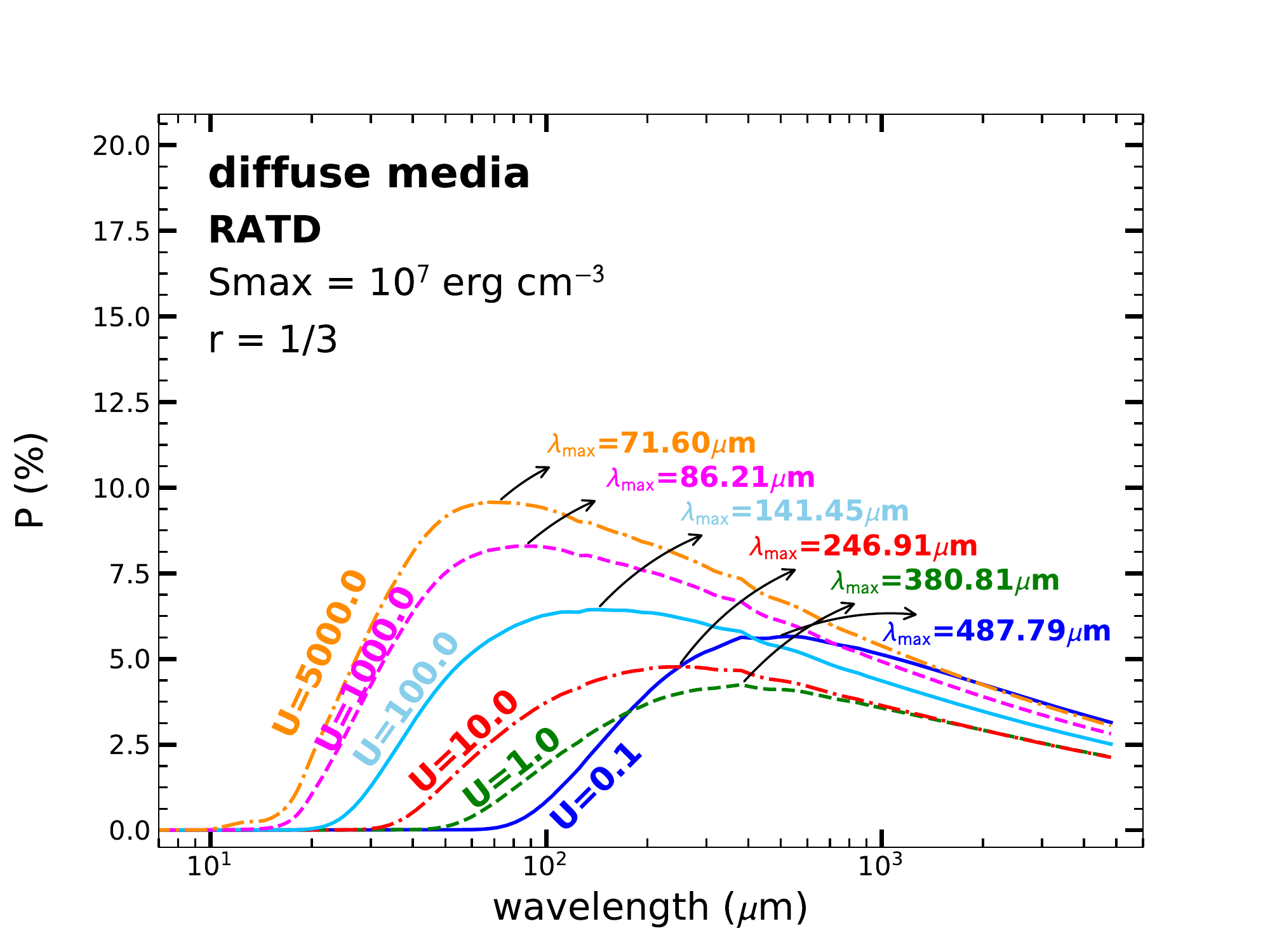}
\caption{Polarization spectrum of thermal emission from aligned grains by RATs with axial ratio $r=1/3$ in the diffuse medium with various radiation field strengths, assuming no grain disruption (left panel) and with disruption by RATD (right panel). The~tensile strength $S_{\max}=10^{7}\erg\cm^{-3}$ is considered. From~\citet{2020ApJ...896...44L}.} 
\label{fig:PemD}
\end{figure}

Figure~\ref{fig:PemD} shows the polarization spectrum of thermal emission from dust grains aligned by RATs in the absence of RATD (left panel) and presence of RATD (right panel) for prolate grains of axial ratio $r=1/3$, assuming the tensile strength $S_{\max}=10^{7}\erg\cm^{-3}$. In~the absence of RATD (left panel), the~maximum polarization increases with increasing the radiation strength $U$ as a result of enhanced alignment of small grains. The~peak wavelength ($\lambda_{\max}$) of the polarization spectrum moves toward short wavelengths as $U$ increases, but~their spectral profiles remain similar. When the RATD mechanism is taken into account, the~polarization degree for $U\gtrsim 1$ is essentially lower than the case without RATD due to the removal of large grains by RATD. Moreover, the~peak polarization degree decreases as the radiation strength increases from $U=0.1$ to $U=1.0$. 

\unskip
\subsection{Microwave Emission from Spinning~Nanoparticles}\label{sec:spindust}
Nanoparticles rapidly spinning are suggested to emit microwave emission. We first describe the spinning dust model which will be used in Section~\ref{sec:appl} to model microwave emission spectrum in the AGB and magnetized shocks where the RATD and METD are~considered.

The rotational emission mechanism is built upon the assumption that nanoparticles own non-zero electric dipole moments. PAH molecules can acquire intrinsic dipole moments due to polar bonds (see \citet{1998ApJ...508..157D}). The~attachment of SiO and SiC molecules to the grain surface gives rise to the electric dipole moment for nanosilicates (\citet{2016ApJ...824...18H}). Let $N$ be the total number of atoms in a spherical nanoparticle of radius $a$. Assuming PAHs with a typical structure C:H=$3:1$ having mean mass per atom $m\approx 9.25$ amu, one obtains $N=545a_{-7}^{3}$ for the mass density $\rho=2\g\cm^{-3}$ (\citet{1998ApJ...508..157D}). Assuming nanosilicates with structure SiO$_{4}$Mg$_{1.1}$Fe$_{0.9}$ having $m=24.15$ amu, one has $N=418a_{-7}^{3}$ for $\rho=4\g\cm^{-3}$ (\citet{2016ApJ...824...18H}).

Let $\beta$ be the dipole moment per atom in the grain. Assuming that dipoles have a random orientation distribution, the~intrinsic dipole moment of the grain can be estimated using the random walk formula:
\bea
\mu^{2}=N\beta^{2}\simeq 86.5(\beta/0.4\D)^{2} a_{-7}^{3} \D^{2},\label{eq:muin}
\ena
for PAHs, and~$\mu^{2}\simeq 66.8(\beta/0.4\D)^{2} a_{-7}^{3} \D^{2}$ for nanosilicates \citet{2016ApJ...824...18H}. The~power emitted by a rotating dipole moment $\mu$ at angular velocity $\omega$ is given by the Larmor formula:
\bea
P(\omega,\mu)=\frac{2}{3}\frac{\omega^4\mu^2\sin^2\theta}{c^3}~~~,
\ena
where $\theta$ is the angle between $\bomega$ and $\bmu$. Assuming an uniform distribution of the dipole orientation, $\theta$, then, $\sin^{2}\theta$ is replaced by $\langle \sin^{2}\theta\rangle=2/3$.

In dense regions where gas-grain collisions dominate rotation dynamics of nanoparticles (e.g.,~in~shocked regions), the~grain angular velocity can be appropriately described by the Maxwellian~distribution: 
\bea
f_{\rm MW}(\omega, T_{\rm rot})=\frac{4\pi}{ (2\pi)^{3/2}}\frac{I^{3/2}\omega^{2}}{(kT_{\rm rot})^{3/2}}\exp\left(-\frac{I\omega^{2}}{2kT_{\rm rot}} \right),\label{eq:fomega}
\ena
where $I$ is the moment of inertia of the spherical nanoparticle of mass density $\rho$, and~$T_{\rm rot}$ is the grain rotational temperature (see \citet{1998ApJ...508..157D}).

The size distribution of nanoparticles is usually described by a log-normal size distribution (\citet{Li:2001p4761}):
\bea
\frac{1}{n_{\H}}\frac{dn_{j}}{da} = \frac{B_{j}}{a}\exp\left(-0.5\left[\frac{\log (a/a_{0,j})}{\sigma_{j}}\right]^{2}\right) ,\label{eq:dnda_log}
\ena
where $j=PAH, sil$ corresponds to PAHs and nanosilicate composition, $a_{0,j}$ and $\sigma_{j}$ are the model parameters, and~$B_{j}$ is a constant determined by
\bea
B_{j}=\frac{3}{(2\pi)^{3/2}}\frac{{\rm exp}(-4.5\sigma_{j}^{2})}{\rho \sigma a_{0,j}^{3}} \left(\frac{m_{X}b_{X}}{1+{\rm erf}[3\sigma/\sqrt{2} + {\rm ln}(a_{0}/a_{\min})/\sigma\sqrt{2}}\right),\label{eq:Bconst}
\ena
where $m_{X}$ is the grain mass per atom X, $b_{X}=X_{\H}Y_{X}$ with $Y_{X}$ being the fraction of $X$ abundance contained in very small sizes and $X_H$ being the solar abundance of element $X$. In~our studies, $X=$ C for PAHs and $X=$ Si for nanosilicates. In~addition, $m_{X}=m_{C}$ for PAHs, and~$m_{X}=m(SiO_{4}Mg_{1.1}Fe_{0.9})$ for nanosilicates of the adopted~composition. 

The peak of the mass distribution $a^{3}dn_{j}/d\ln a$ occurs at $a_{p}=a_{0,j}e^{3\sigma_{j}^{2}}$. Three parameters determine the size distribution of nanoparticles, including $a_{0,j},\sigma_{j}, Y_{X}$. 

The effect of METD completely removes nanoparticles smaller than $a_{\rm disr}$, shifting the lower cutoff $a_{\rm min}$ to $a_{\rm disr}$. On~the other hand, the~effect of RATD enhances the abundance of nanoparticles. So, the~total size distribution of nanoparticles include the log-normal form and the power-law term adjusted by RATD (see \citet{2020ApJ...893..138T}). {Note that the power-law size distribution for nanoparticles is also explored in \citet{2017ApJ...836..179H}.}

Let $j_{\nu}^{a}(\mu, T_{\rm rot})$ be the emissivity from a spinning nanoparticle of size $a$ where $T_{\rm rot}$ depends on local conditions. Thus, one has
\bea
j_{\nu}^{a}(\mu, T_{\rm rot})= \frac{1}{4\pi}P(\omega,\mu) {\rm pdf}(\nu|\omega)=\frac{1}{4\pi}P(\omega,\mu)2\pi f_{\rm MW}(\omega),\label{eq:jem_a}
\ena
where $pdf(\nu|\omega)$ is the probability that the nanoparticle rotating at $\omega$ emits photons at observed frequency $\nu$, and~the relation $\omega=2\pi \nu$ is~assumed.

Here we disregarded the effect of grain wobbling (\citet{Hoang:2010jy}) and assume that nanoparticles are rotating along one axis as in \citet{1998ApJ...508..157D}. This assumption is likely appropriate for shocked regions because suprathermal rotation ($T_{\rm rot}\gtrsim T_{\rm gas}\gg T_{d}$) due to supersonic neutral drift is expected to induce rapid alignment of the axis of maximum inertia moment with the angular momentum (i.e., internal alignment, \citet{1979ApJ...231..404P}).

The rotational emissivity per H nucleon is obtained by integrating over the grain size distribution (see \citet{2011ApJ...741...87H}):
\bea \label{eq:jnu_w}
\frac{j_{\nu}(\mu, T_{\rm rot})}{n_{\H}}=\int_{a_{\min}}^{a_{\max}}j_{\nu}^{a}(\mu,T_{\rm rot})\frac{1}{n_{\H}} \frac{dn}{da} da,\label{eq:jem}
\ena 
where $dn/da = dn_{\rm PAH, sil}/da$ for spinning PAHs and nanosilicates, respectively.

A list of notations used in this review and their meaning are shown in Tables \ref{tab:notations1} and \ref{tab:notations2}.

\section{Applications of RATD in~Astrophysics}\label{sec:appl}
\begin{figure}[H]
\centering
\includegraphics[width=0.9\textwidth]{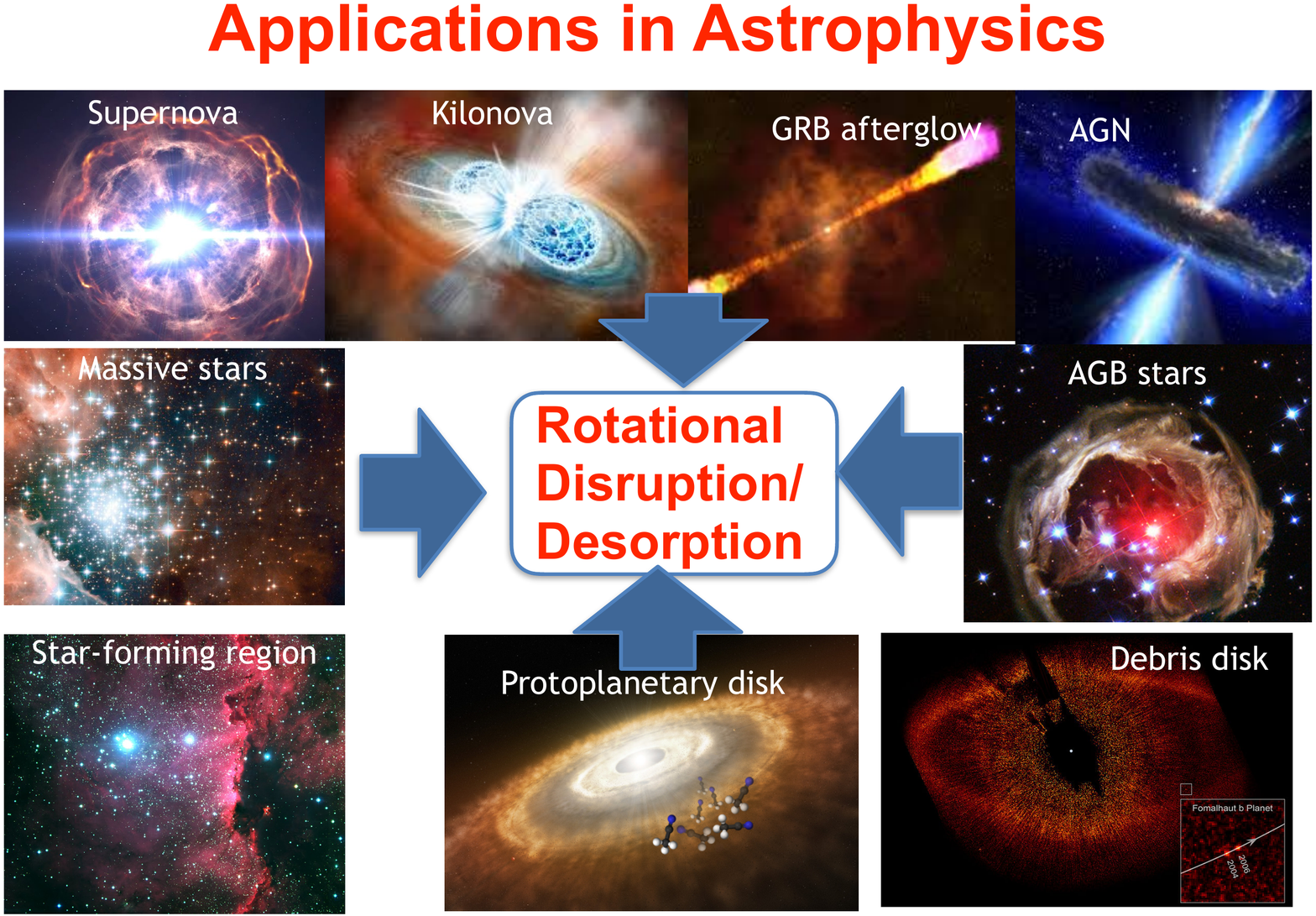}
\caption{Illustration of astrophysical environments where rotational disruption and desorption of dust and ice due to radiative torques are most important thanks to strong radiation fields. The~interstellar medium of our galaxy and other galaxies is not shown here, but~RATD is also important, especially for starburst and high-z~galaxies.}
\label{fig:RATD}
\end{figure}

The rotational disruption and desorption mechanisms induced by RATs that rely on dust-radiation interaction are effective in various environments with considerable radiation fields. Figure~\ref{fig:RATD} shows the selected environments for rotational disruption and desorption. Below, we will review the most important applications of these mechanisms for several astrophysical~environments.

\unskip

\subsection{Effect of RATD on Dust Evolution in the Interstellar~Medium}
The RATD has been used for studying the evolution of the grain size distribution of interstellar dust in our galaxy (\citet{2019ApJ...876...13H}), star-burst galaxies, and~high-redshift \mbox{galaxies (\citet{2020MNRAS.tmp.1093H})}. 

\begin{figure}[H]
\centering
\includegraphics[width=0.9\textwidth]{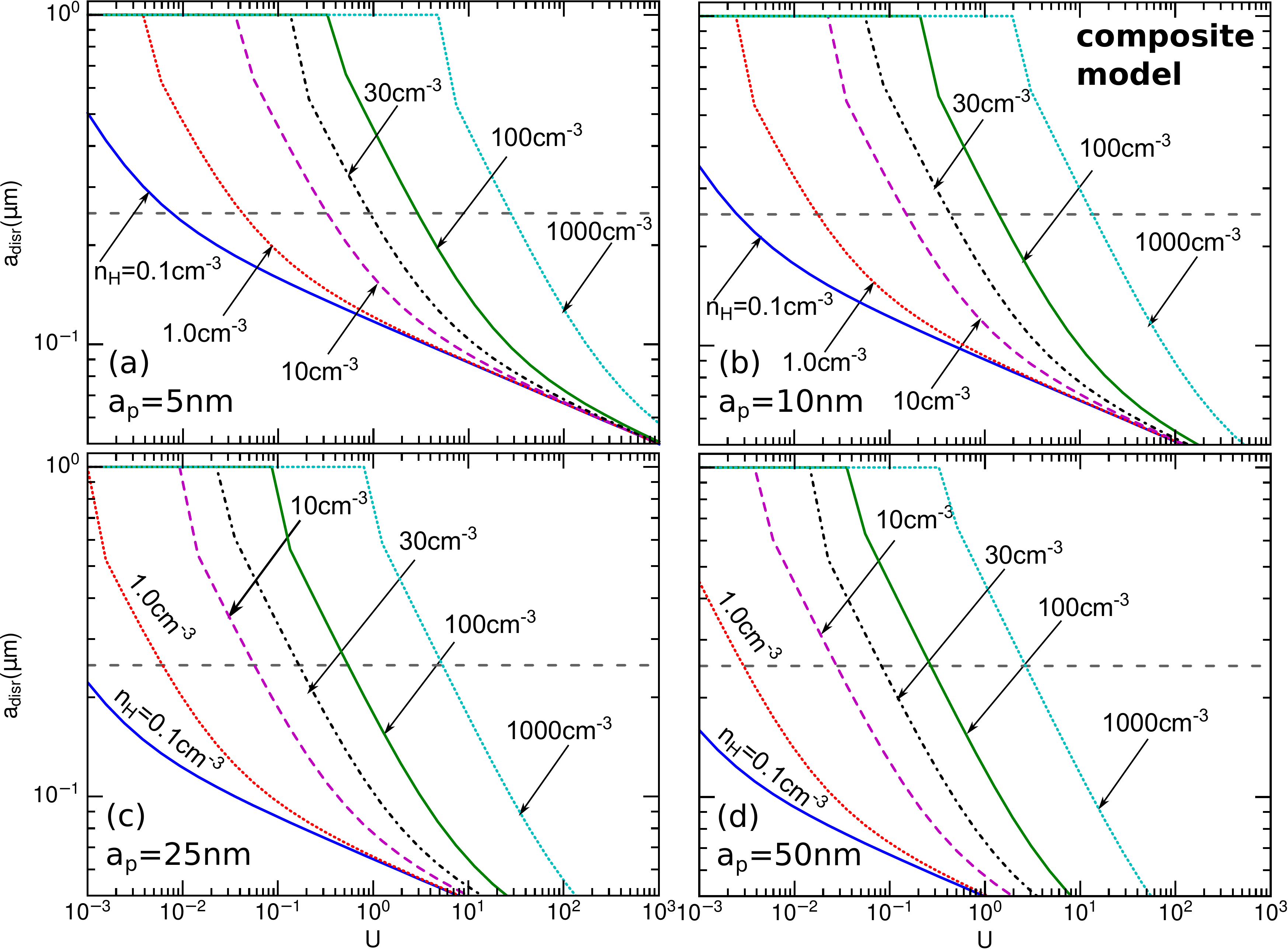}
\caption{Disruption grain size as a function of the radiation strength, $U$, for a composite grain model with porosity $P=0.2$ for the different gas density $n_{\H}$, assuming $a_{p}=5,10, 25$ and 50 nm (panels (a)-(d)). The horizontal dashed line marks the upper limit of the MRN size distribution $a_{\max}=0.25\mu$m. We set $a_{\rm disr}=1.0$ $\upmu$m in case of no disruption. The disruption size decreases with increasing $U$ due to stronger RATs, but it increases with increasing $n_{\rm H}$ due to the increase of rotational damping. Rotational disruption is more efficient for larger $a_{p}$ (panel (d)) due to lower tensile strength (see Eq. \ref{eq:Smax_comp}). From~\citet{2019ApJ...876...13H}.} 
\label{fig:adisr_comp}
\end{figure}

\citet{2019ApJ...876...13H} introduced a {\it dynamical constraint} for dust models using the RATD mechanism for the diffuse ISM. To~characterize the various radiation fields, we consider the different values of the radiation strength $U$, assuming the same radiation spectrum, that is,~$\bar{\lambda}$. Figure~\ref{fig:adisr_comp} shows the grain disruption by RATD for the various physical parameters $n_{\H},U$, assuming the radius of monomers $a_{p}=5,10, 25$ and 50 nm (panels (a)-(d)). The~rapid disruption by RATD compared to other destruction mechanisms (see Table~\ref{tab:destr}) establishes the upper cut-off of the grain size distribution. For~the Galaxy, the~RATD successfully reproduces the upper limit of $a_{\max}\sim 0.25$ $\upmu$m constrained by observations for a composite structure model of large~grains with $a_{p}=5$ nm (see panel (a)).

The effect of RATD on dust evolution in galaxies is carried out in \citet{2020MNRAS.tmp.1093H} where the authors model the grain size distribution for the different ages of the universe. For~the typical starburst model, we examine $U=1000$ (corresponding to $T_\mathrm{d}\sim 60$ K) for an extreme ISRF environment actually observed in starburst galaxies (e.g., \citet[]{Zavala:2018cn,Lim:2019ir}).

\begin{figure}[H]
\includegraphics[width=0.5\textwidth]{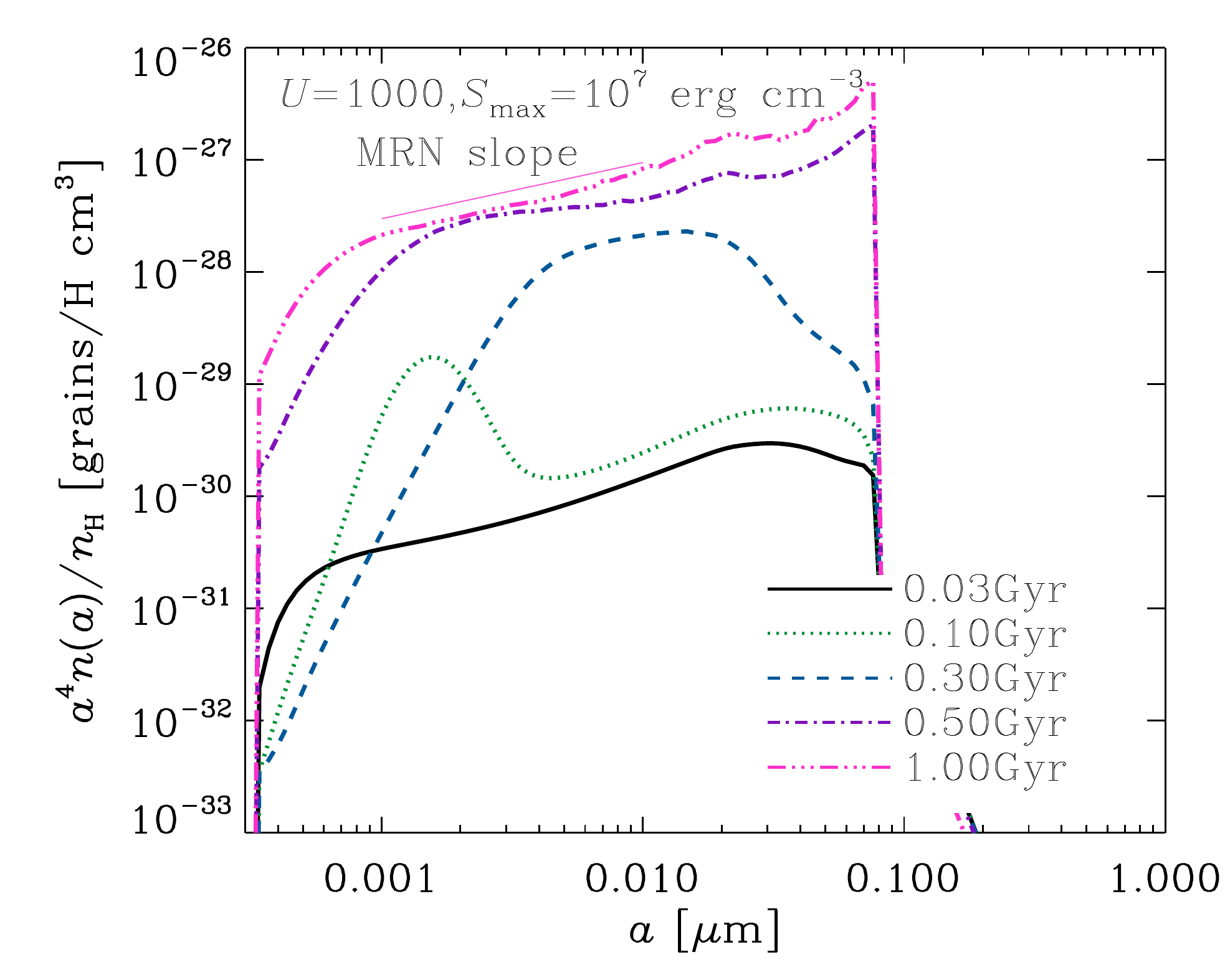}
\includegraphics[width=0.5\textwidth]{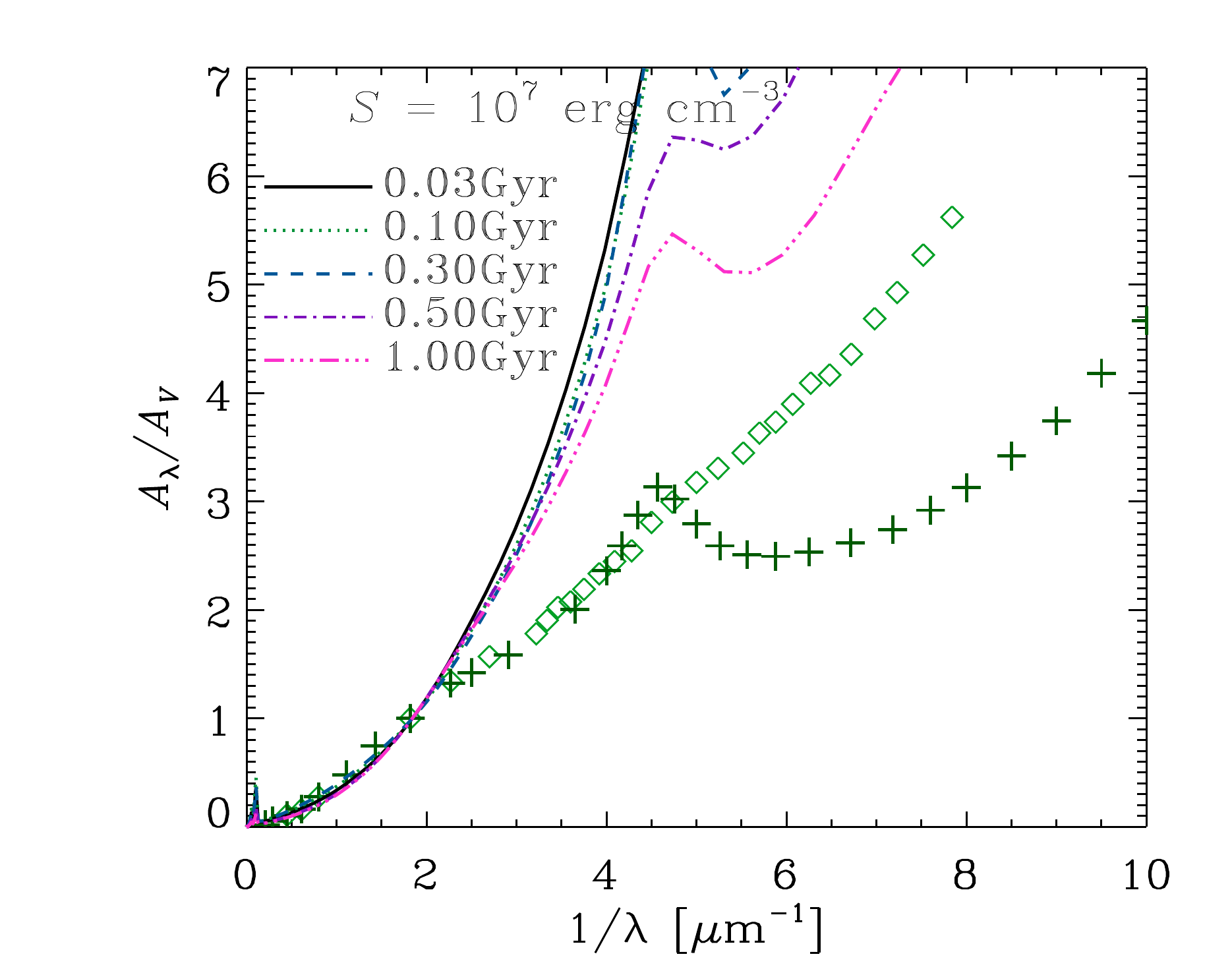}
\caption{Evolution of grain size distribution (left panel) and extinction curve (right panel) for the starburst model. The~solid, dotted, dashed, dot--dashed, and~triple-dot--dashed lines show the grain size distributions at $t=0.03$, 0.1, 0.3, 0.5, and~1 Gyr, respectively. The~thin dotted line shows the MRN 
slope. We adopt a typical value $S_\mathrm{\max}=10^{7}\erg\cm^{-3}$ and fix $U=1000$. In~the right panel, crossed and squared symbols show the MW 
and Small Magellanic Cloud (SMC) extinction curves. From~\citet{2020MNRAS.tmp.1093H}.}
\label{fig:size_sb}
\end{figure}

Figure~\ref{fig:size_sb} (left panel) shows the resulting grain size
distributions. Since the evolutionary time-scale is short,
we show $t=0.03$, 0.1, 0.3, 0.5, and~1 Gyr. Small grains are abundant already at $t<0.1$ Gyr because rotational disruption supplies small~grains.

\unskip
\unskip

Figure~\ref{fig:size_sb} (right panel) shows the extinction curves corresponding to the above grain size distributions for the starburst models. We observe that the extinction curves stay steep for \mbox{$S_\mathrm{\max}=10^7$ erg cm$^{-3}$} because of the small maximum grain radius ($a_\mathrm{\rm disr}$). A~significant steepening of the extinction curve is seen even at $t<0.1$ Gyr because of the efficient small-grain production by rotational disruption. It is interesting to point out that the extinction curves are similar to the Small Magellanic Cloud (SMC) extinction curve at young ages. The~extinction curves become flatter at later stages because of coagulation. Therefore, in~starburst galaxies, the~small-grain production by rotational disruption could put a significant imprint on the extinction curves, especially at young~ages.

For some starburst galaxies, attenuation curves, which include all the radiation transfer effects, are~obtained instead of extinction curves  (e.g., \citet{2001PASP..113.1449C}). The~effects of dust distribution
geometry and of stellar-age-dependent extinction make the attenuation curve significantly different from the original extinction curve
(e.g., \citet{2000ApJ...528..799W,Inoue:2005hf,2016ApJ...833..201S}).
Therefore, the~fact that the extinction curves derived in this paper
are different from the so-called Calzetti attenuation curve is not a~contradiction.

The evolution of grain temperature (radiation strength $U$) with redshift is still being debated. While there have been some observational evidence that the dust temperature tends to be higher in higher redshift galaxies (\citet{Symeonidis:2013ef,Bethermin:2015hn,Schreiber:2017dk,Zavala:2018cn}), the~trend could be driven by an observational bias (\citet{Lim:2019ir}). \citet{Zavala:2018cn} derived the best-fit $T_{d}\sim 12(z+1)+11 \K$, which corresponds to $U\sim (T/16.5)^{6}\sim 0.15(z+2)^{6}$. Thus, for~$z\sim3$, one already has $U\sim 2300$.

At $z\gtrsim 5$, Lyman break galaxies (LBGs) and Lyman $\alpha$ emitters have dust temperatures typically higher than $\sim$35 K, and~some could have dust temperatures as high as $\gtrsim$70 K \mbox{(e.g., \citet{2017arXiv170902526H,Bakx:2020iy})}. \citet{2017MNRAS.471.5018F} also theoretically suggested that the dust temperature in the diffuse ISM of high-redshift LBGs could be as high as 35--60 K. These dust temperatures correspond to $U\sim$ a few tens to a few thousands. There are also some extreme populations of galaxies whose dust temperatures even reach $\sim$90 K (\citet{2019arXiv191205813T}). This indicates that rotational disruption could have a significant imprint on the grain size distributions and the extinction curves in high-redshift~galaxies.

\subsection{Constraining Grain Internal Structures With~Observations}
\begin{figure}[H]
\includegraphics[width=0.5\textwidth]{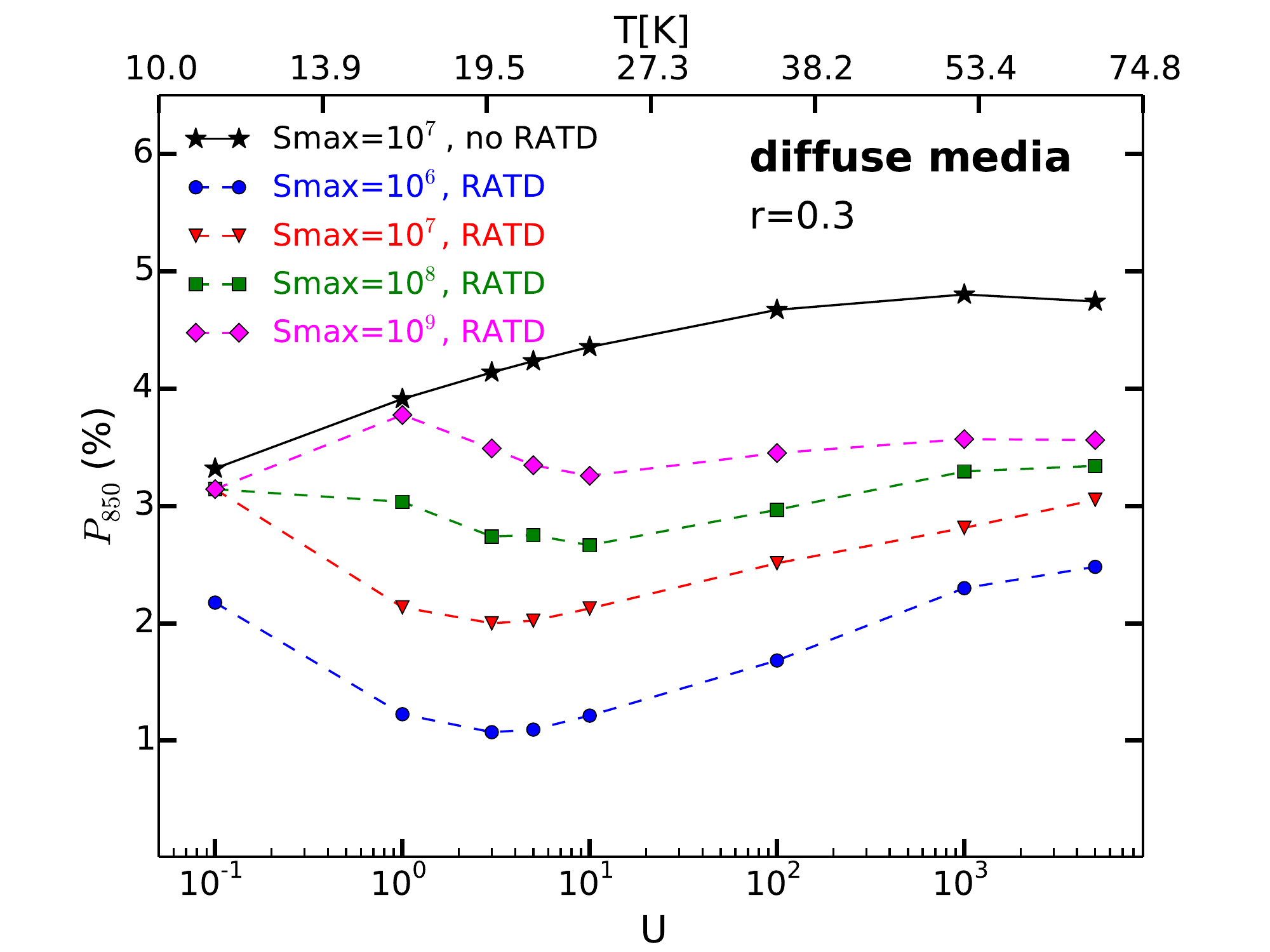} 
\includegraphics[width=0.5\textwidth]{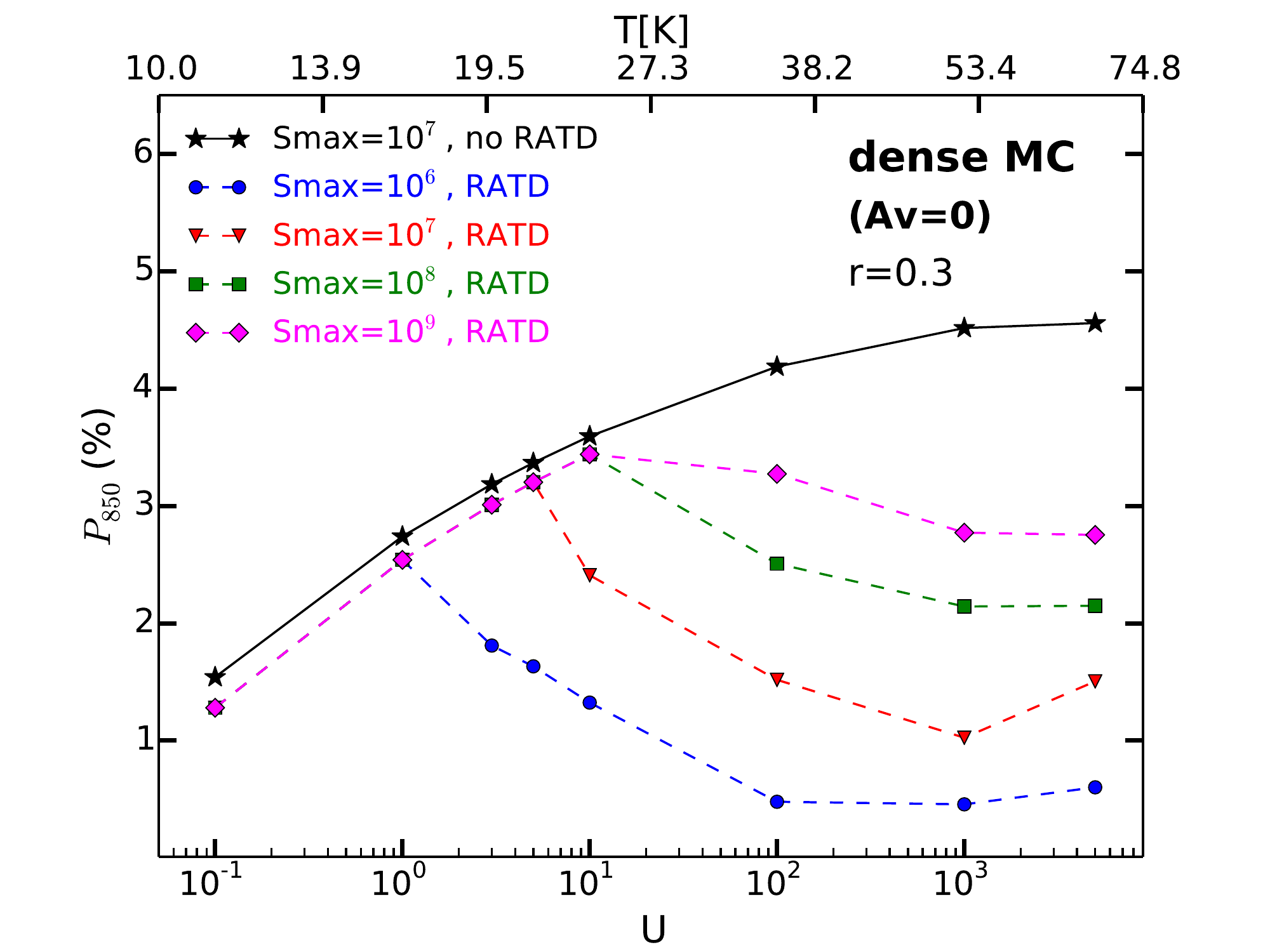} 
\caption{Polarization degree at 850 $\upmu$m as a function of the radiation strength ($U$) or grain temperature ($T_d$, top horizontal axis) for two cases, without~RATD (solid lines) and with RATD (dashed lines), assuming the different tensile strength of grains in the diffuse ISM (left panel) and MC (right panel). Grains with axial ratio of $r=1/3$ are considered. From~\citet{2020ApJ...896...44L}.}
\label{fig:Pem850}
\end{figure}

One of the least known properties of interstellar dust is its internal structure, which determines the grain tensile strength. Using the RATD, \citet{2020ApJ...896...44L} modeled the polarization by aligned grains for various tensile strengths and grain temperatures (radiation strength). Since RATD is most efficient for the largest grains, which dominate the polarized emission in submm/far-IR, it is appropriate to consider how the submm polarization changes with $U$ and $S_{\rm max}$. The~obtained results are shown in Figure~\ref{fig:Pem850}. 

In the absence of grain disruption by RATD, the~polarization at 850 $\upmu$m, denoted by $P_{850}$, increases~monotonically with the radiation intensity (i.e., grain temperature) over the considered range of $U$. The~absence of RATD is equivalent to the situation where grains are made of ideal material without impurity such that the tensile strength is as high as $S_{\max}\sim 10^{11}\erg\cm^{-3}$ (e.g., diamonds). However, when the RATD effect is taken into account for grains made of weaker material ($S_{\max}\lesssim 10^{9}\erg\cm^{-3})$, the~polarization degree $P_{850}$ first increases from a low value of $U$ and then decreases when $U$ becomes sufficiently large. The~critical value $U$ at the turning point is determined by the value $S_{\rm max}$ and local gas density $n_{\H}$ that controls the grain disruption size $a_{\rm disr}$ according to~RATD. 

\citet{Guillet:2017hg} performed a detailed analysis of the variation of $P_{850}$ with the radiation field using {\it Planck data} and discovered that $P_{850}$ first increases with increasing grain temperature from \mbox{$T_{d}\sim$ 16--19 K} and then drops as the dust temperature increases to \mbox{$T_{d}\gtrsim 19\K$}. Such an unusual \mbox{$P_{850}-T_{d}$} relationship cannot be reproduced if large grains are not disrupted (i.e., RATD is not taken into account), as~shown in Figure~\ref{fig:Pem850}. However, the~observed trend is, in~general, consistent with our model with RATD for grains with a tensile strength of $S_{\max}\lesssim 10^{9}\erg\cm^{-3}$. This range of tensile strength favors a composite internal structure of grains over the compact~one.

\unskip

\subsection{Effect of RATD on Colors and Light-Curves of Cosmic~Transients}
Absorption and scattering by foreground dust control the color and light-curves of astrophysical transients (i.e., novae, supernovae, GRBs, kilonovae). In~light of the RATD effect, dust properties are rapidly changed under intense radiation of transients, which are essential for the time-domain astronomy era. In~this subsection, we apply the RATD to model time-varying extinction, color, and~polarization of astrophysical transients. We will focus on Type Ia Supernova and GRB~afterglows.
\subsubsection{Type Ia~Supernovae}

Type Ia supernovae (SNe Ia) have widely been used as standard candles to measure the expansion of the universe due to their stable intrinsic luminosity (\citet{1998AJ....116.1009R}). To~achieve the most precise constraints on cosmological parameters, the~effect of dust extinction on the SNe Ia's intrinsic light curve must be accurately characterized. Optical to near-infrared photometric observations of SNe Ia during the early phase (i.e., within~a few weeks after maximum brightness) reveal unusual properties of dust extinction, with~unprecedented low values of $R_{\rm V}\lesssim 2$ (Refs. \citet{2008A&A...487...19N,2014ApJ...789...32B}), much lower than the standard Milky Way value of $R_{\rm V}\sim 3.1$ (Ref. \citet{2003ARA&A..41..241D}). Moreover, polarimetric observations also report unusually low wavelengths of the maximum polarization\mbox{ ($\lambda_{\rm max}<0.4$ $\upmu$m)} for several SNe Ia (\citet{Kawabata:2014gy,Patat:2015bb}). Numerical modeling of dust extinction (\citet{2016P&SS..133...36N}) and polarization curves (\citet{2017ApJ...836...13H}) toward individual SNe Ia demonstrate that the anomalous values of $R_{\rm V}$ and $\lambda_{\rm max}$ can be reproduced by the enhancement in the relative abundance of small grains to large grains in the host galaxy. The~RATD mechanism could resolve this puzzle, as~proposed by \citet{Hoang:2019da}.

\begin{figure}[H]
\includegraphics[width=0.33\textwidth]{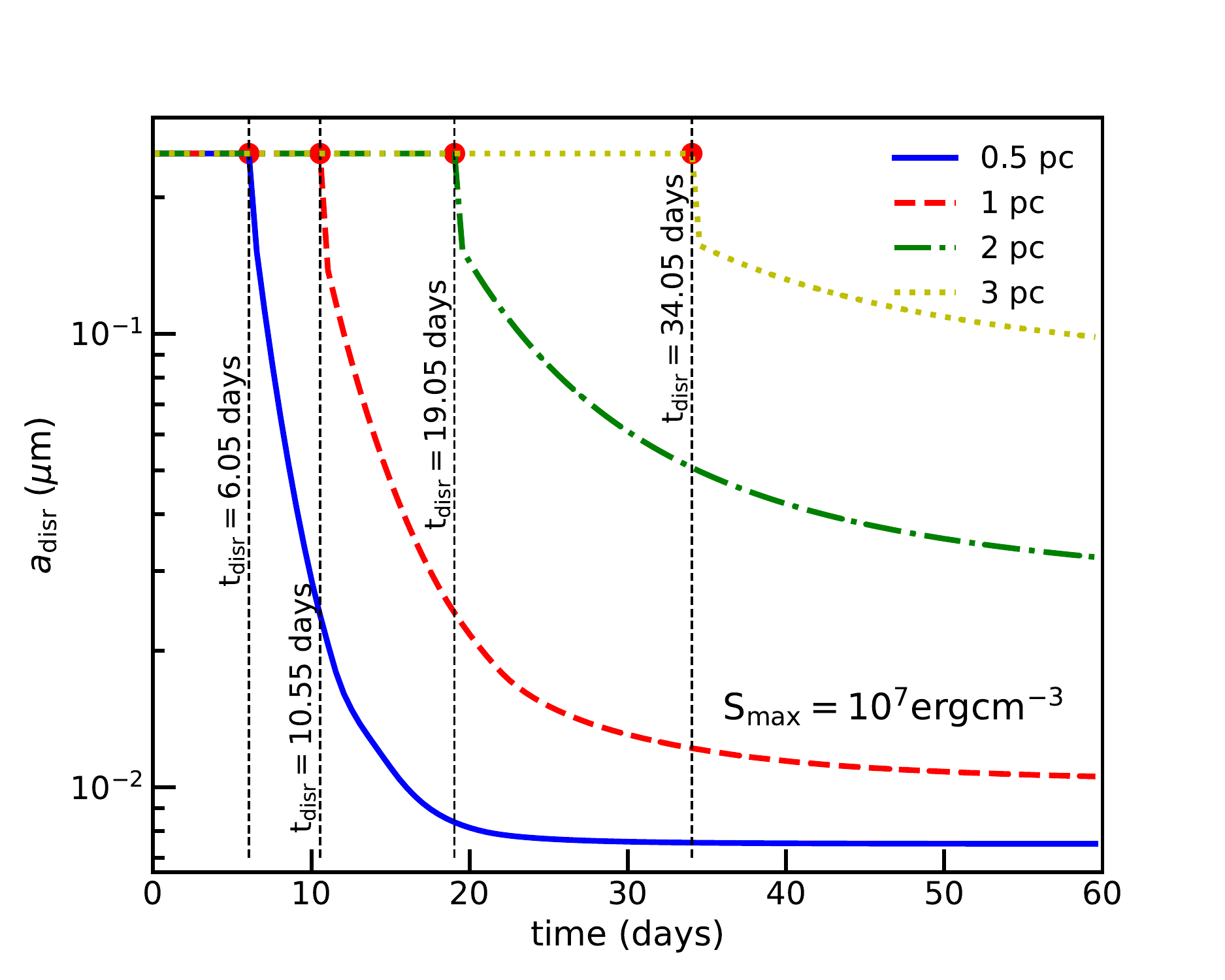}           
\includegraphics[width=0.33\textwidth]{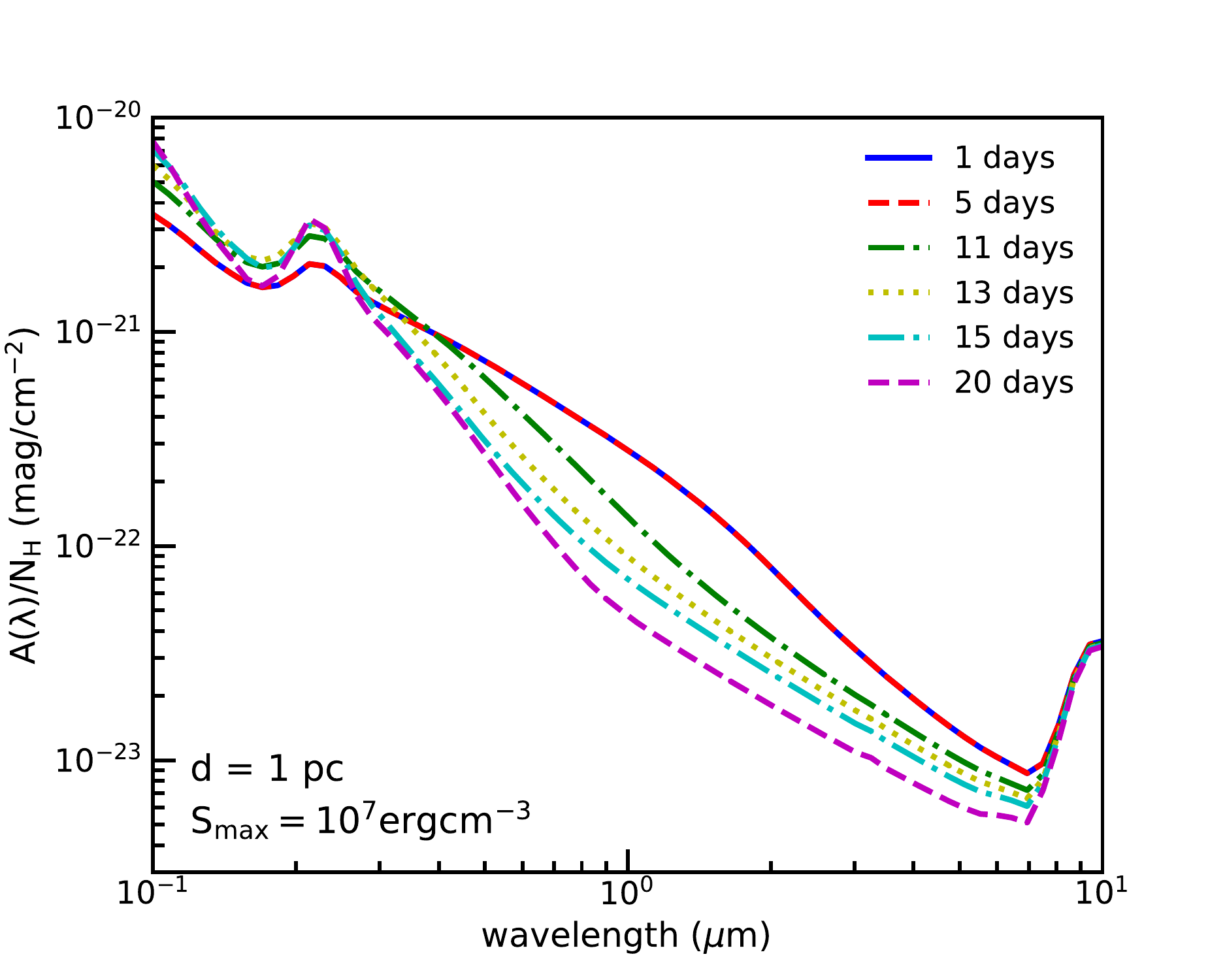}            \includegraphics[width=0.33\textwidth]{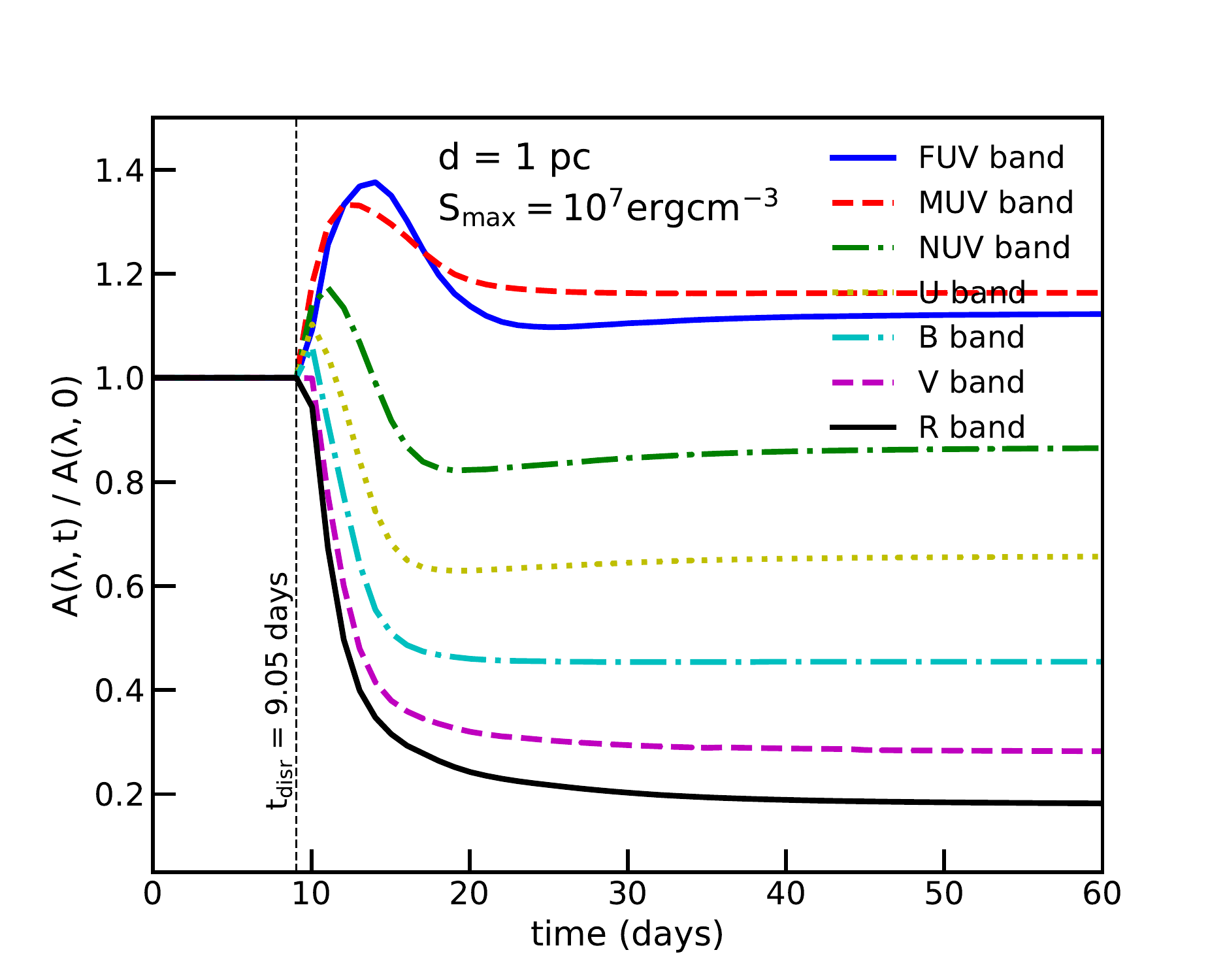}
\caption{Left panel: Disruption sizes vs. time for dust grains at different distances. Middle panel: Extinction curves evaluated at different times for $S_{\rm max}=10^{7}\erg\cm^{-3}$, assuming the dust cloud at 1 pc from SNe Ia. Optical-NIR extinction decreases while UV extinction increases over time due to RATD. Right panel: Variation of $A(\lambda,t)/A(\lambda,0)$ from FUV-R bands with time. Vertical lines mark the disruption time of graphite which occurs earlier than silicates. The~ratio is constant initially and starts to vary with time when RATD begins at $t_{\rm disr}$. After~the disruption ceases, the~ratio is constant again. From~\citet{2020ApJ...888...93G}.}
\label{fig:Alamda_t}
\end{figure}

Using the RATD theory, \citet{2020ApJ...888...93G} performed modeling of time-varying disruption, extinction, and~polarization. Figure~\ref{fig:Alamda_t} shows our results for dust grains having a maximum tensile strength $S_{\rm max}=10^{7}\erg\cm^{-3}$. The~disruption occurs on a short time of tens of days for dust clouds located within pc (left panel). Figure~\ref{fig:Alamda_t} (middle panel) shows the extinction curves (see Section~\ref{sec:ext}) evaluated at different times for dust grains located at distance $d= 1$ pc from the source. The~extinction curve at $\rm t = 5$ days (red dashed line) is the same as the extinction at $t=1$ days because $t<t_{\rm disr}$ \mbox{(see Equation~(\ref{eq:tdisr1}))}. For~$t> t_{\rm disr}\sim 10$ days, the~optical-NIR extinction decreases rapidly with time due to the removal of large grains by RATD. On~the other hand, the~UV extinction is increased due to the enhancement in the abundance of small grains by RATD. The~extinction at $\lambda > 7$ $\upmu$m is essentially unchanged because of the wavelength is much larger than the grain radius, that is,~$\lambda\gg (2\pi a)$.

Figure~\ref{fig:Alamda_t} (right panel) shows the time-dependence of the ratio $ A(\lambda,t)/A(\lambda,0)$ for the different photometric bands (FUV to R bands) and various cloud distances. Here we choose $ \lambda=0.15$ $\upmu$m for the far-UV band (FUV), $\lambda=0.25$ $\upmu$m for the mid-UV band (MUV) and $ \lambda=0.3$ $\upmu$m for the near-UV band (NUV). As~shown, the~ratio $A(\lambda,t)/A(\lambda,0)$ is constant during the initial stage of $t<t_{\rm disr}$ before grain disruption, and~it starts to rapidly change when RATD begins at $t_{\rm disr}$. 

\begin{figure}[H]
\includegraphics[width=0.5\textwidth]{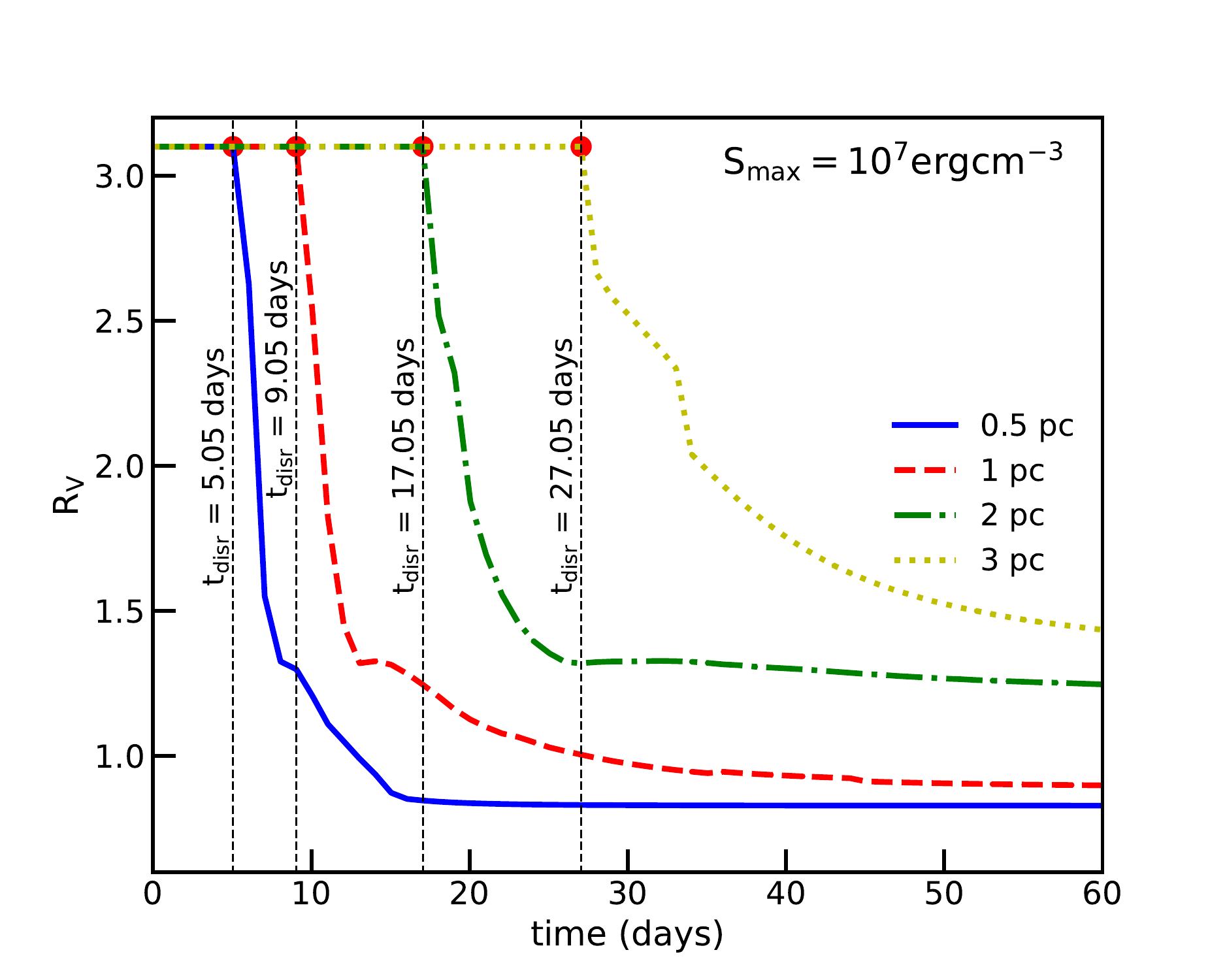}      
\includegraphics[width=0.5\textwidth]{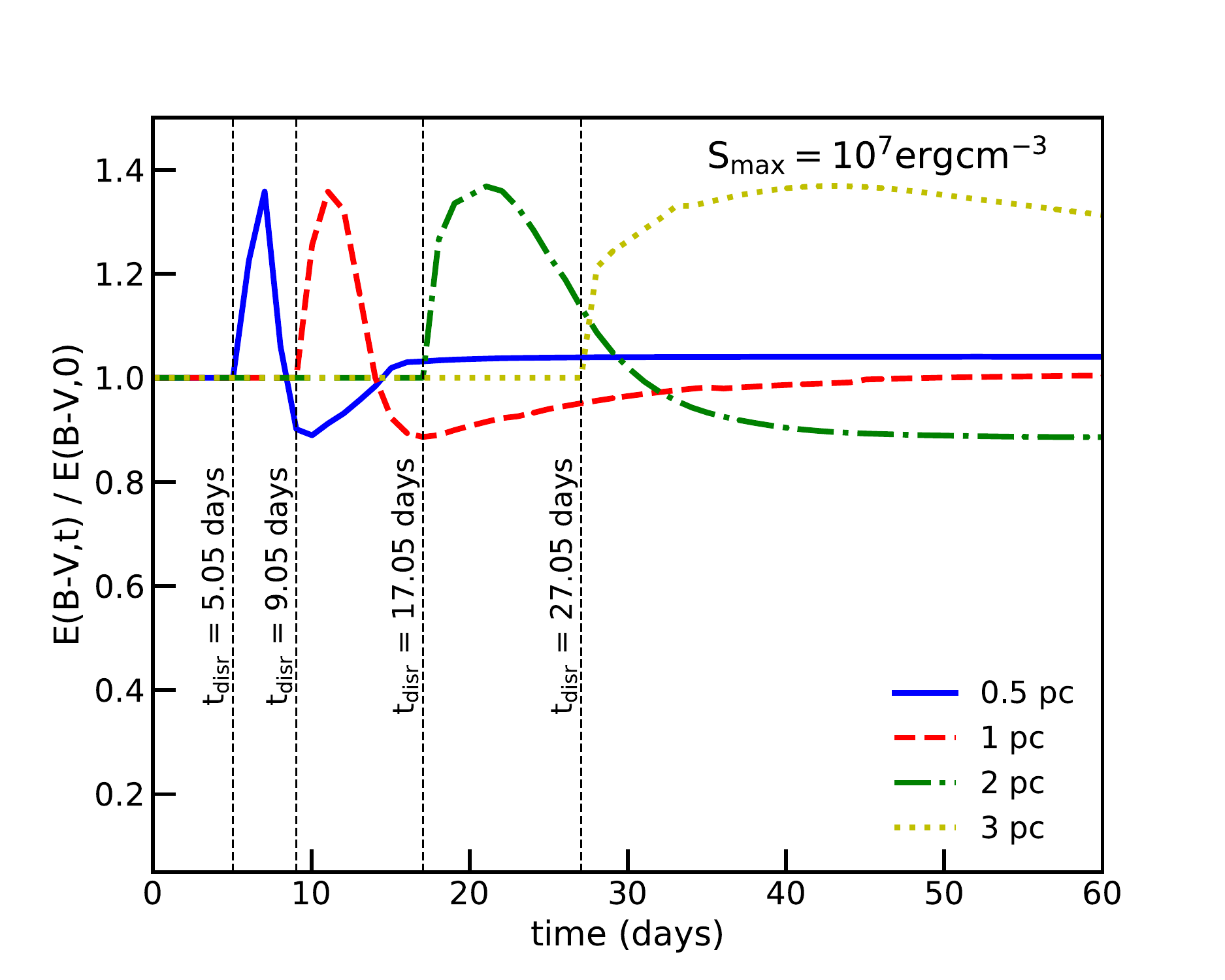}
\caption{Time variation of $R_{\rm V}$ (left panel) and $E(B-V)$ (right panel) for different cloud distances and $S_{\rm max}=10^{7} \erg \cm^{-3}$. Both $E(B-V)$ and $R_{V}$ begin to change when grain disruption starts at $t\sim t_{\rm disr}$ (marked by vertical dotted lines). $R_{\rm V}$ decreases rapidly from their original values from $t=t_{\rm disr}$ to 40~days and then almost saturates when RATD ceases. From~\citet{2020ApJ...888...93G}.} 
    \label{fig:Rv}
\end{figure}

From $A(\lambda,t)$, we can calculate the color excess $E(\rm B-V)=A_{\rm B}-A_{\rm V}$ and $R_{\rm V}=A_{\rm V}/E(B-V)$ to understand how these quantities vary with time due to RATD. Figure~\ref{fig:Rv} (left panel) shows the variation of $R_{\rm V}$ with time. The~value $R_{\rm V}$ starts to rapidly decrease from the initial standard value of $R_{\rm V}=3.1$ to $R_{V}\sim$ 1--1.5 after less than $40$ days. The~moment where $R_{V}$ starts to decline is similar to the grain disruption time $t_{\rm disr}$. The~time required to decrease $R_{\rm V}$ from its original value is shorter for grains closer to the source, and~the terminal value of $R_{V}$ is also smaller. Figure~\ref{fig:Rv} (right panel) shows the variation of $E(B-V,t)/E(B-V,0)$ with time for the different cloud distances. For~a given cloud distance, the~color excess remains constant until grain disruption begins at $t\sim t_{\rm disr}$. Subsequently, the~ratio increases rapidly and then decreases to a saturated level when RATD~ceases. 

\begin{figure}[H]
\includegraphics[width=0.5\textwidth]{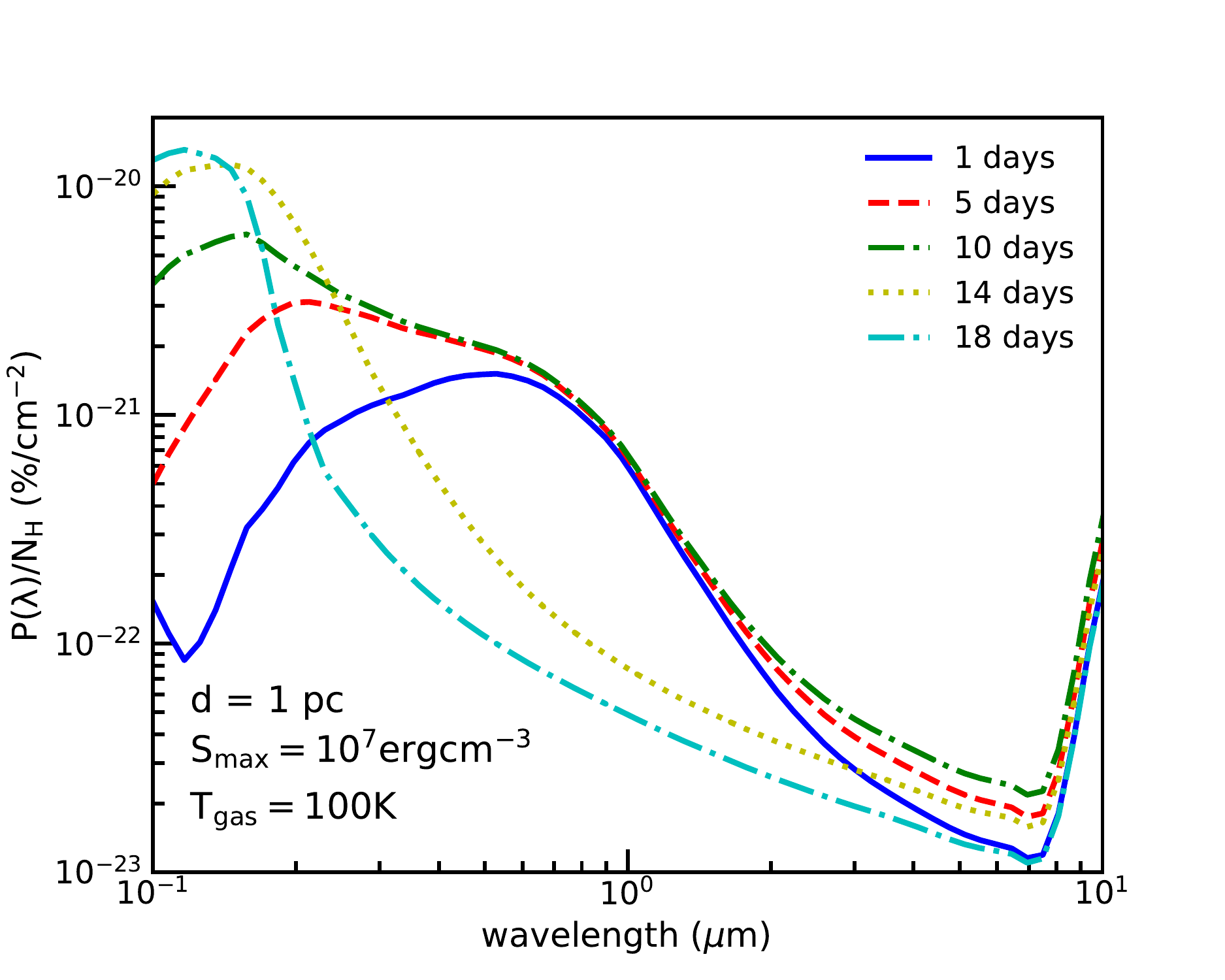}
\includegraphics[width=0.5\textwidth]{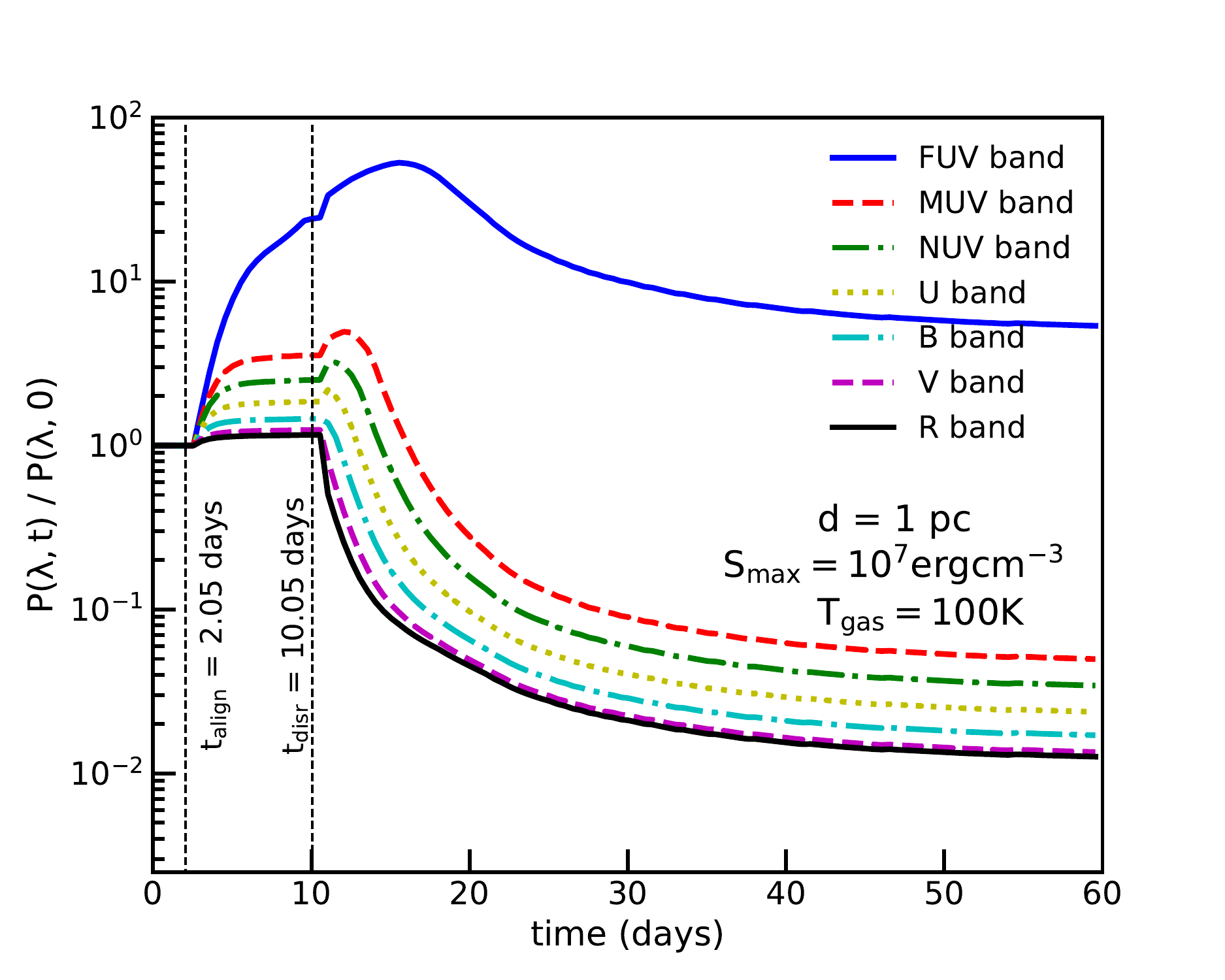}
\caption{Left panel: Polarization curves evaluated at different times for a  dust cloud at 1 pc, \mbox{assuming $S_{\rm max} = 10^{7} \erg\cm^{-3}$}. Enhanced~alignment of small grains induces the blueshift of the peak wavelength. The~RATD effect reduces polarization at $\lambda>0.3$ $\upmu$m, and~the efficiency is weaker for higher $S_{\rm max}$. Polarization~curves evaluated at $t=20$ days for different cloud distances, assuming $S_{\rm max} = 10^{7} \erg\cm^{-3}$. Right~panel: Ratio $P(\lambda,t)/P(\lambda,0)$ vs. time from far-UV band (FUV) to V band for different cloud distances assuming \mbox{$S_{\rm max}=10^{7}\erg\cm^{-3}$}. Optical/NIR polarization degree first increases due to enhanced alignment by RATs and then declines when grain disruption by RATD starts. Dotted vertical lines mark alignment time ($t_{\rm align}$) and disruption time ($t_{\rm disr}$) of silicate grains. From~\citet{2020ApJ...888...93G}.}
\label{fig:Plamda_t} 
\end{figure}

Figure~\ref{fig:Plamda_t} (left panel) shows the polarization curve produced by aligned grains with the magnetic field (see Section~\ref{sec:ext}) computed at different times for a dust cloud at 1 pc, assuming $S_{\rm max}=10^{7}\erg\cm^{-3}$. At~$t\lesssim1$ days, dust grains are aligned by the average diffuse interstellar radiation, so the maximum polarization occurs at $\lambda_{\rm max}\sim 0.55$ $\upmu$m. After~that, SNe radiation dominates and makes smaller grains to be aligned. As~a result, the~UV polarization is increased rapidly, and~the peak wavelength of $\lambda_{\rm max}$ is decreased. The~degree of optical-NIR polarization ($\lambda > 0.5$ $\upmu$m) is slightly increased. After~$t\sim 10$ days, grain disruption by RATD begins, reducing the abundance of large grains. Therefore, the~degree of optical-NIR polarization decreases substantially, which results in a narrower polarization profile compared to the original polarization~curve. 

Figure~\ref{fig:Plamda_t} (right panel) shows the temporal variation of $P(\lambda,t)/P(\lambda,0)$ from FUV to R bands for the different cloud distances. During~the initial stage, the~ratio $P(\lambda,t)/P(\lambda,0)$ is constant, however, this stage is rather short, between~1--5 days, corresponding to the alignment timescale $t_{\rm align}$ (see also \citet{2017ApJ...836...13H}). After~that, the~polarization degree increases gradually, and~this rising period continues until $t\sim$ 5--30 days when RATD begins (i.e., at~$t=t_{\rm disr}$) for $d=$ 0.5--3 pc. After~that, the~polarization degree declines rapidly and achieves a saturated level when RATD ceases, which occurs after $t\sim 20$ days for $d=0.5$ pc, $40$ days for $d=1$ pc, respectively. In~summary, due to RAT alignment and RATD, the~polarization degree increases from $t_{\rm align}$ to $t_{\rm disr}$, and~it decreases rapidly at $t> t_{\rm disr}$.

\begin{figure}[H]
\includegraphics[width=0.5\textwidth]{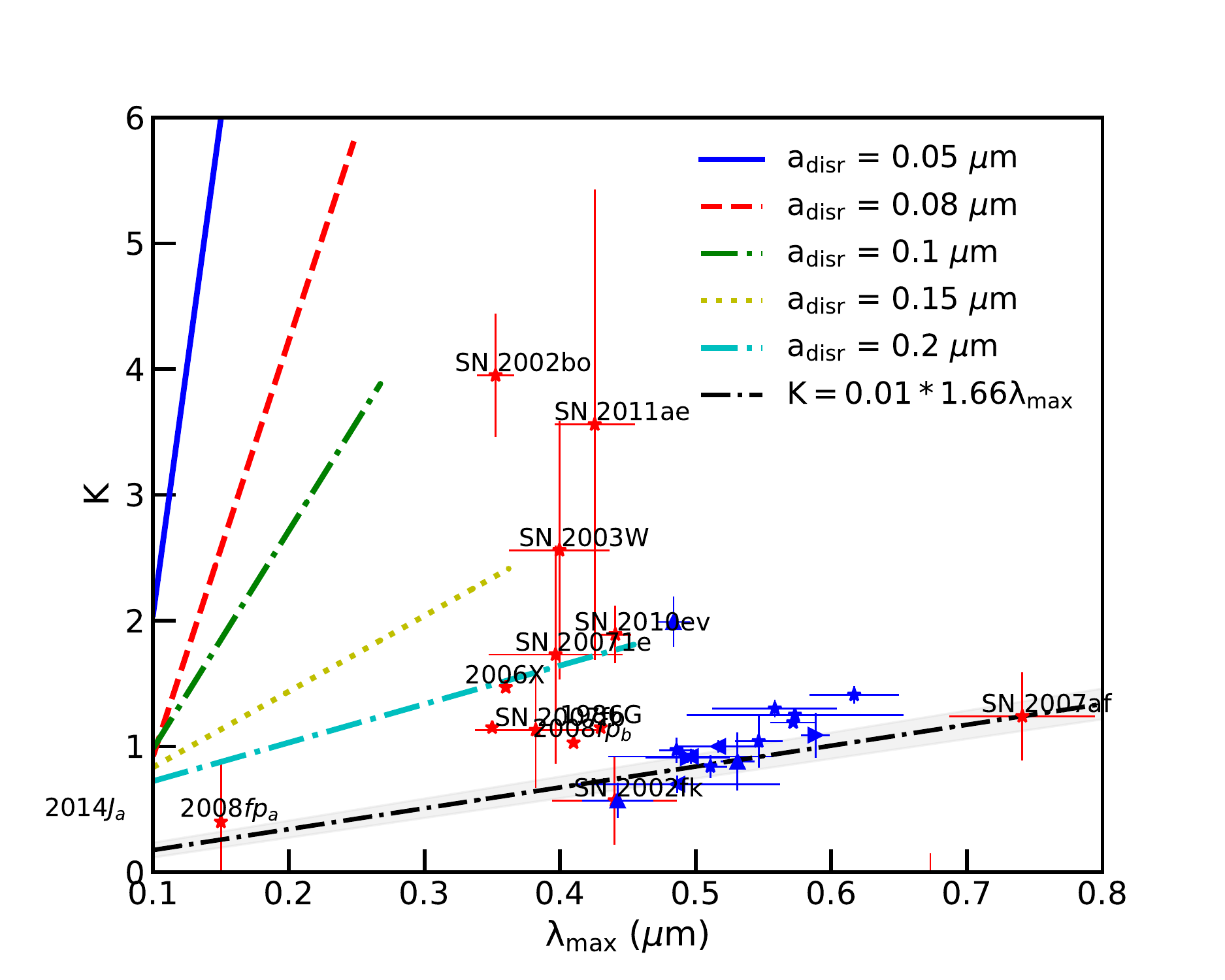}               
\includegraphics[width=0.5\textwidth]{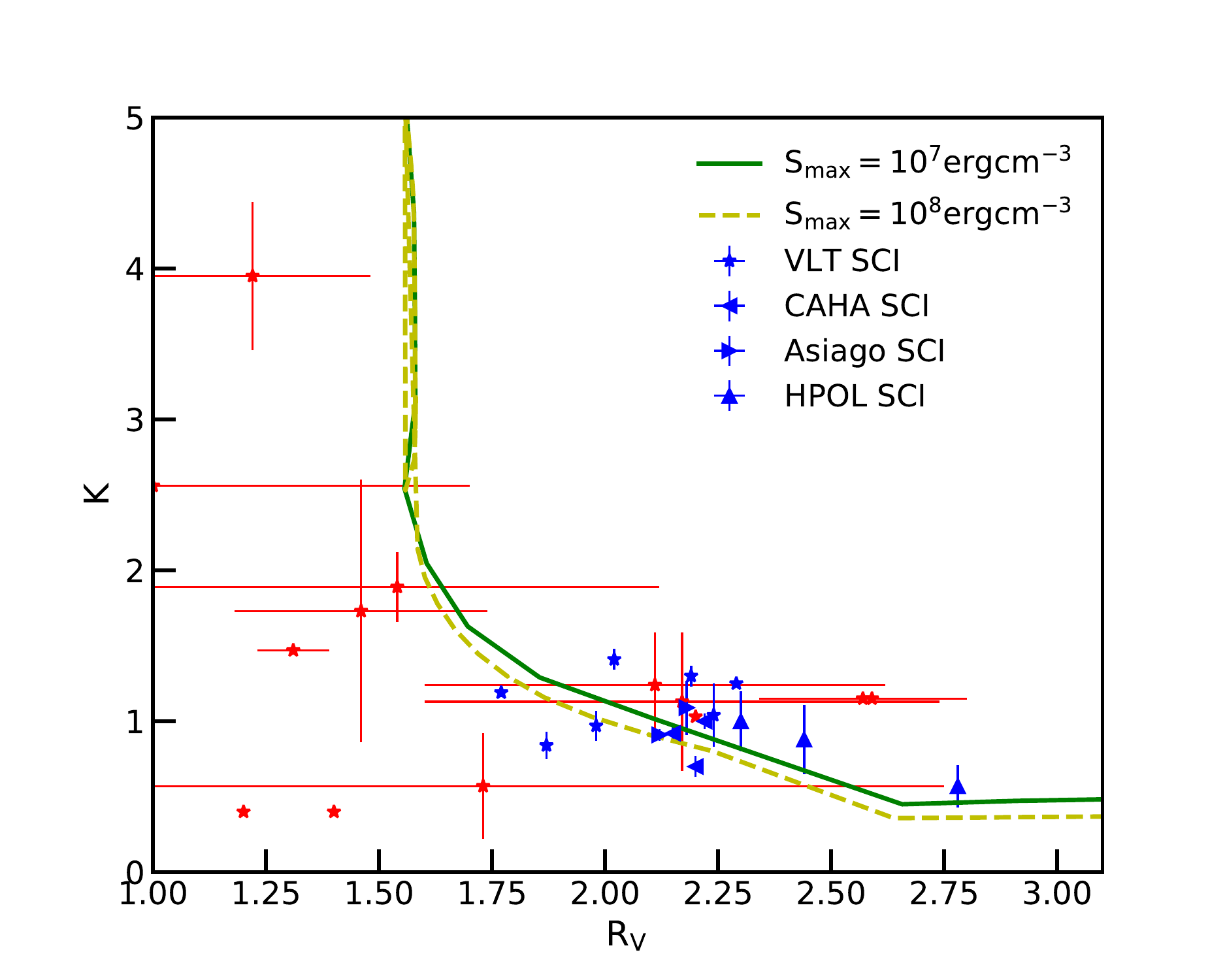}
\caption{Left panel: Relationship between $K$ and $\lambda_{\rm max}$ predicted by our model for different disruption sizes and alignment size. The~black line is the standard $K-\lambda_{\rm max}$ relationship. Right panel: $K$ vs. $R_{V}$ from our models compared with observational data for SNe Ia. Symbols show observational data for SNe Ia presented (red symbols) and normal stars in our galaxy (blue symbols). From~\citet{2020ApJ...888...93G}.}
\label{fig:K_max}
\end{figure}

Figure~\ref{fig:K_max} compares our modeling results for $K-\lambda_{max}$ and $K-R_{V}$ with observational data where $K,\lambda_{\max}$ are two parameters present in the Serkowski law (\citet{Serkowski:1975p6681}):
\begin{align} \label{eq:22}
P_{Serk}({\lambda}) = P_{\rm max} \exp\left[-K \ln^{2}\left(\frac{\lambda_{\rm max}}{\lambda}\right)\right],
\end{align}
where $P_{\rm max}$ is the maximum degree of polarization, $\lambda_{\rm max}$ is the wavelength at the peak wavelength, and~$K$ is a parameter (\citet{1980ApJ...235..905W,Whittet:1992p6073}). Here, the~model parameters  $K$ and $\lambda_{\rm max}$ are obtained by fitting the Serkowski law (Equation (\ref{eq:22})) to the polarization curve calculated by Equation~(\ref{eq:16}).

The left panel of Figure~\ref{fig:K_max} shows the peak wavelength $\lambda_{\rm max}$ and the parameter $K$ for several values of the grain disruption size from $a_{\rm disr}=0.2$ $\upmu$m to $0.05$ $\upmu$m. For~each value $a_{\rm disr}$, the~alignment size is varied from $a_{\rm align}=0.05$ $\upmu$m to $\rm 0.002$ $\upmu$m to account for the effect of enhanced alignment by SNe light for more details). We see that, for~a given $a_{\rm disr}$, $K$ decreases rapidly with decreasing $\lambda_{\rm max}$ due to the decrease of $a_{\rm align}$. Moreover, for~a given $a_{\rm align}$, $\lambda_{\rm max}$ tends to decrease with decreasing $a_{\rm disr}$. In~particular, we see that the high $K$ values of SNe Ia could be reproduced by our models with RATD with different $a_{\rm disr}$. The~right panel of Figure~\ref{fig:K_max} shows the variation of $K-R_{V}$ where a decrease of $K$ with $R_{V}$ is observed. The~model is in good agreement with observational~data.

\subsubsection{GRB~Afterglows}
GRBs are thought to explode in dusty clouds. Thus, their intense radiation fields are expected to have a dramatic impact on the surrounding dust, which in turn affects the color and light-curves of the GRB afterglow. In~\citet{2020ApJ...895...16H}, we model the effect of RATD on the extinction, polarization, and~light-curves of GRB~afterglows.

\begin{figure}[H]
\includegraphics[width=0.5\textwidth]{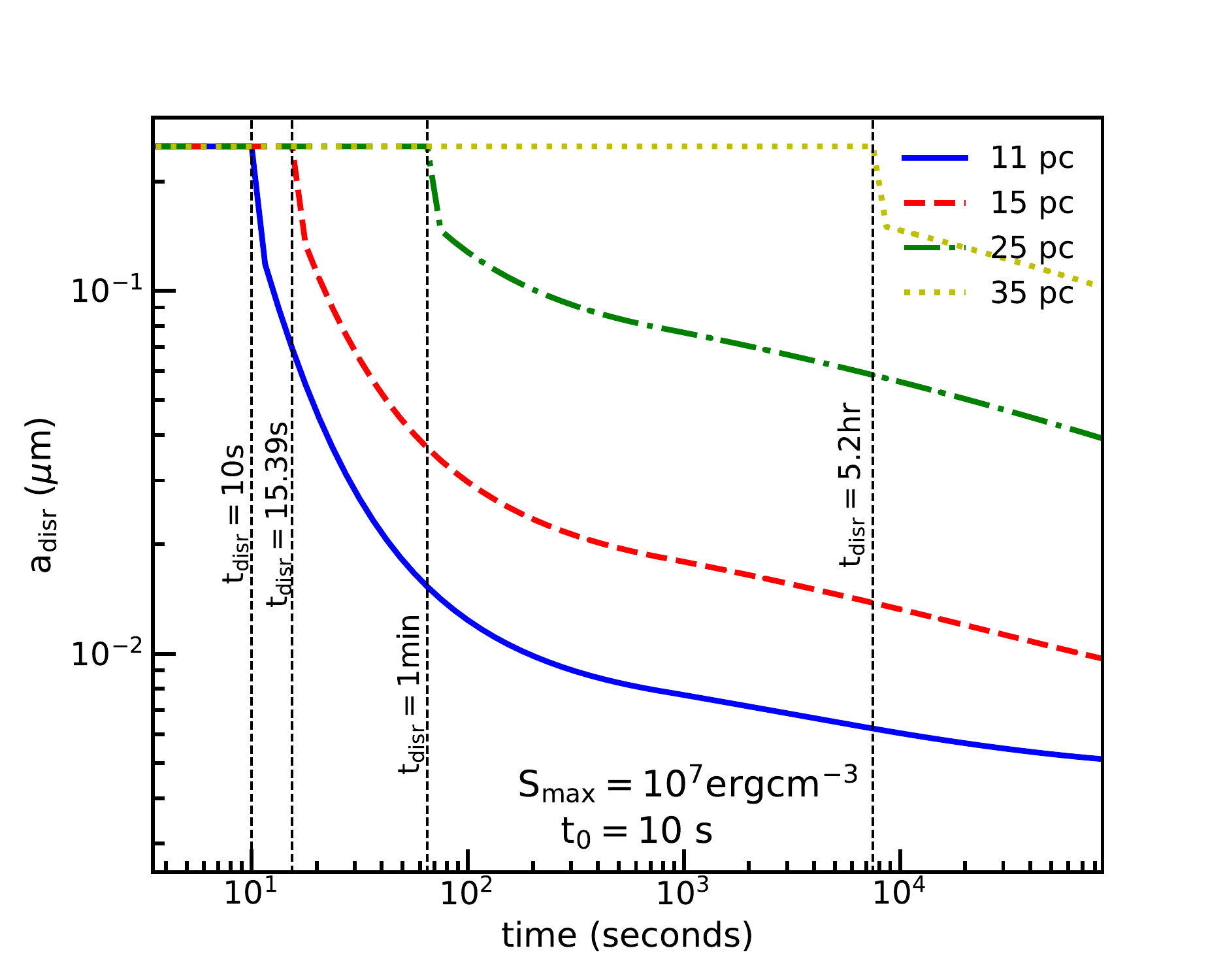}
\includegraphics[width=0.5\textwidth]{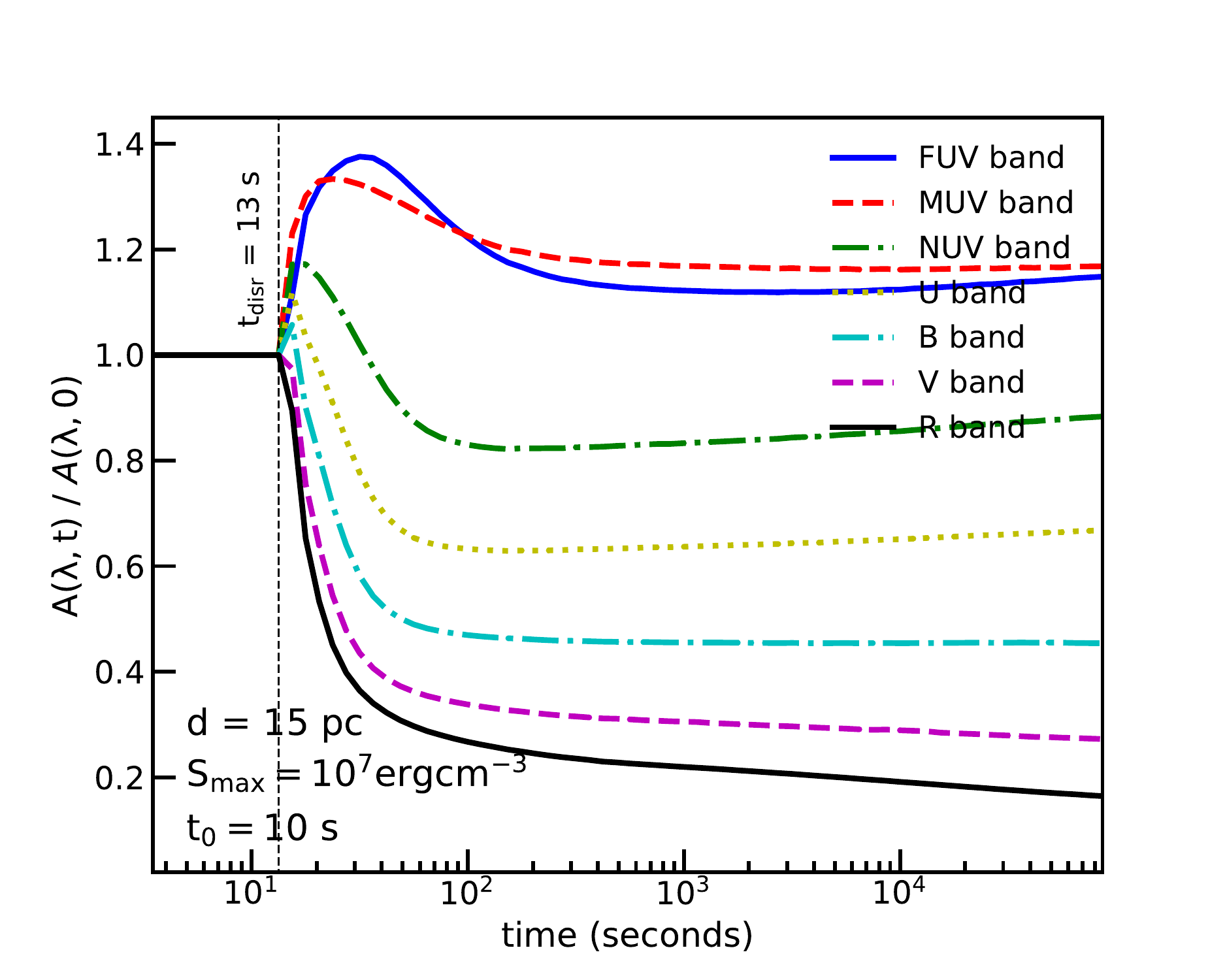}
\caption{Left panel: Variation of grain disruption size by RATD with time for the different cloud distance, assuming $S_{\rm max}=10^{7}\erg\cm^{-3}$. RATD occurs earlier (see vertical lines) and $a_{\rm disr}$ can achieve smaller values for smaller $t_{0}$. Right panel: Variation of the extinction in the different bands with time. From~\citet{2020ApJ...895...16H}.}
\label{fig:GRB_adisr}
\end{figure}

Figure~\ref{fig:GRB_adisr} (left panel) shows the grain disruption size due to the RATD effect as a function of time for the different cloud distance, assuming $S_{\rm max}=10^{7}\erg\cm^{-3}$ and $t_{0}=10$ s. For~a given cloud, the~grain disruption size decreases with time due to the increase of RATD. Figure~\ref{fig:GRB_adisr} (right panel) shows the variation of the extinction in the different bands with time. When the RATD begins, the~optical-IR extinction decreases rapidly, whereas the UV extinction increases then decreases due to the decrease of grain sizes via~RATD.

\begin{figure}[H]
\includegraphics[width=0.5\textwidth]{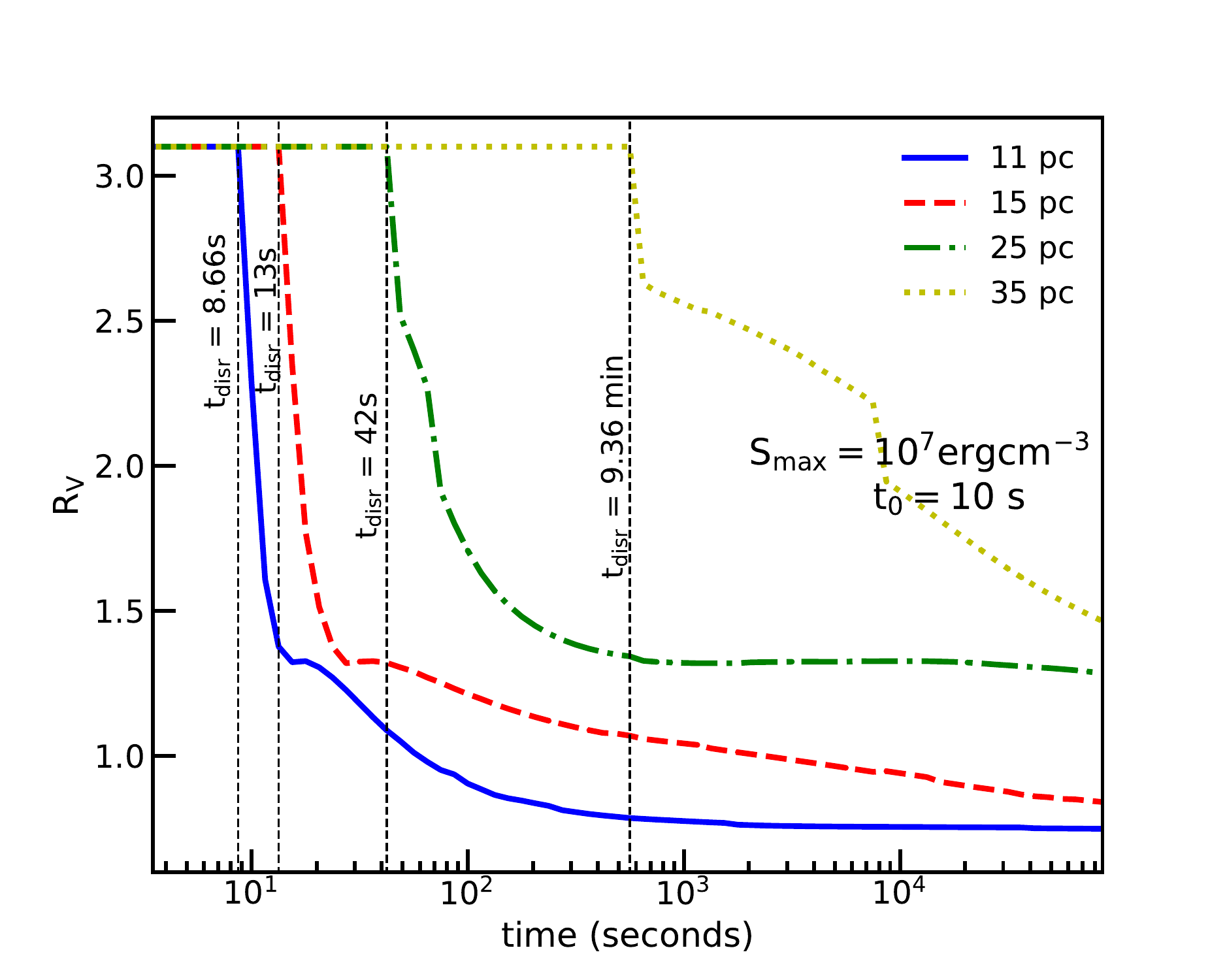}
\includegraphics[width=0.5\textwidth]{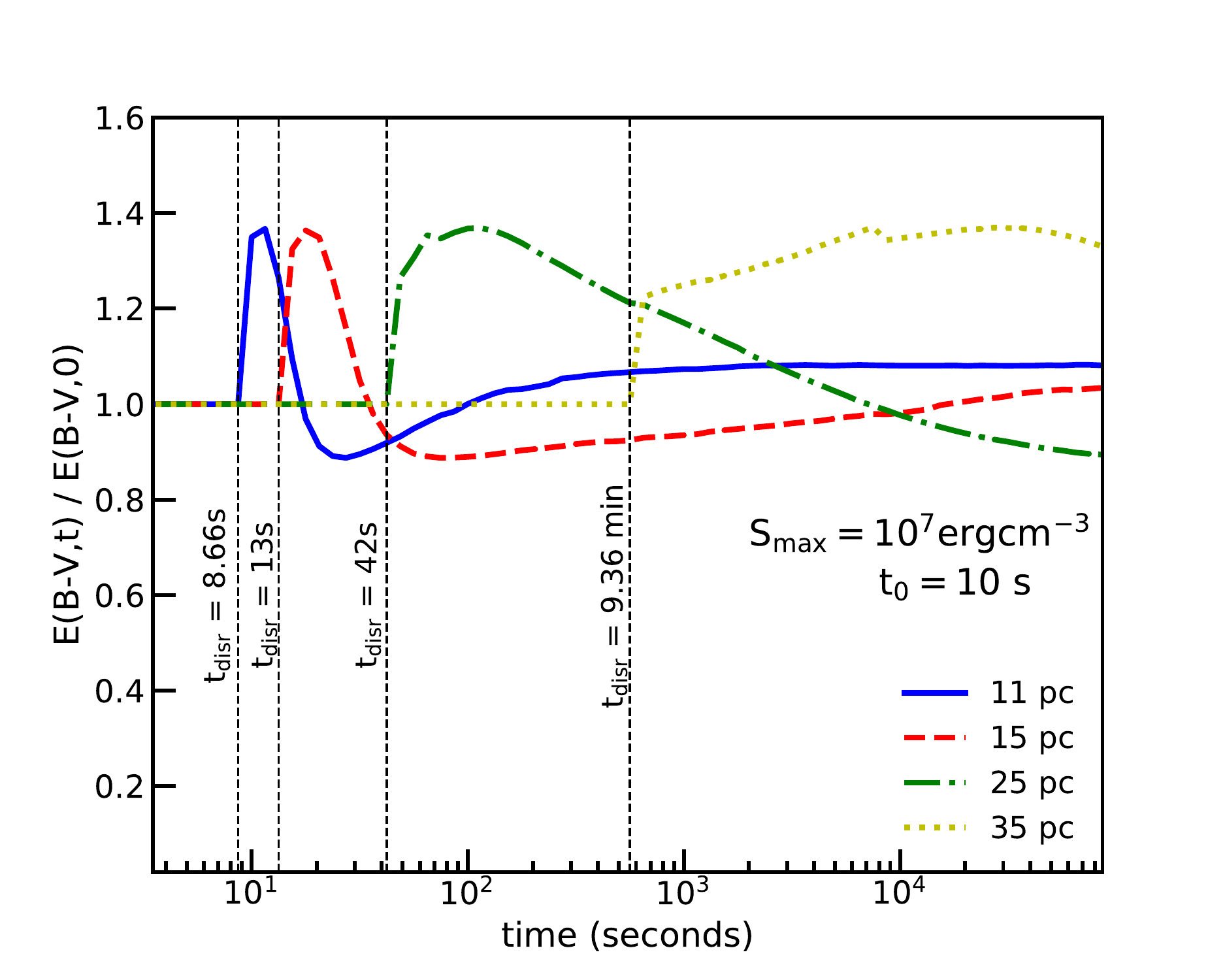}
\caption{Time variation of $R_{\rm V}$ (left panel) and $E(B-V)$ (right panel) and  for different cloud distances and $S_{\rm max}=10^{7} \erg \cm^{-3}$. Both $R_{V}$ and $E(B-V)$ begin to change when grain disruption begin at $t\sim t_{\rm disr}$ (marked by vertical dotted lines). From~\citet{2020ApJ...895...16H}.}
\label{fig:GRB_peaktime}
\end{figure}

Figure~\ref{fig:GRB_peaktime} shows the time-variation $R_{V}$ (left) and $E(B-V)$ (right) assuming that the dust cloud is located at different distances from the source. The~value $R_{V}$ is found to decrease gradually due to RATD that removes large grains over~time.

{The application of RATD for studying evolution of dust and ice in comets (a different kind of transients) is recently explored in \citet{Hoang:2020ws}. We find that large aggregate grains rapidly disrupt into small fragments, resulting in the change of dust properties within the cometary coma.}

\subsection{Circumstellar Envelopes of AGB~Stars}
In \citet{2020ApJ...893..138T}, we apply the RATD and METD mechanisms to study dust evolution in the envelope of AGB stars. Subject to a strong radiation field from the central star, large grains formed in dense clumps are disrupted by RATD into smaller ones, including nanoparticles. At~the same time, such nanoparticles are moving outward by radiation pressure through the gas and are disrupted by stochastic mechanical torques. Rapidly spinning nanoparticles produced by RATD produce strong microwave emission, as~shown in Section~\ref{sec:spindust}.

\begin{figure}[H]
\includegraphics[width=0.45\textwidth]{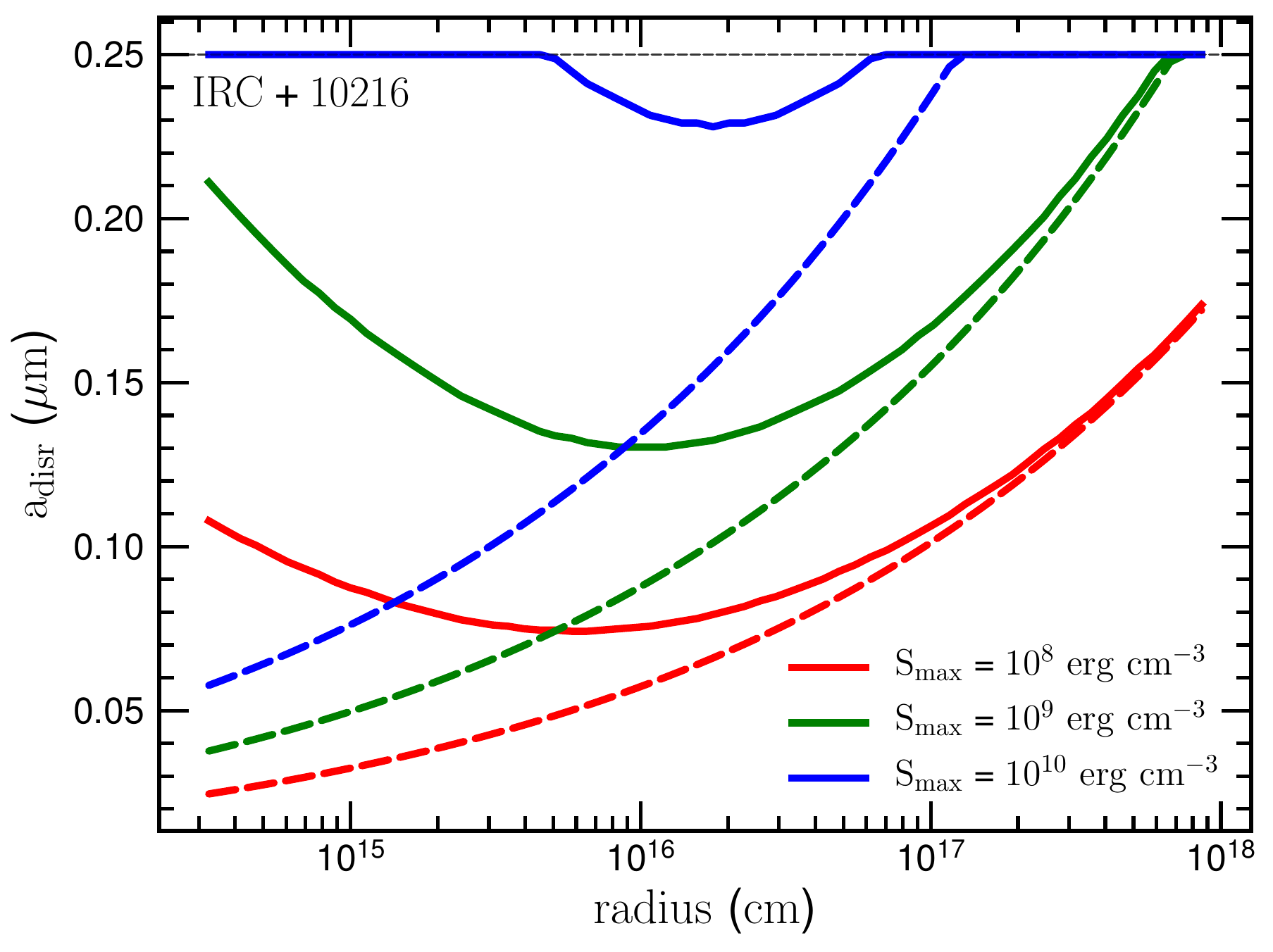}
\includegraphics[width=0.45\textwidth]{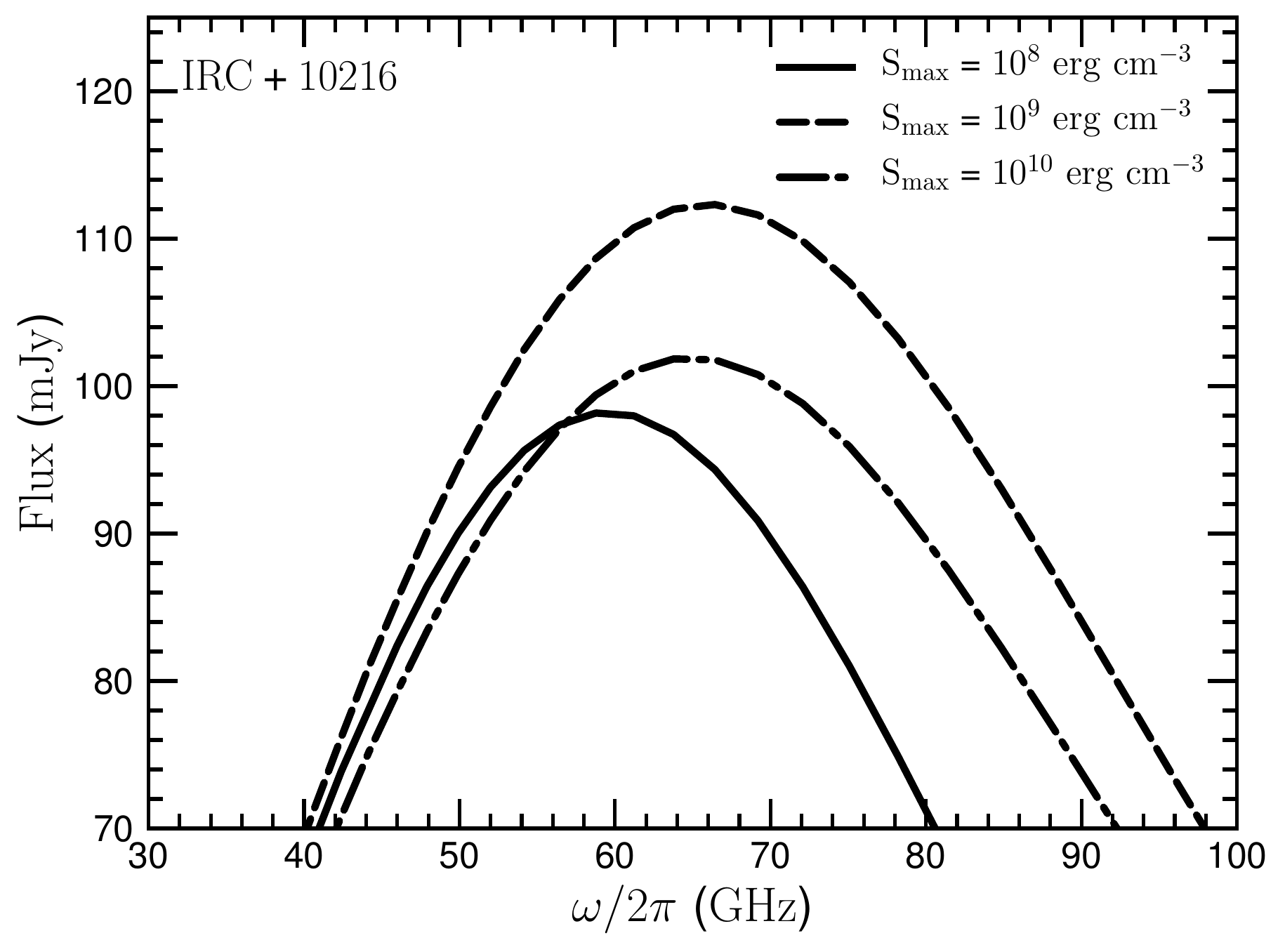}
\caption{Left panel: Disruption size of dust grains versus distance from the star in the circumstellar envelope (CSE) of IRC 
+ 10216 for the different tensile strengths ($S_{\rm max}$). The~initial maximum grain size is chosen as 0.25 $\upmu$m. Right panel: Emission flux from spinning nanoparticles in CSEs around C-rich star (IRC + 10216, top). We adopt $M_{d/g}=0.01$, $\eta=-3.5$, $\beta=0.4\D$, $Y_C \sim 0.05$ \mbox{(\citet{2007ApJ...657..810D}), and~$Y_{Si}=0.2$ (\citet{2016ApJ...824...18H})}.}
\label{fig:AGB_RATD}
\end{figure}

Figure~\ref{fig:AGB_RATD} shows the disruption size of grains by RATD (left panel) in the AGB and its microwave emission from spinning dust (right panel). The~disruption size is smaller for weaker grain materials (i.e., lower $S_{\rm max}$). 

The early detection of cm-wave observations toward to the AGB stars, that is,~at 15 GHz (or 2 cm) and 20 GHz (or 1.5 cm) from IRC + 10216 (\citet{1989A&A...220...92S}), and~at 8.4 GHz (or 3.57 cm) from 4 AGB stars over 21 samples (\citet{1995ApJ...455..293K}) cannot be explained by thermal dust emission. Recently, \citet{2007MNRAS.377..931D} presented the SED observations from a large sample of O-rich and C-rich AGB stars envelopes and showed emission excess at cm wavelengths for many stars, including some post-AGB and supergiants with circumstellar~shells. 

\begin{figure}[H]
\centering
 \includegraphics[width=0.44\textwidth]{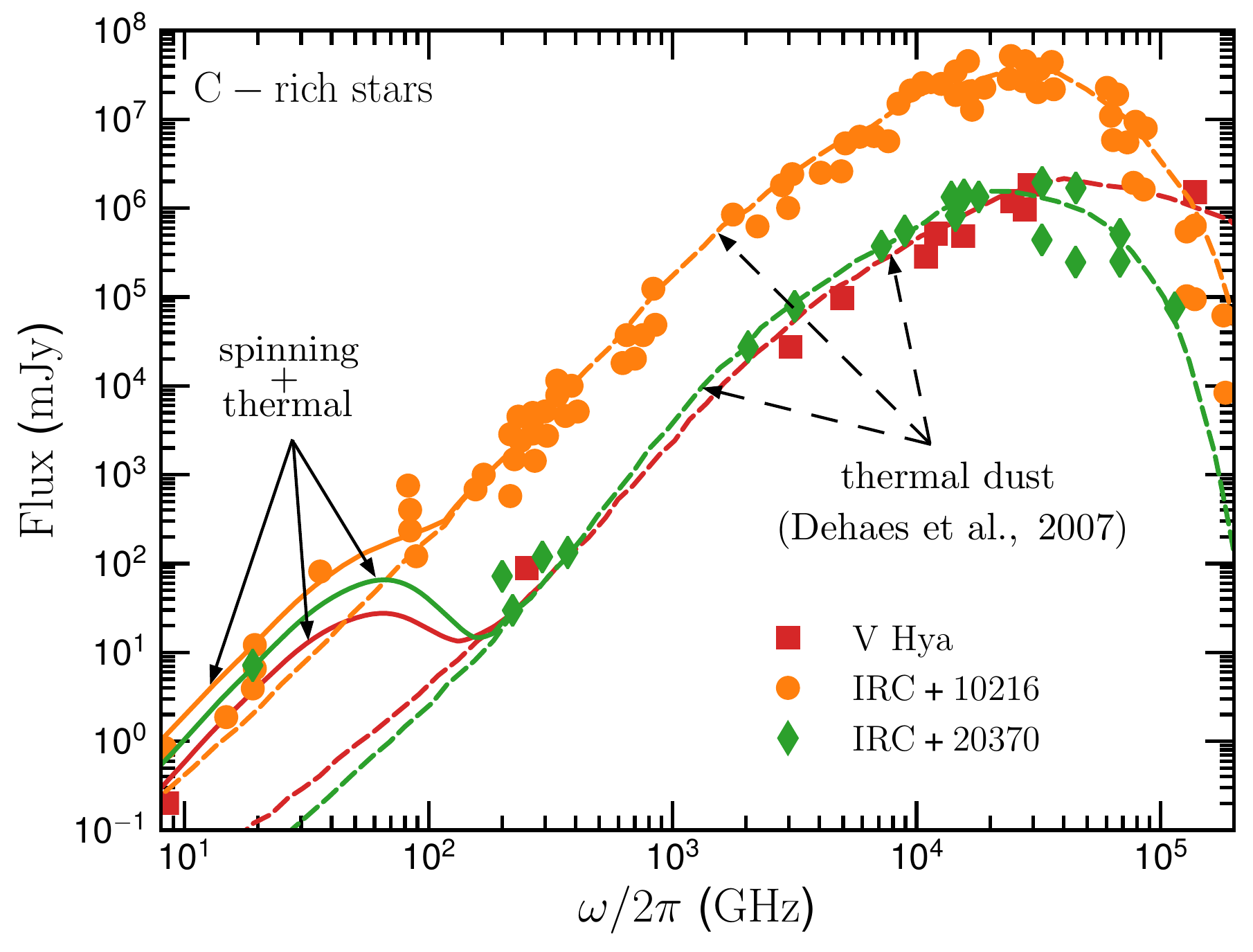}
 \includegraphics[width=0.45\textwidth]{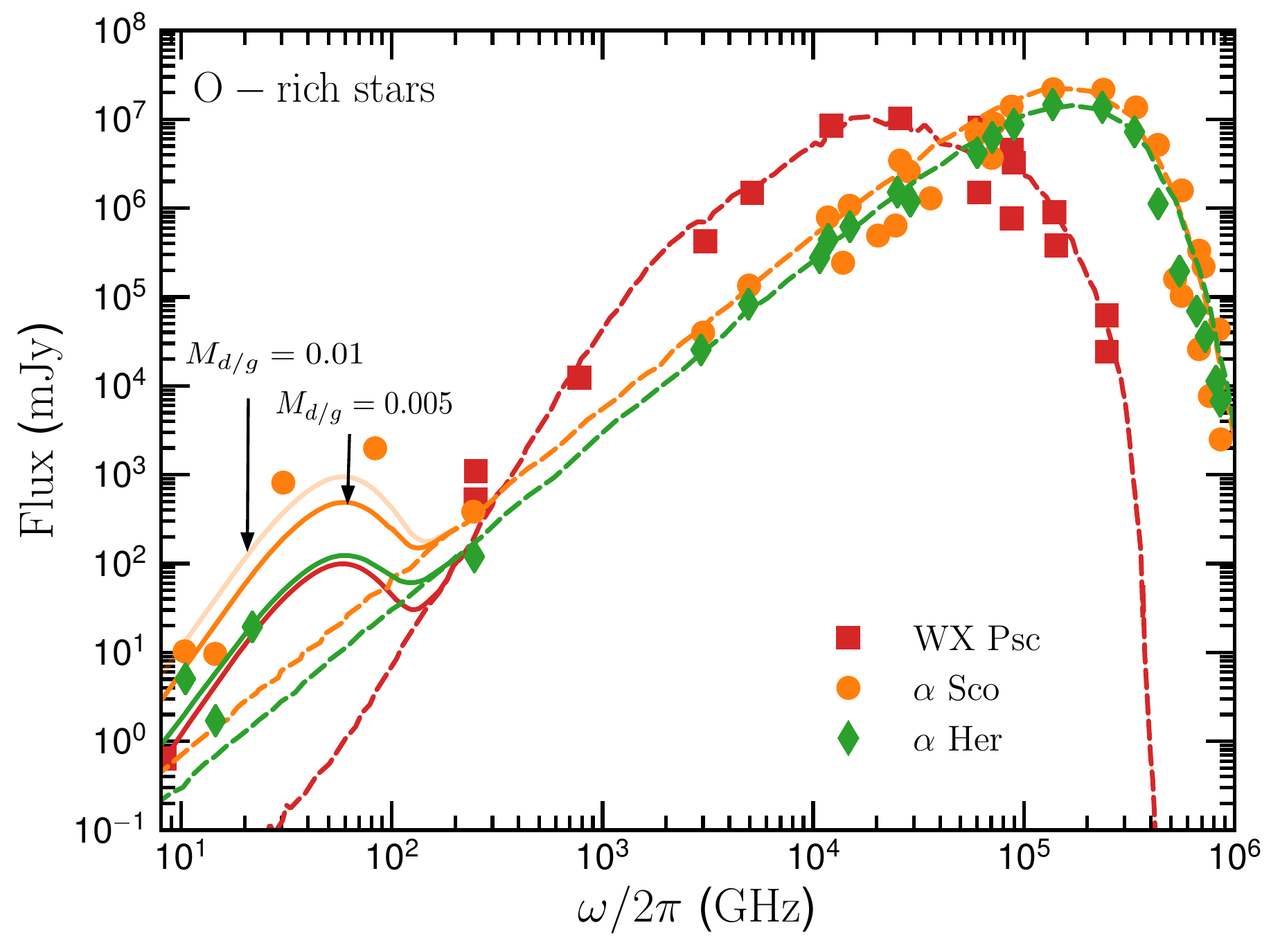}
 \caption{Comparison of spinning dust (solid lines, this work) and thermal dust (dashed lines, taken~from \citet{2007MNRAS.377..931D}) model to radio observational data from the literature (filled~symbols). The~solid lines show that the spinning dust model can reproduce well the CSEs of both C-rich (left panel) and O-rich stars (right panel). \mbox{C-rich stars - $S_{\max} \geq 10^{10}\erg\cm^{-3}$}, $\beta=0.4\D$: \mbox{$\eta=-3.7$ (IRC + 10216)}, $\eta=-3.6$ (IRC + 20370) and $\eta=-3.2$ (V Hy$\alpha$). \mbox{O-rich stars - $S_{\max} \geq 10^{9}\erg\cm^{-3}$: $\beta=2.3\D$}, $\eta=-5.0$ ($\alpha$ Sco); $\beta=1.6\D, \eta \sim -4.4$ ($\alpha$ Her); $\beta=1.2 \D, \eta=-4.8$ (WX Psc). Note: the dust-to-gas mass ratio is fixed as $M_{d/g}=0.005$ as in \citet{2007MNRAS.377..931D}. The~faint orange line likely indicates a higher dust-to-gas-mass ratio of $M_{d/g}\simeq$ 0.01 in the case of $\alpha$ Sco.}
 \label{fig:AMEfit}
 \end{figure}

\citet{2020ApJ...893..138T} attempt to fit the data with spinning dust. The~authors vary three parameters $S_{\max}$, $\beta$ and $\eta$ while fixing other physical parameters until a best-fit model is achieved. In~purpose of showing the whole SED, we combine with the best model of the thermal dust emission provided by \citet{2007MNRAS.377..931D}, which were modeled by DUSTY code (\citet{1999ascl.soft11001I}) with $M_{d/g}=0.005$. Figure~\ref{fig:AMEfit} shows our best-fit models to observational data for three C-rich (left panel) and three O-rich (right panel) stars with the corresponding set of fit parameters in the caption. Apparently, thermal dust and spinning dust are able to reproduce the mm-cm emission for both C-rich star (top panel, Figure~\ref{fig:AMEfit}) and O-rich stars (bottom panel, Figure~\ref{fig:AMEfit}).

\unskip
\subsection{Rotational Disruption of Dust and Ice in Protoplanetary~Disks}
The widespread presence of PAHs/nanoparticles in the surface layer of protoplanetary disks (PPDs) around Herbig Ae/Be stars, even in the inner gap (e.g., \citet{2019A&A...623A.135B}) is difficult to explain because PAHs/nanoparticles in the surface layer are expected to be destroyed by extreme UV photons and the X-ray component of the star's radiation spectrum \mbox{(\citet{2010A&A...511A...6S})}. To~explain the observations of PAHs in the disk, \citet{2010A&A...511A...6S} suggested that PAHs are transported from the disk interior to the surface via turbulent mixing in the vertical direction, which requires the presence of PAHs in the disk interior. This scenario is difficult to reconcile with the fact that PAHs are expected to be depleted due to condensation into the ice mantle of dust grains in cold, dense clouds (\citet{1999Sci...283.1135B,2014A&A...562A..22C,2015ApJ...799...14C}). 

In \citet{Tung:2020tz}, we applied the RATD mechanism to study the evolution of ice by radiation from young stars in protoplanetary disks (PPDs). We propose a top-down mechanism in which PAHs/nanoparticles are produced from the disruption of dust grains when being transported from the disk interior to the surface~layers.

\begin{figure}[H]
\includegraphics[width=0.5\textwidth]{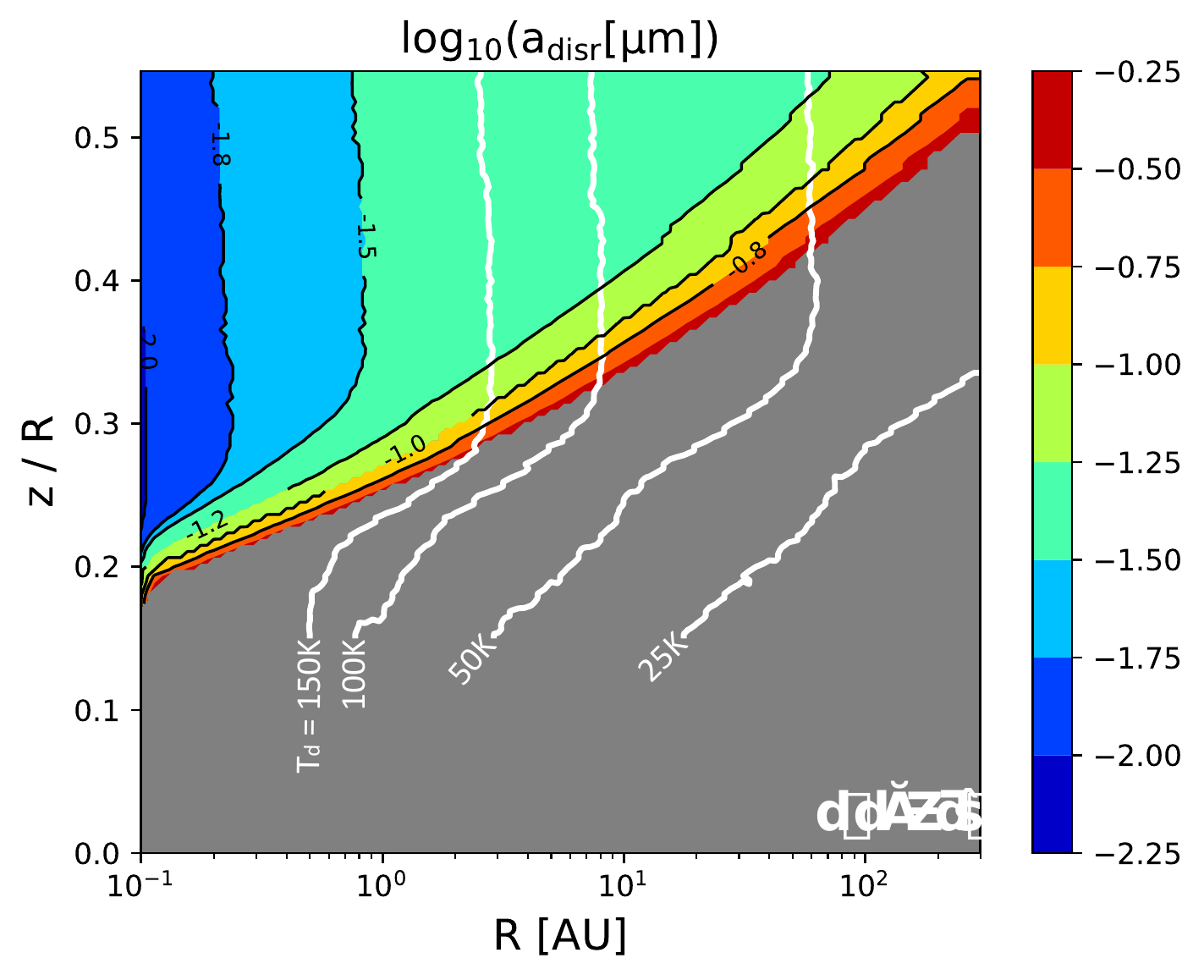}
\includegraphics[width=0.5\textwidth]{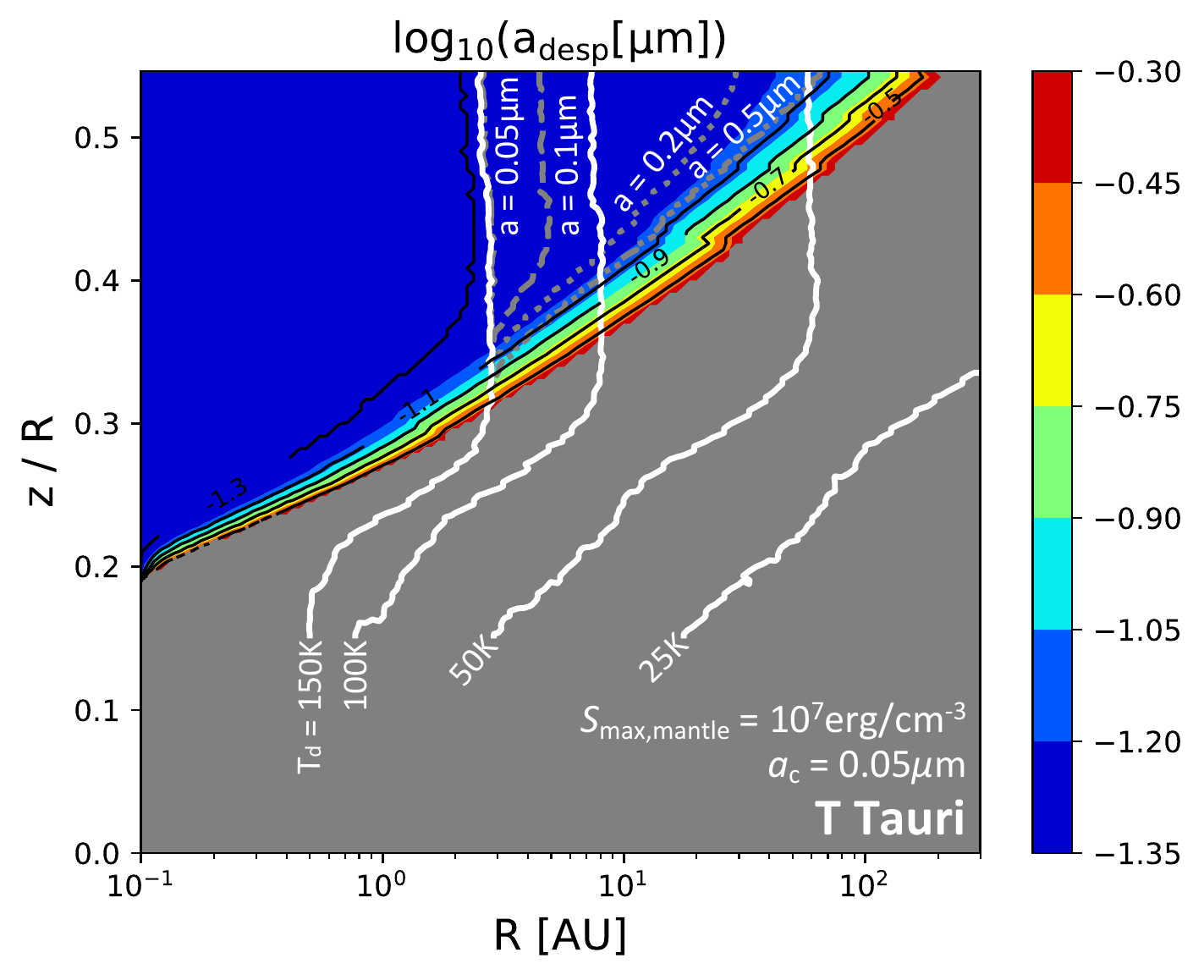}

\caption{Left panel: Disruption sizes of composite grains for a protoplanetary disk around T-Tauri stars as a function of $R$ and $z/R$. White lines show the locations of the disk that have $T_d = 150,100,50,25 \K$. Right panel: Desorption sizes of ice mantles for a PPD around T-Tauri stars as a function of $R$ and $z/R$, assuming~a fixed core radius $a_{\rm c} = 0.05$ $\upmu$m and the varying mantle thickness. Gray dashed and dotted lines illustrate the region where $\tau_{\rm sub,rot}^{-1}(T_d) = \tau_{\rm sub,0}^{-1}(T_{\rm sub})$ for different grain sizes $a$, assuming water ice of $T_{\rm sub}=150\K$.}
\label{fig:a_disr_TTauri}
\end{figure}

Figure~\ref{fig:a_disr_TTauri} (left panel) shows that large grains can be disrupted into nanoparticles in a vast area of the surface and intermediate layers by the RATD mechanism. Moreover, PAHs/nanoparticles frozen in the icy grain mantles can desorb from the ice mantle via the ro-thermal desorption mechanism (\citet{2019ApJ...885..125H}; right panel). Thus, if~grains are small enough ($a\lesssim$ 1 $\upmu$m) to be mixed with the gas by turbulence, they can escape grain settling and coagulation in the disk mid-plane and are transported to the surface layer (\citet{2004A&A...421.1075D,2009A&A...496..597F}). When exposed to stellar radiation, large grains within the range size [$a_{\rm disr}$, $a_{\rm disr,max}$] could be disrupted by RATD into PAHs/nanoparticles (see Figure~\ref{fig:a_disr_TTauri}, left panel). Furthermore, our results suggest an increase in the relative abundance of nanoparticles with decreasing the distance to the star. Therefore, in~the framework of rotational disruption and desorption, PAHs and nanoparticles would be continuously replenished in the surface layer, with~the support of turbulent transport. This can resolve the longstanding puzzle about the ubiquitous presence of PAHs/nanoparticles in the hot surface layer of~PPDs. 

The water snowline characterized by the water sublimation temperature of $T_{d}\sim150\K$ divides the inner region of rocky planets from the outer region of gas giant planets.  Nevertheless, the~precise location of the snowline is a longstanding problem in planetary science \mbox{(\citet{2006ApJ...640.1115L,Min:2011id})}. Using the standard model of PPDs, we show that ice mantles from micron-sized grains ($a<100 $ $\upmu$m) are disrupted and identify the snowline's location in the presence of rotational desorption. We find that the water snowline is extended outward (\citet{Tung:2020tz}). 

\subsection{Circumsolar Dust and The~F-Corona}
\begin{figure}[H]
\includegraphics[width=0.5\textwidth]{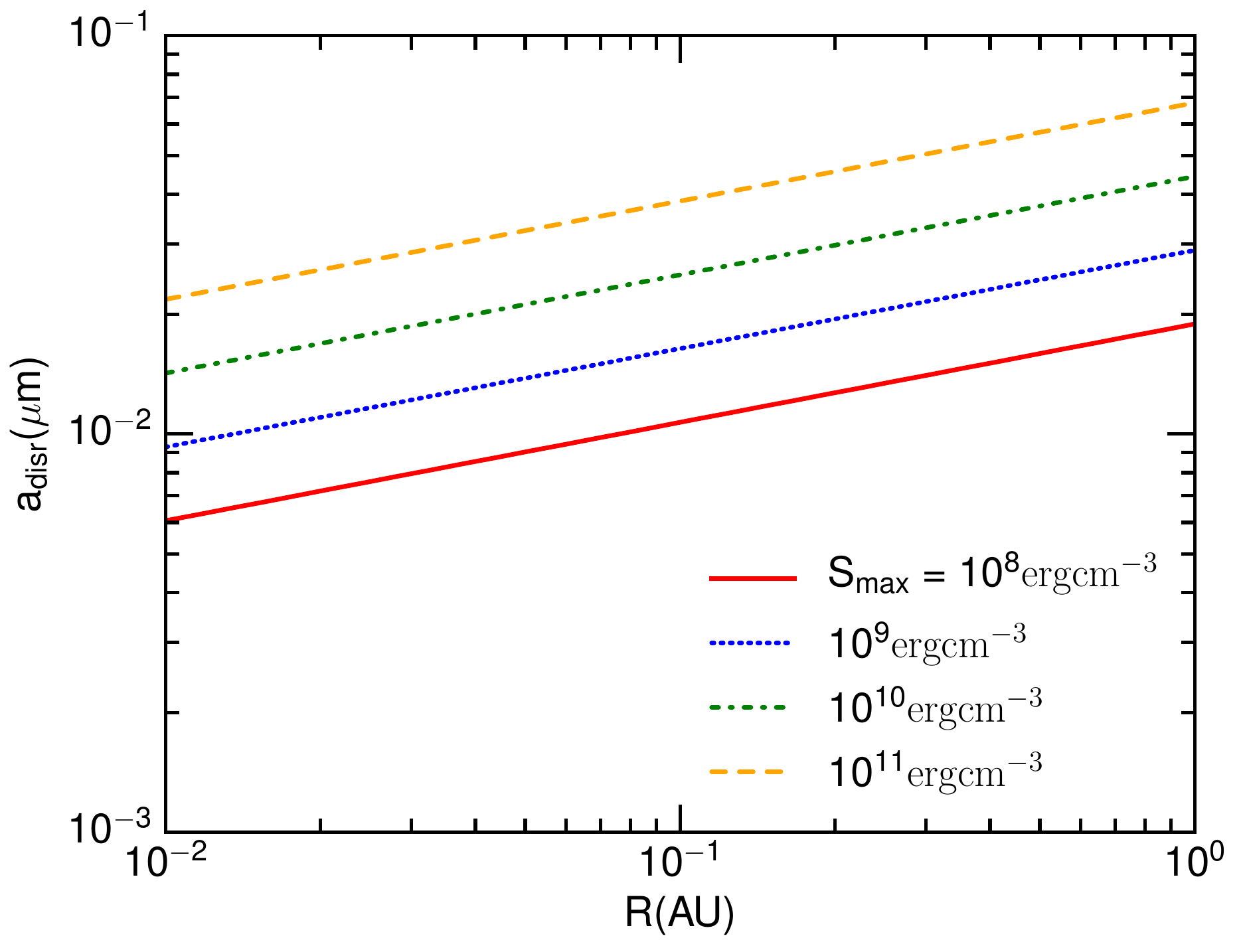}
\includegraphics[width=0.5\textwidth]{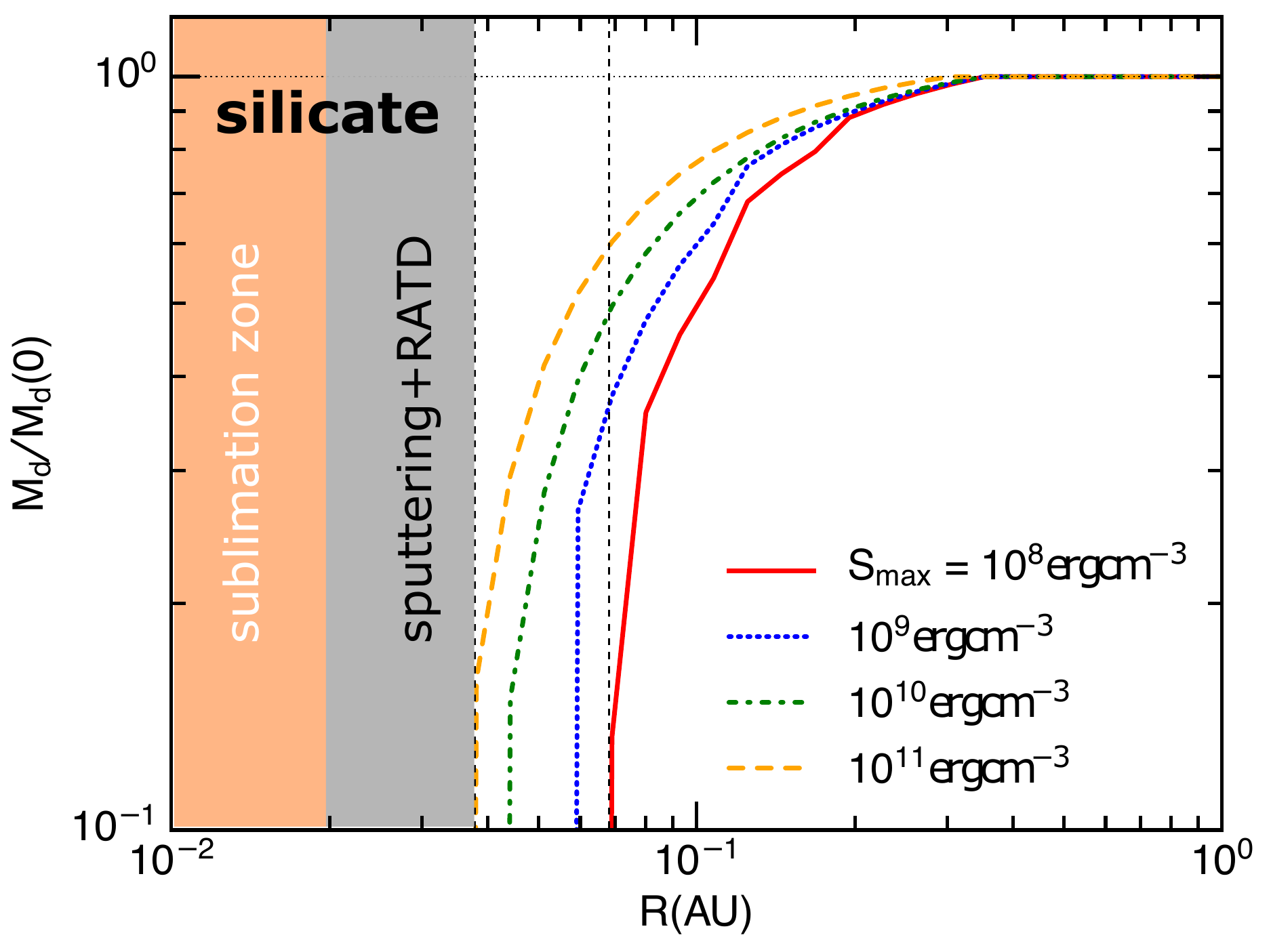}
\caption{Left panel: Grain disruption size by RATD as a function of heliodistance for the different tensile strength. Large grains are disrupted by RATD into very small grains. Right panel: Decrease of the dust mass ($M_{d}/M_{d}(0)$) with heliocentric distance due to RATD and nonthermal sputtering by the solar wind for silicate grains where $M_{d}(0)$ is the original dust mass. Four values of the tensile strength $S_{\max}=10^{8}$--$10^{11}\erg\cm^{-3}$ are considered. The~dust mass decreases toward the Sun, starting from $\sim0.2$~\AU (for highest strength) and reaches the new dust-free-zone predicted by RATD and sputtering effects (gray shaded area).}
\label{fig:solar_Md}
\end{figure}

In \citet{Hoang:2020uf}, we applied the RATD for circumsolar dust (F-corona). As~shown in Figure~\ref{fig:solar_Md} (left panel), the~intense solar radiation can efficiently disrupt large grains into nanoparticles. We also found that energetic protons from the solar wind can efficiently destroy smallest nanoparticles via nonthermal sputtering, which decreases the~F-corona.

We can calculate the decrease of the volume mass of dust as a function of heliocentric distance as~follows:
\bea
\frac{M_{d}(R)}{M_{d}(R,0)}=\frac{\int_{a_{\rm nsp}}^{a_{\rm disr}} (4\pi a^{3}\rho /3) (dn/da)da}{M_{d}(R,0)},
\ena
where $a_{\rm nsp}$ be the critical size of nanoparticles that are destroyed by nonthermal sputtering, and~$M_{d}(R,0)$ is the original dust mass in the absence of sputtering
\bea
M_{d}(R,0)=\int_{a_{\rm min}}^{a_{\rm disr}} \left(\frac{4\pi a^{3}\rho }{3}\right) \left(\frac{dn}{da}\right)da,
\ena
where a power-law grain size distribution of $dn/da = Ca^{-3.5}$ with $a_{\rm min}=3$~\AA~ is~adopted. 

Figure~\ref{fig:solar_Md} (right panel) shows the variation of the relative dust mass $M_{d}/M_{d}(0)$ vs. the heliocentric distance as a result of RATD and sputtering for the different tensile strength. We see that $M_{d}/M_{d}(0)$ starts to decrease considerably from $R\sim 0.2$ AU ($\sim 42R_{\odot}$), and~a significant mass loss occurs at $R\lesssim 0.03\AU (6R_{\odot})$, which suggests a new dust-free-zone. Grains made of weak material are more efficiently disrupted by RATD and experience larger mass~loss. 

The increase in the abundance of small grains near the Sun predicted by RATD is consistent with previous studies (\citet{1979PASJ...31..585M,1992A&A...261..329M}). Our results show that F-corona dust as well as dust in the inner solar system ($R<1\AU$) mostly contains nanoparticles of size $a\sim$1--10 nm (see Figure~\ref{fig:solar_Md}, left). This is a natural explanation for nanodust detected by in-situ measurements (see, e.g.,~ \citet{Mann:2007em,Ip:2019ek}). This RATD mechanism is more efficient than collisional fragmentation previously thought (see e.g.,~\citet{Mann:2007em} for discussion of various mechanisms to form nanodust in the inner solar system).   

In-situ measurements by dust detector onboard the Helio spacecraft reported a F-corona decrease at heliocentric distances between $D=0.3-1$ AU (\citet{1985ASSL..119..105G}). This is thought to be due to mutual collisions that makes grains smaller and decrease of the forward scattering cross-section. However, the~RATD appears to be more efficient in producing small grains due to its short~timescale.

Note that RATD is valid for grains of $\bar{\lambda}/a>0.1$. For~larger grains of $a>\bar{\lambda}/0.1\sim 9$ $\upmu$m, RATs~of such very large grains are not yet available due to the lack of numerical calculations because it requires expensive computations to achieve reliable results of RATs for grains of $2\pi~a/\lambda\gg 1$ (see~e.g.,~\citet{Draine:2004p6718}). Expecting the decrease of RATs with increasing $a$, the~disruption may still be important when the decrease of RATs is compensated by the increase of the radiation energy~density.

\begin{figure}[H]
\centering
\includegraphics[width=0.9\textwidth]{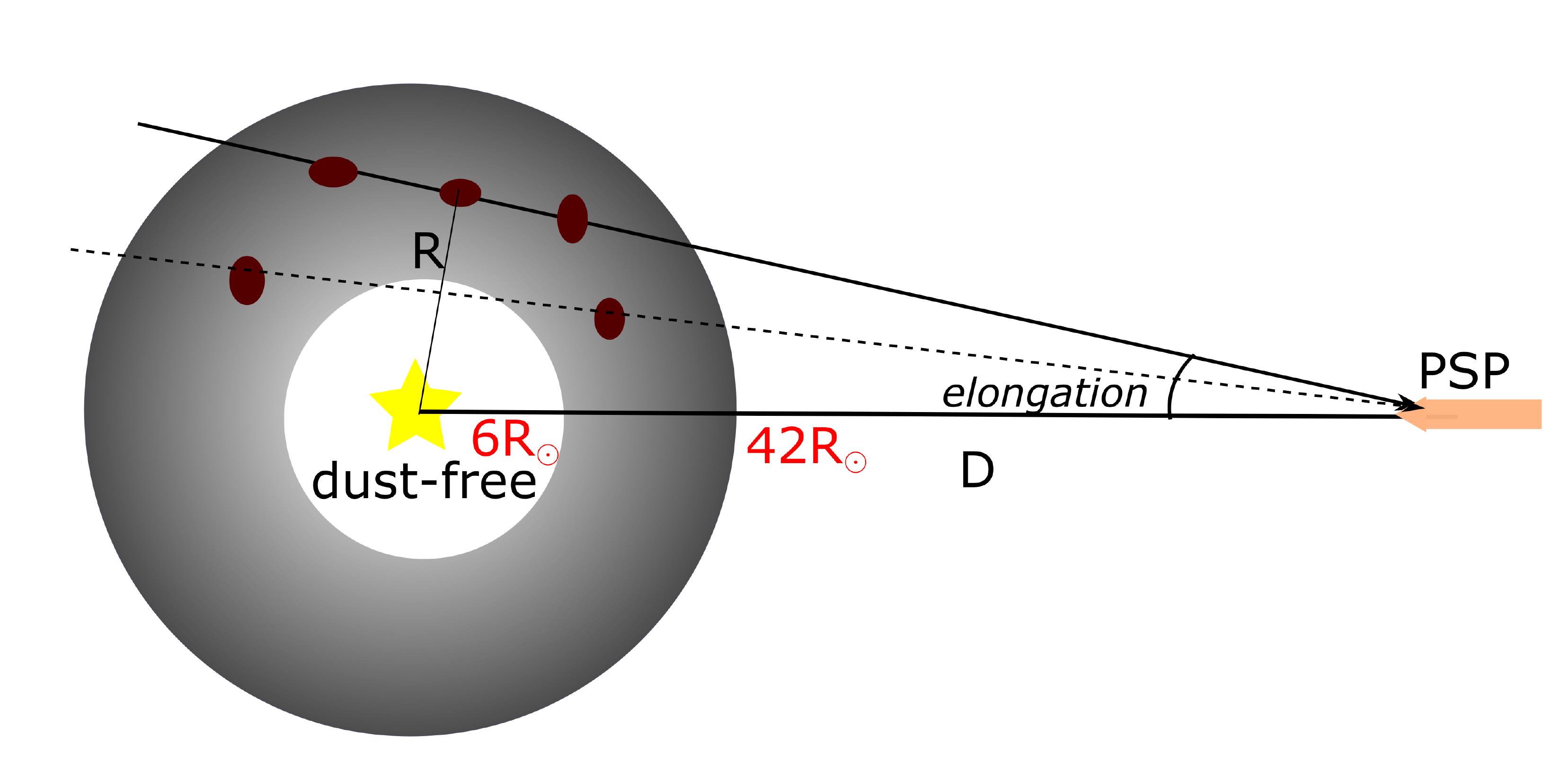}
\caption{Schematic illustration of the F-corona predicted by RATD and nonthermal sputtering which would be observed by the Parker Solar Probe (PSP) at heliocentric distance of $D=0.3$ AU and various elongation angles. The~F-corona decrease is illustrated by a radial gradient, starting from $42R_{\odot}$ to the edge of the new dust-free-zone at $6R_{\odot}$.}
\label{fig:Fcorona}
\end{figure}

Figure~\ref{fig:Fcorona} illustrates the F-corona as a result of RATD and nonthermal sputtering, which would be observed with the PSP. The~new dust-free-zone is located at $R=6$ $R_{\odot}$. Beyond~this radius, the~F-corona decreases with the radius, starting from 42 $R_{\odot}$ to the~dust-free-zone.
 
The first-year results from the PSP at heliocentric distances of $D=0.16-0.25$ AU ($34.3- 53.7R_{\odot}$) reveal the gradual decrease the F-corona \citet{Howard:2019ih}. With~the elongation $\epsilon\sim$ 15--20$^{\circ}$, one can estimate the corresponding elongation in solar radii $R_{\epsilon}\sim$ (0.166--0.336)$\sin(15^{\circ})\AU\sim$ 9--19$R_{\odot}$. 

Indeed, the~sublimation radius reveals a range of the dust-free-zone between 4--5 $R_{\odot}$ for silicate grains of sizes $a\sim$ 0.1--0.001 $\upmu$m. Therefore, even with the effect of RATD, thermal sublimation alone cannot explain the thinning-out of circumsolar dust observed from $R_{\epsilon}\lesssim 19$ $R_{\odot}$. However, our results shown in Figure~\ref{fig:solar_Md} (right panel) indicate that the joint effect of RATD and nonthermal sputtering could successfully explain the gradual decrease of F-corona toward the Sun. Moreover, the~PSP's observation is consistent with our result with the largest value of $S_{\max}$ because this model predicts the F-corona decrease from a closest distance of 19 $R_{\odot}$.

The PSP is planned to undergo 24 orbits around the Sun. The~latest orbits (22--24) will reach the closest distance of 10.74 $R_{\odot}$ \citet{Szalay:2019uc}. Previous studies predict that the dust-free-zone is between 4--5 $R_{\odot}$, which cannot be confirmed with the PSP. We found that the joint action of RATD and sputtering increase the radius of dust-free-zone to 7--11 $R_{\odot}$ (see Figure~\ref{fig:solar_Md}). This would be tested with the next orbits of the~PSP. 

\unskip
\subsection{Rotational Desorption of Ice in Star-Forming Regions Around~Ysos}
Dust grains surrounding YSOs are expected to contain ice mantles. In~these regions, water and COMs are detected around the hot cores/corinos,  which are inner regions surrounding high-mass protostars (\citet{1987ApJ...315..621B}) and low-mass protostars (\citet{2004ApJ...615..354B,2018arXiv180608137B}).. Previous studies rely on thermal sublimation of ice when grains are heated to above 100 K. We now apply our theory in the previous section to study the desorption of ice mantles from grains in hot cores and hot corinos and demonstrate that rotational desorption is more efficient and can work at $T<100\K$.

Let $L$ be the bolometric luminosity of the central protostar. The~radiation strength at distance $r$ from the source is given by
\bea
U(r)=\left(\frac{L}{4\pi r^{2}c u_{\rm ISRF}}\right)=U_{\rm in}\left(\frac{r_{\rm in}}{r}\right)^{2},
\ena
where $U_{\rm in}$ denotes the radiation strength at inner radius $r_{\rm in}$. 

The gas density and temperature can be approximately described by power laws:
\bea
n_{\gas}=n_{\rm in}\left(\frac{r_{\rm in}}{r}\right)^{p},\\
T_{\gas}=T_{\rm in}\left(\frac{r_{\rm in}}{r}\right)^{q},
\ena
where $n_{\rm in}$ and $T_{\rm in}$ are gas density and temperature at radius $r_{\rm in}$, and~$q=2/(4+\beta)$ with $\beta$ the dust opacity index (see e.g.,~\citet{2000ApJ...530..851C}). The~typical density profile in the inner hot region is $p\sim 1.5$. A~more detailed model of hot cores is presented in \citet{Nomura:2004jt}.

From Equation~(\ref{eq:adisr_low}), one obtains the disruption size of ice mantles as follows
\bea
a_{\rm disr}(r)\simeq 0.13\gamma^{-1/1.7}\bar{\lambda}_
{0.5}(S_{\max,7}/\hat{\rho}_{\rm ice})^{1/3.4}(1+F_{\rm IR})^{1/1.7}
\left(\frac{n_{\rm in}T_{\rm in}^{1/2}}{100U_{\rm in}}\right)^{1/1.7}\left(\frac{r_{\rm in}}{r}\right)^{(p+q/2-2)/1.7}\upmu \text{m},~~~\label{eq:adisr_core}
\ena
which slowly decreases with radius $r$ as $r^{(p+q/2-2)/1.7}\sim r^{-0.1}$ for typical~slopes.

For low-mass protostars, one can assume $r_{\rm in}=25\AU$ and $n_{\rm in}\sim 10^{8}\cm^{-3}$ and $L=36L_{\odot}$ \citet{Visser:2012km}, one gets $a_{\rm disr}\sim 0.29$ $\upmu$m at $r=r_{\rm in}$ and $a_{\rm disr}\sim 0.63$ $\upmu$m for $r=10r_{\rm in}$. For~hot cores, we adopt a typical luminosity of $L=10^{5}L_{\odot}$ and typical parameters $r_{\rm in}\sim 500 \AU, n_{\rm in}\sim 10^{8}\cm^{-3}, U_{\rm in}\sim 2\times 10^{7}$ and $T_{\rm in}\sim 274\K$ (see e.g.,~\citet{Bisschop:2007cu}). Therefore, Equation~(\ref{eq:adisr_core}) gives $a_{\rm disr}=0.1$ $\upmu$m and $0.16$~$\upmu$m at $r=r_{\rm in}, 10r_{\rm in}$ respectively. The~results for $n_{\rm in}=10^{7}\cm^{-3}$ as usually assumed (\citet{1999MNRAS.305..755V}) are even more~promising. 

\begin{figure}[H]
\centering
\includegraphics[width=0.8\textwidth]{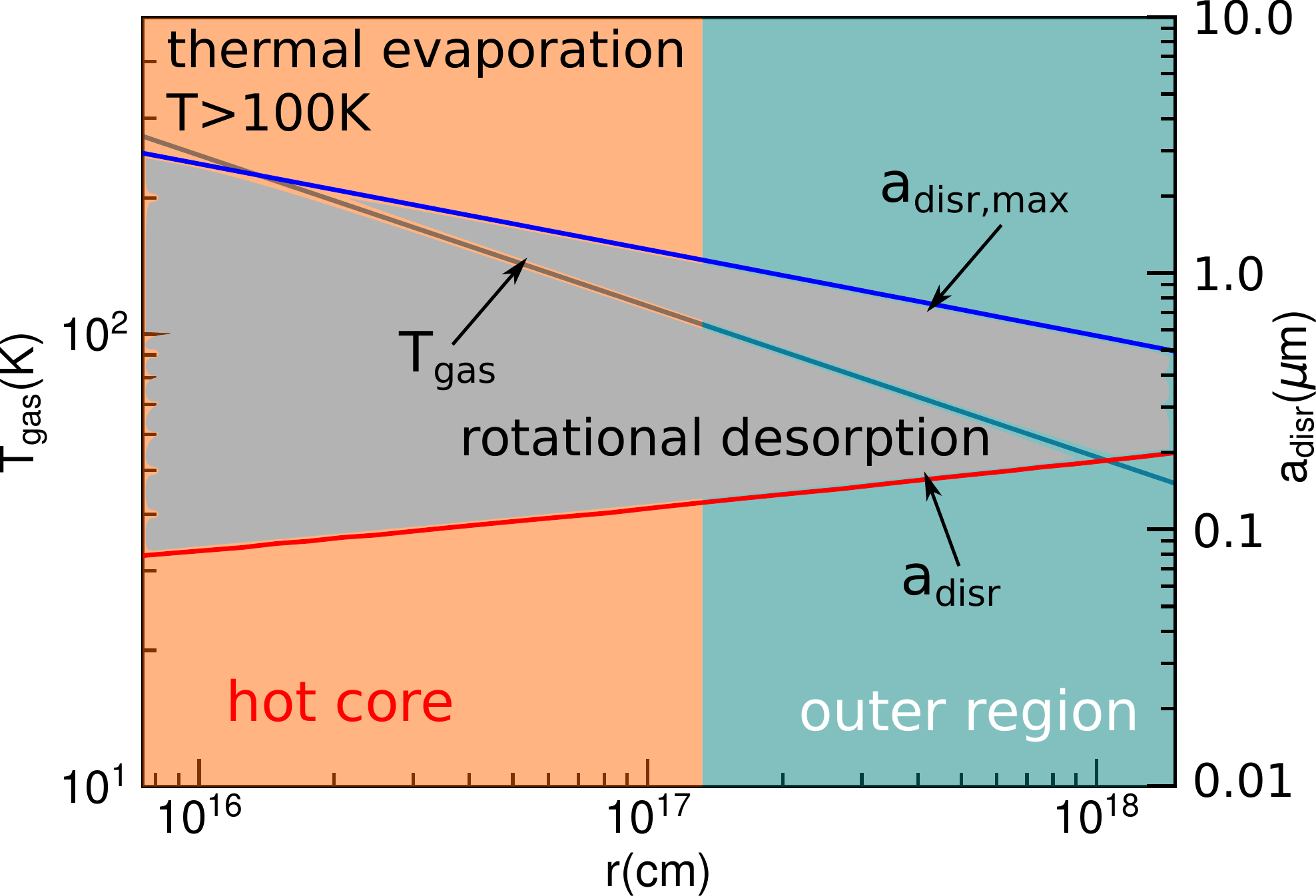}
\caption{Gas temperature and grain disruption size vs. radius for an envelope around a high-mass protostar. Classical thermal sublimation is important only for the inner region (hot core), whereas~rotational desorption is important for both hot core and outer~region.}
\label{fig:adisr_Tn}
\end{figure}

Figure~\ref{fig:adisr_Tn} illustrates the importance of rotational desorption vs. classical thermal sublimation of ice mantles around a protostar of $L=10^{5}L_{\odot}$ and $\bar{\lambda}=0.5$ $\upmu$m. Thermal evaporation is important only in the inner regions where $T_{\gas}>100\K$, whereas rotational desorption can be efficient at larger radii with low temperatures of $T_{\gas}\sim$ 40--100 K. 

We note that even in the hot inner region where thermal sublimation is active, rotational desorption and ro-thermal desorption (\citet{2019ApJ...885..125H}) are more efficient than the classical sublimation for molecules with high binding energy such as water and~COMs.

The efficiency of rotational desorption in cold extended regions shown in Figure~\ref{fig:adisr_Tn} could explain the presence of COMs from outer extended regions around hot cores by \citet{Fayolle:2015cu}. Furthermore, this mechanism can explain the presence of HCOOH, CH$_3$CHO from cold regions with $T_{\gas}\sim$ 40--70 K (below the sublimation threshold of these molecules) and low column density by \citet{Bisschop:2007cu}. Future high-resolution observations by ALMA would be unique to test our prediction of an extended regions of~COMs.

\subsection{Rotational Disruption of Nanoparticles in C- and~CJ-Shocks}
Interstellar shocks are ubiquitous in the ISM, which are induced by outflows around young stars and supernova remnants. If~the shock speed is lower than the magnetosonic speed, physical~parameters are continuous throughout the shock, for~which the term C-type shocks are introduced (\citet{1980ApJ...241.1021D}; also \citet{2004ApJ...610..781C}). For~the C-type shock in dense magnetized molecular clouds, in~the shock reference frame, the~ambient pre-shock gas flows into the shock such that their physical parameters change smoothly with the distance in the shock. At~the shock interface, the~neutral and ion velocities are the same as the shock velocity considered in the shock reference frame. Due to the deceleration when colliding with the shock matter, neutrals and ions are slowed down until they move together with the shock front, that is,~$v_{n}=v_{i}=0$. Due to magnetic forces, ions and charged grains are coupled to the ambient magnetic field and move slower than neutrals, resulting in $v_{n}> v_{i}$ or drift of neutral gas with respect to charged grains and~ions. 

In \citet{2019ApJ...877...36H,2019ApJ...886...44T}, we studied rotational dynamics of grains in shocks and found that the supersonic relative motion could spin nanoparticles up to to suprathermal rotation. As~a result, centrifugal stress can disrupt them into tiny fragments (see Section~\ref{sec:nano}). This effect is most efficient for smallest nanoparticles and increases the lower cutoff of the grain size~distribution.

\begin{figure}[H]
\includegraphics[width=0.5\textwidth]{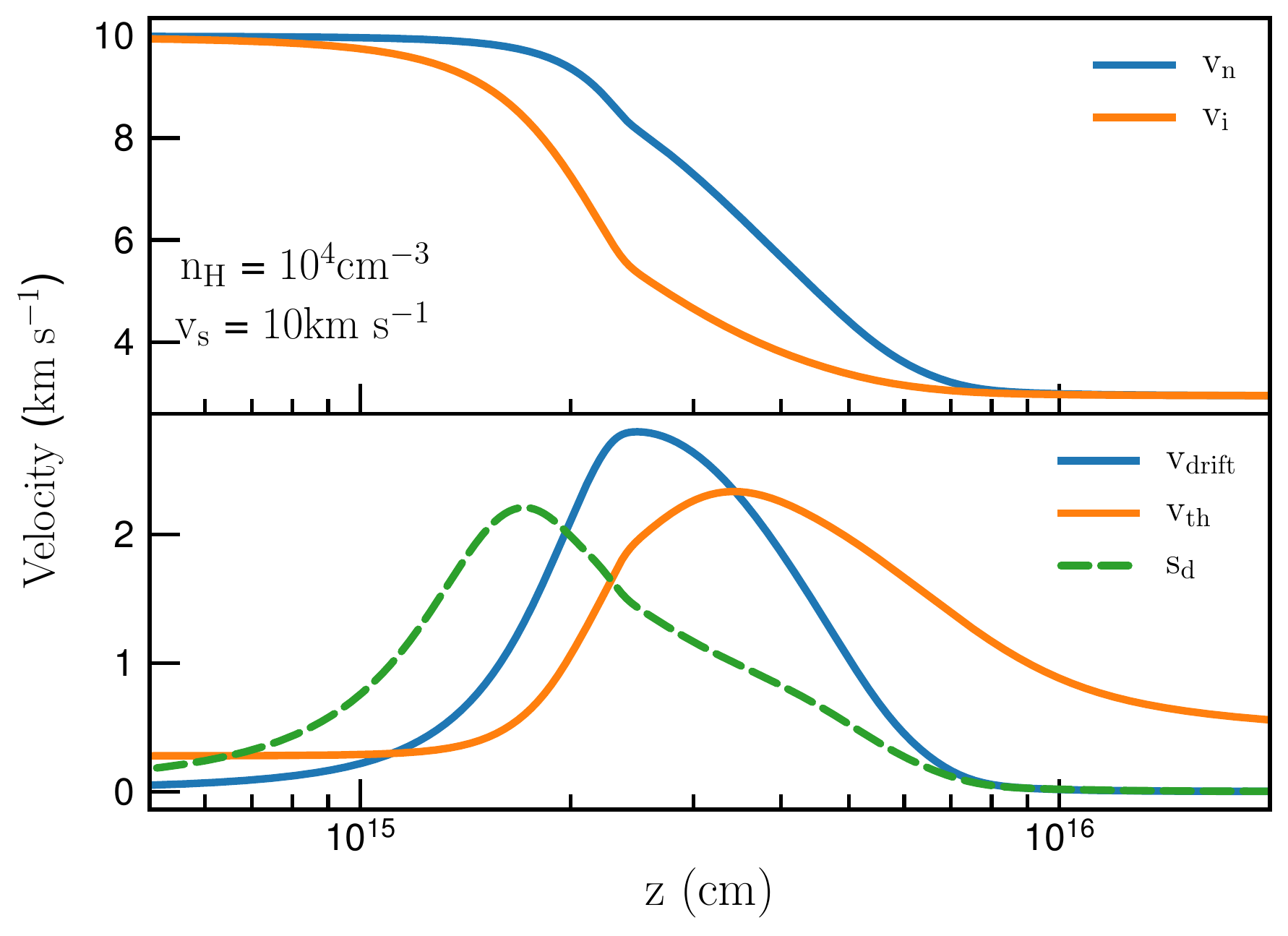}
\includegraphics[width=0.5\textwidth]{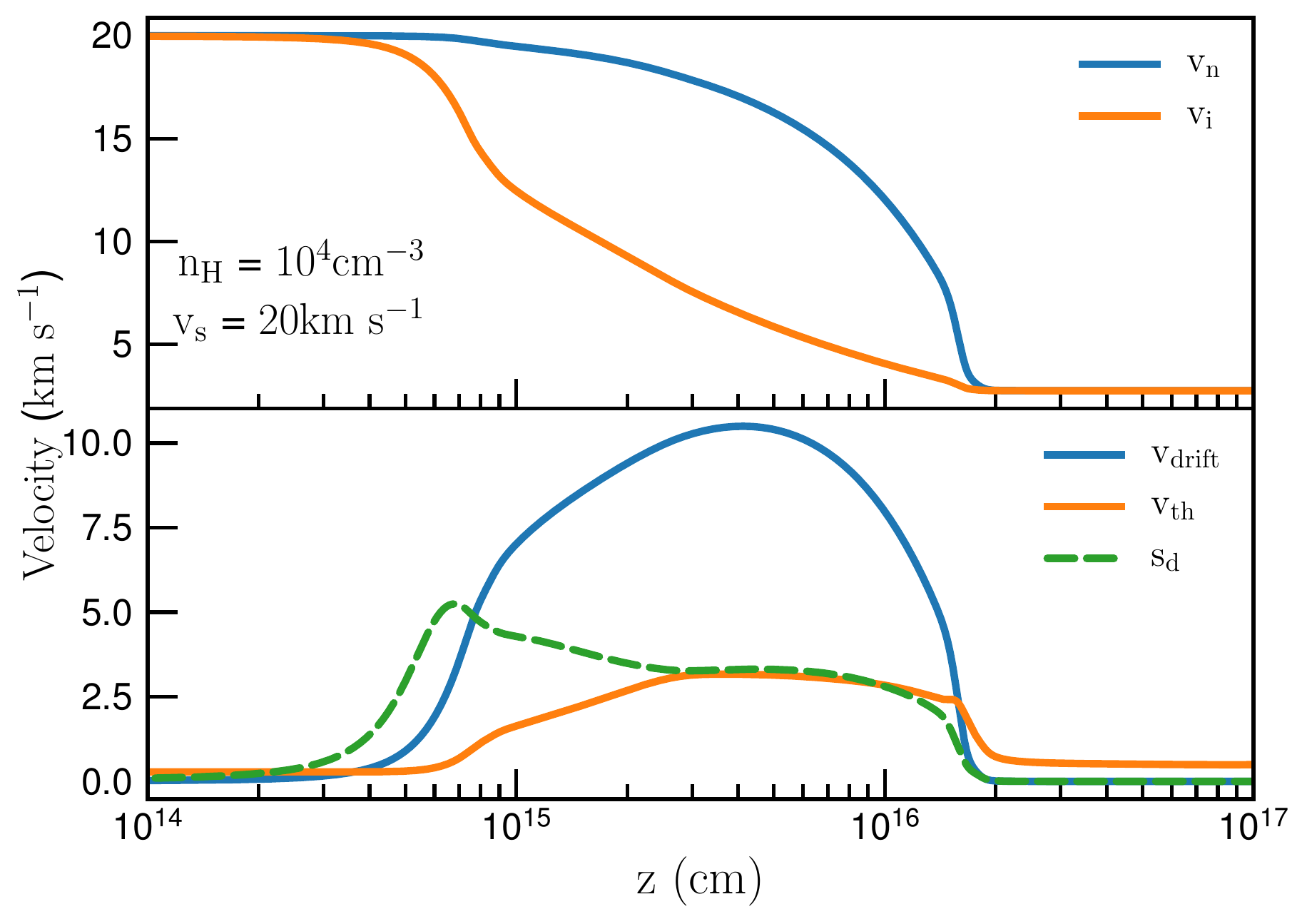}
\caption{Velocity profiles of neutral ($v_{n})$, ion and charged grains ($v_{i}$), and~their relative velocity ($v_{\rm drift}=v_{n}-v_{i})$ in the C-shocks for $v_{s}=10\km\s^{-1}$ (left) and $20\km\s^{-1}$ (right). The~dashed line shows the drift parameter $s_{d}=v_{\rm drift}/v_{th}$ which is dimensionless. The~drift velocity increases with the shock velocity, but~$s_{d}$ is slightly changed due to an increased thermal velocity. From~\citet{2019ApJ...877...36H}.}
\label{fig:Cshockn4-velo}
\end{figure}

\citet{2019ApJ...877...36H} calculate the ion and neutral velocities for the different shock velocities using the one-dimensional plane-parallel Paris-Durham shock \mbox{model \citet{2015A&A...578A..63F}}. Our initial elemental abundances in the gas, grain cores, ice mantle, and~PAHs are the same as in the previous studies (\citet{2003MNRAS.343..390F,2013A&A...550A.106L,2018MNRAS.473.1472T}). Figure~\ref{fig:Cshockn4-velo} shows the velocity structure of neutral ($v_{n}$) and ions ($v_{i}$), {as well as the drift velocity of neutrals relative to ions ($v_{\rm drift}=v_{n}-v_{i}$)}, assuming $n_{\H}=10^{4}\cm^{-3}$. 

The drift velocity $v_{\rm drift}$ rises and reaches the maximum value in the middle of the shock and then declines to zero. The~drift parameter, $s_{d}$, increases rapidly with $z$, and~then declines when the gas is heated to high temperatures. Note the peak of $s_{d}$ does not coincide with the peak of $v_{\rm drift}$ due to the effect of $v_{\rm th}$ or $T_{\rm gas}$. To~calculate the smallest size $a_{\rm min}$ that nanoparticles can withstand the rotational disruption, we compute $\langle \omega^{2}\rangle$ using the rotational temperature $T_{\rm rot}$ at each shock location for a grid of grain sizes from 0.35--10 nm and compare it with $\omega_{\rm cri}$.

\begin{figure}[H]
\includegraphics[width=0.5\textwidth]{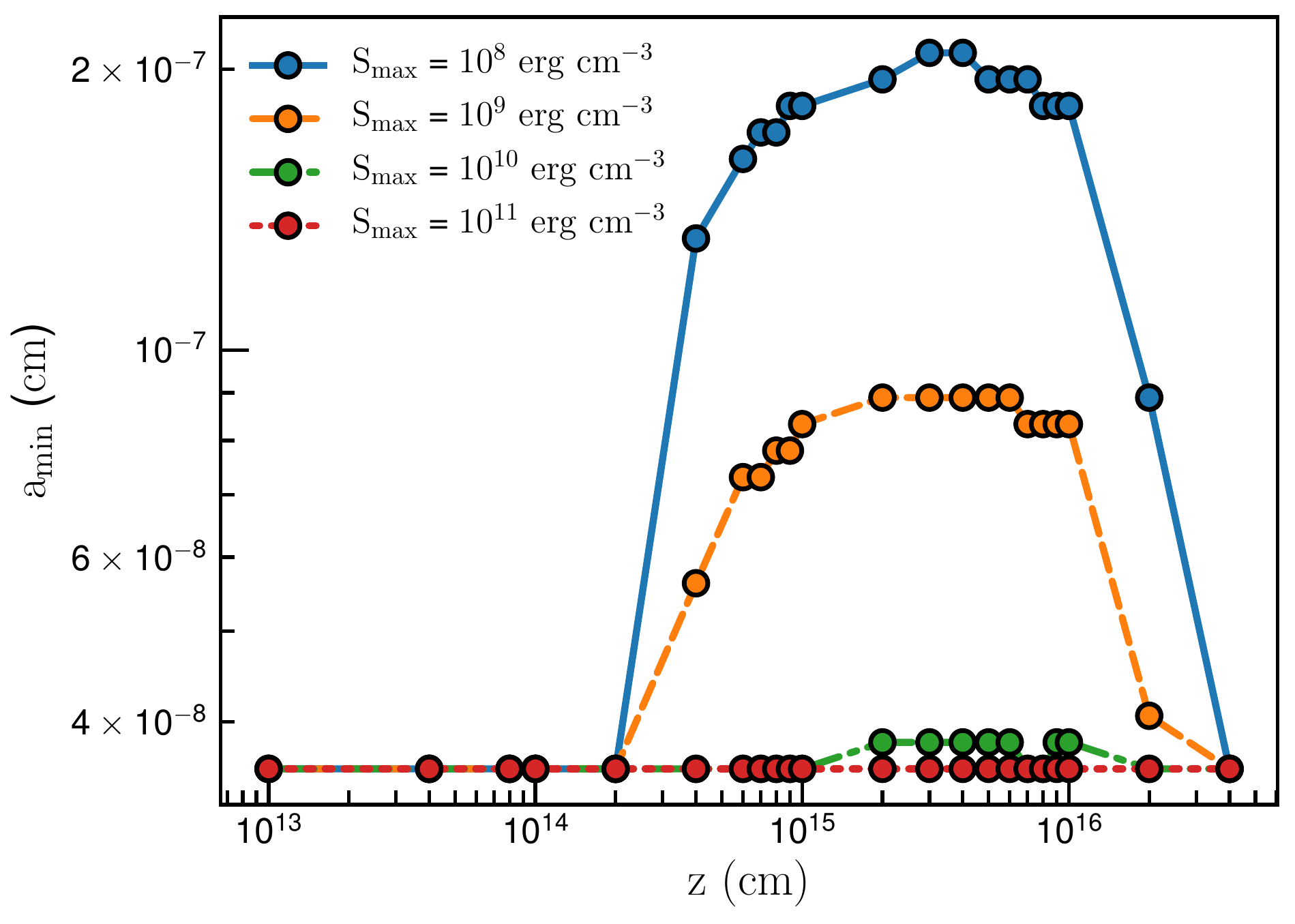}
\includegraphics[width=0.5\textwidth]{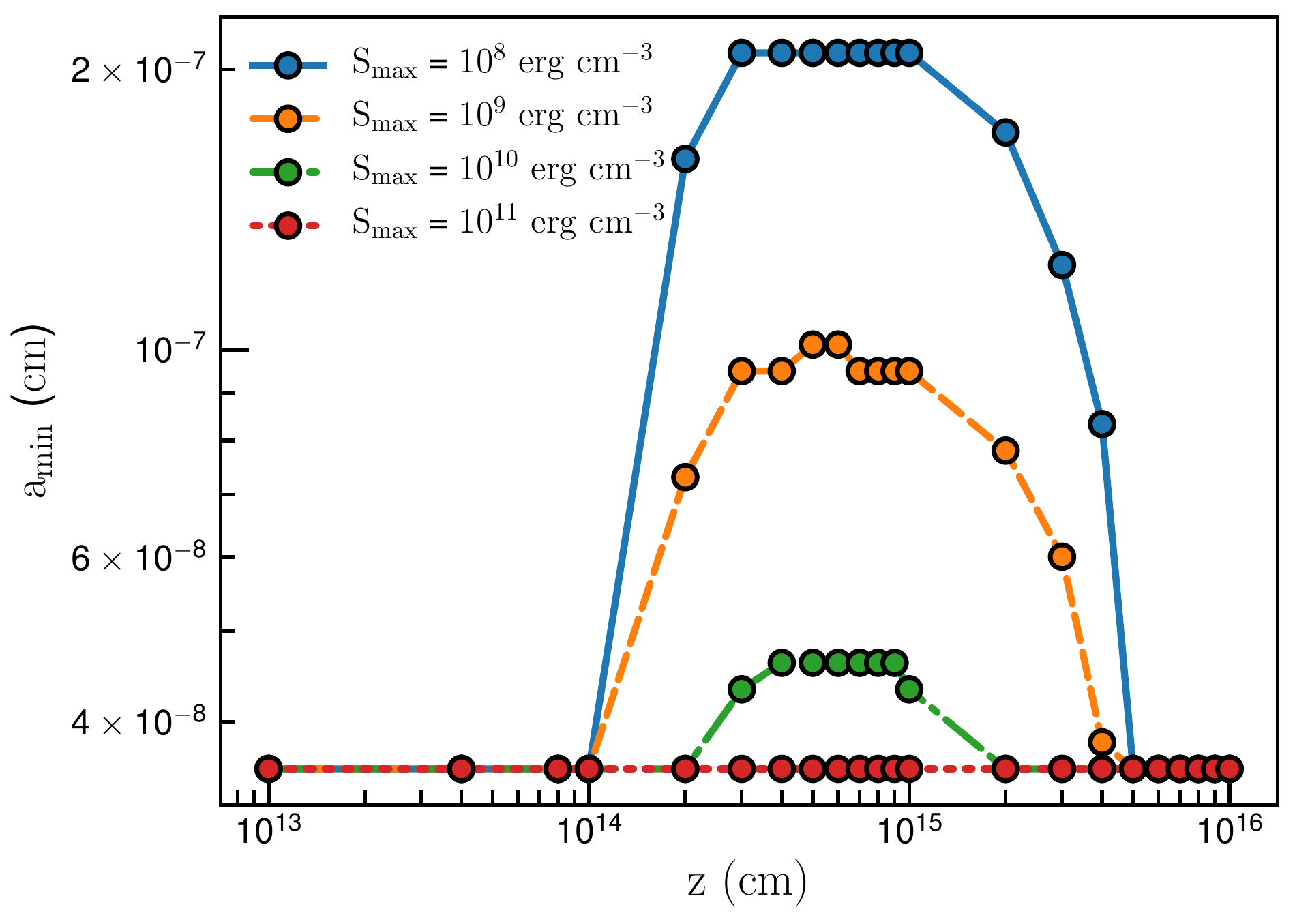}
\caption{Minimum size below which polycyclic aromatic hydrocarbons (PAHs) are destroyed by rotational disruption vs. distance in the shock, assuming the different material tensile strengths for $v_s=30\km\s^{-1}$. Two shock models with gas density $n_{\H}=10^{4}\cm^{-3}$ (left panel) and
$n_{\H}=10^{5}\cm^{-3}$ (right panel). From~\citet{2019ApJ...877...36H}.}
\label{fig:a_cri}
\end{figure}

Figure~\ref{fig:a_cri} shows the obtained minimum size $a_{\rm min}$ as a function of distance in the shock for different values of $S_{\rm max}$ and two shock models. Strong nanoparticles can survive the shock passage (red line), while weak nanoparticles can be destroyed. Grain disruption size increases toward the middle of the shock and then rapidly declines, which resembles the temperature and velocity profile of shocks (see~e.g.,~Figure~\ref{fig:Cshockn4-velo}). The~disruption size is below 0.5 nm for the typical $S_{\rm max,10}=1$, but~it can be increased to $2.0$ nm for weaker materials (see blue, orange and green lines). { Disruption is stronger for the shock model with higher gas density (right panel)}.

\begin{figure}[H]
\includegraphics[width=0.45\textwidth]{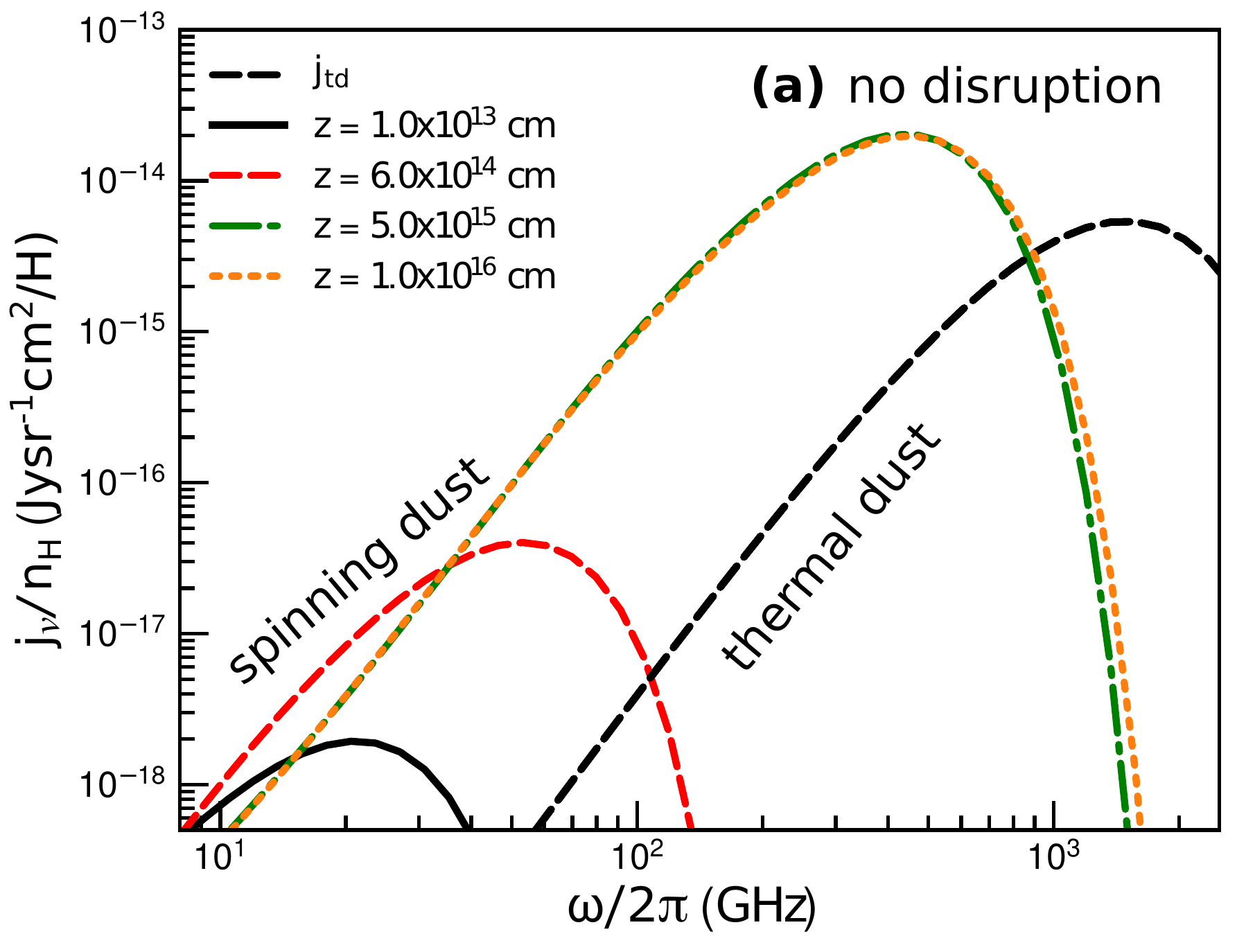}
\includegraphics[width=0.45\textwidth]{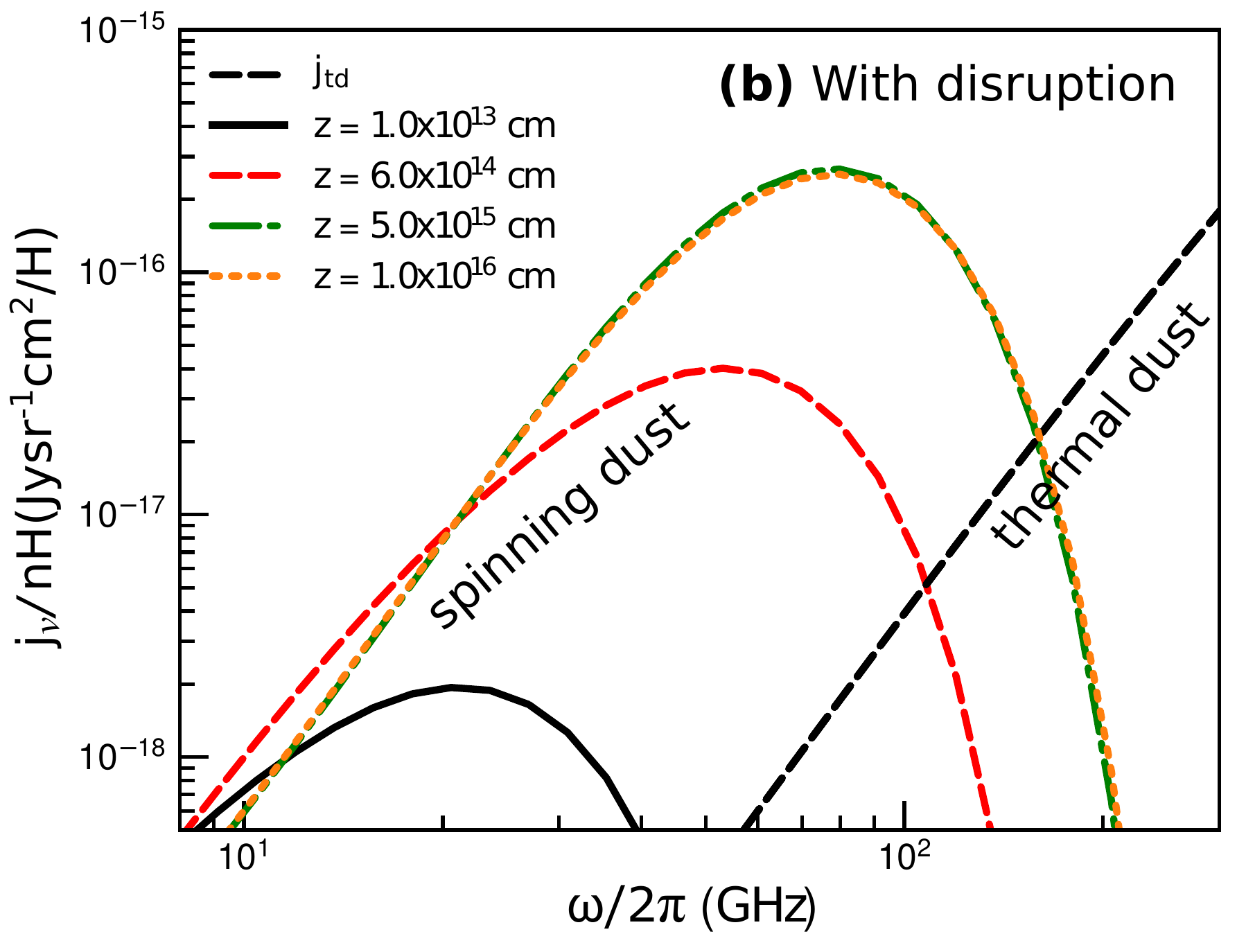}
\caption{Rotational emission spectrum from spinning nanoparticles for $n_{\H}=10^{4}\cm^{-3}$ and \mbox{$v_{s}=20 \km\s^{-1}$} computed at several positions in the shock without rotational disruption (panel (\textbf{a})) and with rotational disruption for $S_{\rm max}=10^{10}\erg\cm^{-3}$ (panel (\textbf{b})). Thermal dust emission from large grains is shown in dashed black line for comparison. From~\citet{2019ApJ...877...36H}. Emissivity is reduced significantly in the presence of~disruption.}
\label{fig:spindust_e}
\end{figure}

Using Equation~(\ref{eq:jnu_w}) we calculate the spinning dust emissivity at various locations inside the shock. The~emissivity is calculated assuming that dust is composed of 90$\%$ PAHs and 10$\%$ nanosilicates. In~the absence of rotation disruption, a$_{\rm min}$ is taken to be equal to 3.56~\AA. When the rotational disruption effect is taken into account, a$_{\rm min}$ is determined by $a_{\rm disr}$. We fix the abundance of PAHs and nanosilicates throughout the shock, although~their abundance should vary in the shock due to grain shattering (\citet{2011A&A...527A.123G}). Figure~\ref{fig:spindust_e} shows spinning dust emissivity from nanoparticles computed at different location $z$ in the shock. The~black dashed line shows thermal dust emission from large grains. When~the rotational disruption effect is not taken into account, spinning dust emissivity is very strong and can peak at very high frequencies of $\nu\sim 500$ GHz for some locations (see panel (a)). When~accounting for rotational disruption (panel (b)), both rotational emissivity and peak frequency are reduced significantly due to the destruction of the smallest nanoparticles via rotational disruption. The~effect of rotational disruption is clearly demonstrated through emission spectrum at locations $z=5\times 10^{15}\cm$ and $z=10^{16}\cm$, where the peak frequency is reduced from $\nu\sim 500$ GHz (panel (a)) to $\nu\sim 80$ GHz (panel (b)). In~both cases, spinning dust is still dominant over thermal dust at frequencies $\nu<100$ GHz (lower panel).

The grain size distribution of PAHs and nanoparticles in the shocked dense regions is poorly known due to the lack of observational constraints. In~dense cold clouds, due to the lack of UV photons, both sublimation and thermal sputtering are not effective, such that one can expect a much smaller lower cutoff of the grain size distribution compared to the diffuse ISM. In~studies of grain shattering, the~smallest size of nanoparticles is usually fixed to $a_{\min}=0.5$ nm without physical justification (\citet{1996ApJ...469..740J,2011A&A...527A.123G}). \citet{2010A&A...510A..36M} studied the destruction of PAHs in shocks by sputtering and found that PAHs can be efficiently destroyed by shocks of velocities $v_{s}>100\km\s^{-1}$. For~lower shock velocities, PAHs and smallest nanoparticles (i.e., nanoparticles smaller than several nanometers) are expected to survive the shock passage. As~a result, constraining the lower size cutoff and abundance of nanoparticles is of great importance. We suggested that spinning dust from nanoparticles could be used to constrain the abundance of nanoparticles in~shocks.

\unskip

\section{Discussion}\label{sec:discuss}

\vspace{-6pt}
\subsection{Radiative Torques of Irregular~Grains}
The interaction of dust grains and radiation results in four fundamental effects: grain heating, grain charging, radiation pressure, and~radiative torque. Radiative torque is a new notion introduced about 40 years ago in dust astrophysics by \citet{1976Ap&SS..43..291D}. Numerical calculations for realistic grain shapes are only available 20 years by \citet{1996ApJ...470..551D}, whereas an experimental test of the grain spin-up by RATs was conducted in \citet{2004ApJ...614..781A}. An~analytical model of RATs was proposed by \citet{2007MNRAS.378..910L}. With~extensive calculations of RATs for many shapes by \citet{Herranen:2019kj}, it is now established that RATs are a fundamental property of dust grains exposed to a radiation field. As~a result, it is crucially important to understand the effect of RATs on dust~properties. 

Previously, RATs are successfully used to explain the ubiquitous alignment of dust grains (\citet{1997ApJ...480..633D,2007MNRAS.378..910L}). The~RAT alignment theory has become a popular theory of grain alignment (\citet{LAH15,Andersson:2015bq,2016ApJ...831..159H}). The~rotational disruption (RATD) effect induced by RATs was discovered by \citet{Hoang:2019da} where the authors found that RATs could spin up grains to extremely fast rotation such that the centrifugal stress exceeds the material strength, disrupting dust grains into small fragments (see Section~\ref{sec:theory}). The~RATD mechanism is based on a fundamental property, {\it radiative torques}, arising in the interaction of radiation and dust grains of irregular shapes, which is natural in~astrophysics. 

\subsection{Ubiquitous Application of Rotational Disruption by Radiative~Torques}
Subject to intense radiation sources in astrophysics (e.g., quasars and GRBs, dust grains are known to be destroyed by thermal sublimation and Coulomb explosions \mbox{(see e.g.,~\citet{2000ApJ...537..796W,2006ApJ...645.1188W})}. While these traditional mechanisms require high radiation energy density and thus work only in a limited region, the~RATD effect can be effective in an average radiation field (e.g., ISRF) and becomes more efficient with stronger radiation fields. This makes RATD work in a vast range of astrophysical environments, with~the grain temperature $T_{d}\sim 20-T_{\rm sub}\sim 1500\K$. This corresponds to a radiation energy density $U$ from 1 to $(1500/20)^{6}\sim 10^{11}$, assuming $U\sim (T_{d}/20\K)^{6}$, comprising the diffuse ISM, star-forming regions, and~circumstellar regions, environments surrounding cosmic transients, starburst and high-z galaxies. The~RATD mechanism is only ineffective in dense molecular clouds shielded of the ISRF. The~dominance of rotational disruption over thermal sublimation can be understood by means of energy consideration. Indeed, in~order to heat the dust grain to the sublimation temperature, $T_{\rm sub}$, the~radiation energy must be $E_{\rm rad}\sim T_{\rm sub}^{4}$. On~the other hand, in~order to spin-up dust grains to the critical rotation rate $\omega_{\rm disr}$, the~radiation energy required is $E_{\rm rad}\sim \omega_{\rm disr}$. Due to the fourth-order dependence, the~sublimation energy is much higher than the energy required for grain~disruption. 

\subsection{Relationship between Rotational Disruption and Grain~Alignment}
Throughout this review, we have assumed that RATs spin up grains to their maximum rotational angular velocity and ignore the effect of grain alignment. However, RATs not only spin up but also align grains with the magnetic field. The~maximum angular velocity depends on the relative angle between the magnetic field and the radiation field (\citet{2009ApJ...697.1316H}). 

We have also assumed that all aligned grains have their maximum angular velocity of $\omega_{\rm RAT}$ (Equations  (\ref{eq:omega_RAT0}) and  (\ref{eq:omega_RAT}), which corresponds to the situation that all grains are driven to high-J attractors (\citet{2016ApJ...831..159H,Hoang:2008gb}). In~general, the~fraction of grains on high-J attractors, denoted by $f_{\rm high-J}$, depends on the grain properties (shape, size, and~magnetic properties), and~$0<f_{\rm high-J}\le 1$. \citet{Herranen:2020im}~ {calculated} RATs for aggregates and {found $f_{\rm high-J}\sim$} 0.35--1  { for aggregate grains.} \mbox{The~presence of iron inclusions} is found to increase $f_{\rm high-J}$ to unity (\citet{2016ApJ...831..159H}). Moreover, in~the absence of high-J attractors (e.g., high-J repellors), grains may still be disrupted because gas collisions randomize their orientation in the phase space, and~the grains would spend a significant amount of time in the vicinity of high-J repellors (\citet{Hoang:2008gb}). As~a result, grains can still be disrupted if their instantaneous angular velocity exceeds $\omega_{\rm disr}$. A~detailed study of grain disruption for this case should follow the rotational dynamics of grains induced by RATs for the GRB radiation field (e.g., \citet{2016ApJ...831..159H}) and compares the instantaneous centrifugal stress with the tensile strength of the grain. More details about the relation of alignment and the RATD are given in Lazarian \& Hoang (2020). 

Second, in~the case of intense radiation fields, grains can be aligned along the radiation direction (\citet{2007MNRAS.378..910L,2016ApJ...831..159H,2017ApJ...839...56T}). Carbonaceous grains with diamagnetic properties are expected to be aligned along the radiation direction. Thus, the~RATD effect would work at their maximum level if high-J attractors are~present.

Third, since the fraction of high-J depends on the grain magnetic susceptibility, which is determined by the abundance of iron embedded, the~RATD effect would be useful to constrain the iron fraction in dust (\citet{2008ApJ...676L..25L,2016ApJ...831..159H,2019ApJ...883..122L}).

\subsection{Dust in the Time-Domain Astronomy~Era}
With substantial investment to construct new observational facilities at optical-NIR wavelengths, including GMT, SPHEREx, LSST, TMT, E-ELT, JWST and WFIRST, we are entering the golden age of the time-domain astronomy. Millions of transients (SNe Ia, CCSNe, GRBs) would be detected (e.g., \citet{2019BAAS...51c.339G,2019BAAS...51c.305F}). UV-Optical-NIR observations by these instruments, including colors and light curves of astrophysical transients are powerful to study progenitors, explosion mechanism, and~their environments. In~the light of RATD, dust properties within tens of pc are rapidly changing under the effect of the intense radiation from transients themselves. As~a result, the~observed color and light-curves of transients could change over time just due to dust variation. The~RATD effect is particularly important for CCSNe and GRBs because they are expected to explode in dusty star-forming regions. An~accurate understanding of cosmic transients as well as using them as a standardized candle for cosmological studies must take into account the effect of time-varying dust properties by the RATD effect. The~decrease of optical-NIR extinction/polarization due to RATD would be tested with LSST and~WFIRST.

\subsection{Astrochemistry on Rotating Grain~Surfaces}
\textls[-19]{Grain surfaces play a crucial role in astrochemistry \mbox{(\citet{Herbst:2009go,2018IAUS..332....3V})}}. The~foundations of surface astrochemistry assume dust grains at rest and disregard the fact that grains are rapidly rotating due to radiative or mechanical~torques. 

Grain surfaces and ice mantles are believed to play a crucial role in the formation and desorption of molecules, including H$_{2}$, H$_{2}$O, and~COMs (see e.g.,~\citet{Herbst:2009go}). Grain surface chemistry in general involves four main physical processes: (1) accretion of gas atoms/molecules to the grain surface, (2) mobility of adsorbed species on or in the ice mantle, (3) probability to form molecules upon collisions, and~(4) desorption of newly formed molecules from the grain surface (see \citet{2018IAUS..332....3V}). 

Suprathermal rotation by RATs is found to be important for surface chemistry in star-forming and photodissociation regions. It is found that suprathermal rotation can assist thermal desorption of molecules from the ice mantle, which enables molecule desorption at temperatures $T_{d}<100\K$, lower~than classical thermal sublimation threshold (Section \ref{sec:des}, \mbox{\citet{2020ApJ...891...38H,2019ApJ...885..125H}}). The~rate of surface chemical reactions depends on the mobility of adsorbed species on the grain surface (\citet{1972ApJ...174..321W,1992ApJS...82..167H}). The~mobility of adsorbed molecules on the surface could also be enhanced, resulting in an increase of molecular formation rate (\citet{Hoang:2019wb}). Therefore, the~effect of grain suprathermal rotation is important and dramatically changes the current paradigm of surface astrochemistry where grain rotation is~disregarded.

\subsection{Dust Polarization and Molecular~Tracer}

In the rotational desorption paradigm, the~release of COMs from ice mantles is accompanied by the decrease in the abundance of large grains because the intense radiation field that disrupts the ice mantles also disrupts large dust aggregates, and~the disruption of ice mantles reduces the grain size. This effect has a unique signature on observations. {First, we~expect the abundance of COMs increases with decreasing the degree of dust polarization at long wavelengths which are most sensitive to the abundance of large grains.} Second, the~depletion of large aggregate grains may result in the change in the polarization pattern because very large grains are expected to experience efficient self-scattering \mbox{(\citet{2015ApJ...809...78K})}, whereas smaller ones are aligned along the radiation~direction or magnetic field direction \mbox{(\citet{2007MNRAS.378..910L,2016ApJ...831..159H,2017ApJ...839...56T,2019ApJ...883..122L})}. {In our on-going study (Tram~et~al. to be submitted), we found some evidence for rotational desorption using observational data of COMs and dust polarization.}

\section{Conclusions and~Outlook}\label{sec:summ}

We have reviewed the rotational disruption of dust grains by radiative torques and mechanical torques due to dust-radiation and dust-gas interactions and presented various applications. Our main conclusions are summarized as follows:\\

\begin{itemize}

\item
Radiative torques are a fundamental property of dust-radiation interaction. Dust grains could be spun up to suprathermal rotation by RATs such that resulting centrifugal stress exceeds the tensile strength of grain material, resulting in the disruption of grains into fragments. 
Because~the RATD mechanism does not require an intense radiation field to be effective, it has ubiquitous application for most astrophysical environments, from~the diffuse ISM to star-forming regions, protoplanetary disks, circumstellar regions, and~high-z galaxies (see Section~\ref{sec:appl}).

\item
The RATD mechanism could successfully explain some longstanding puzzles in astrophysics, including the anomalous dust properties observed toward SNe Ia and H II regions around massive stars, steep extinction curves toward GRBs, and~microwave emission excess in AGB~envelopes.

\item
The RATD mechanism changes the grain size distribution and abundance, which affects many astrophysical observations, including dust extinction, emission, and~polarization. This RATD mechanism thus opens a new dimension into dust physics and offers new diagnostics of astrophysical phenomena. For~instance, one of the longstanding puzzles of astrophysical dust is its internal structure, that is,~compact vs. fluffy/composite vs. core-mantle. The~RATD provides a theoretical basis for probing the internal structure of dust~grains. 

\item
In the time-domain astronomy era, intrinsic light-curves and colors of astrophysical transients are required to understand progenitors, explosion mechanisms, and~transient's environments. In~light of the RATD effect, the~rapid variation of grain size distribution on a timescale of minutes or days results in the decrease of optical-NIR dust extinction, but~the increase of UV extinction. This reduces the value of $R_{V}$ and color $E_{B-V}$ observed toward transients. Thus, the~effect of time-varying dust properties by RATD must be considered for accurate transient~astrophysics.

\item
The centrifugal force arising from grain suprathermal rotation induced by RATs plays a crucial role in ice evolution. It can desorb ice mantles from the grain surface in star-forming and photodissociation regions. Moreover, water ice can be desorbed with the expense of smaller energy compared to thermal sublimation. As~a result, the~snow-line in the protoplanetary disk is pushed outward compared to the classical~snow-line.

\item
Suprathermal rotation by RATs plays a critical role in surface astrochemistry in star-forming and photodissociation regions. It is found that suprathermal rotation can assist thermal desorption of molecules from the ice mantle, enabling molecule desorption at temperatures $T_{d}<100\K$, lower~than classical thermal sublimation thresholds. The~mobility of adsorbed molecules on the surface could also be enhanced, increasing molecule formation rate. This could dramatically change the current paradigm of surface astrochemistry, where grain rotation is~disregarded.

\item
Nanoparticles can be spun-up to suprathermal rotation by the relative supersonic motion of dust and gas. As~a result, the~smallest nanoparticles can be disrupted by centrifugal stress due to their small inertia moment. The~METD mechanism is efficient in magnetized shocks and grains drifting through the gas by radiation~pressure.

\item
In addition to optical-IR wavelengths, rotational disruption also affects microwave emission via the spinning dust mechanism. Nanoparticles play an important role in gas heating and dynamics, shock dynamics. Thus, observations in microwave are unique to trace nanoparticles and test the RATD and METD~mechanisms.

\end{itemize}

\vspace{6pt} 



\funding{This research is supported by the National Research Foundation of Korea (NRF) grants funded by the Korea government (MSIT) through the Basic Science Research Program (2017R1D1A1B03035359) and Mid-career Research Program (2019R1A2C1087045).

}

\acknowledgments{We are grateful to two anonymous referees for thorough reading and helpful comments that improved the content of this~review. We thank Vietnam National Space Center (VNSC) and Phenikaa Institute of Advances Studies (PIAS), Phenikaa University, Vietnam, for~their hospitality during which the review is written.} 

\conflictsofinterest{The authors declare no conflict of~interest.} 

\newpage
\abbreviations{The following abbreviations are used in this manuscript:\\
\noindent 
\begin{tabular}{@{}ll}
NIR & Near-infrared\\
FIR & Far-infrared\\
AMO & Analytical MOdel\\
DDA & Discrete Dipole Approximation\\
DDSCAT & Discrete Dipole Scattering \\
ISM  & Interstellar Medium\\
ISRF & Interstellar Radiation Field\\
RATs & RAdiative Torques\\
RATA & RAdiative Torque Alignment\\
RATD & RAdiative Torque Disruption\\
ROTD & Rotational Desorption\\
METD & MEchanical Torque Disruption\\
COM & Complex Organic Molecule\\
SNIa & Type Ia Supernova \\
CCSN & Core-Collapse Supernova\\
GRB & Gamma-Ray Burst\\
MC  & Molecular Cloud \\
SMC & Small Magellanic Cloud \\
PDR & Photodissociation Region\\
YSO & Young Stellar Object\\
YMSC & Young Massive Stellar Cluster\\
PPD & Protoplanetary disk\\
AGN & Active Galactic Nuclei\\
AGB & Asymptotic Giant Branch\\
CSE & Circumstellar Envelope\\
PSP & Parker Solar Probe\\
VSG & Very Small Grain\\
PAH & Polycyclic Aromatic Hydrocarbon
\end{tabular}}

\appendixtitles{yes} 
\appendix

\begin{table}[H]
\begin{center}
\caption{Glossary of Notations and~Meaning}\label{tab:notations1}
\begin{tabular}{l l } \toprule
{\textbf{Notation}} & {{\textbf{Meaning}}}\cr
\midrule
$a$ & grain radius\cr
$\rho$ & grain mass density\cr
$\rho_{\rm ice}$ & mass density of ice\cr

$m_{\rm gr}$ & grain mass\cr
$I_{1}$ & inertia momentum around the axis of maximum moment of inertia \cr
$Q_{\rm ext}$ & extinction efficiency \cr
$Q_{\rm abs}$ & absorption efficiency \cr
$Q_{\rm pol}$ & polarization efficiency \cr

$n_{\H}$ & proton number density\cr
$N_{\H}$ & proton column number density\cr
$m_{\H}$ & proton mass\cr

$T_{\gas}$ & gas temperature \cr
$T_{d}$ & dust grain temperature \cr

$\tau_{\rm gas}$ & rotational damping time by gas collisions\cr
$\tau_{\rm IR}$ & rotational damping time by infrared emission\cr
$F_{\rm IR}$ & dimensionless coefficient of IR rotational damping\cr

$\lambda$ & radiation wavelength\cr
$u_{\lambda}$ & specific energy density of the radiation field\cr
$\bar{\lambda}$ & mean wavelength of the radiation spectrum\cr
$\gamma_{\rm rad}$ & anisotropy degree of the radiation field \cr
$B_{\lambda}$ & Planck function\cr
$A_{\lambda}$ & dust extinction\cr
$E(B-V)=A_{B}-A_{V}$ & color excess\cr
$R_{V}=A_{V}/E(B-V)$ & ratio of total-to-selective extinction \cr
$P_{\rm ext}(\lambda)$ & polarization of starlight by dust extinction\cr
$P_{\rm max}$ & maximum value of $P_{\rm ext}$\cr
$\lambda_{\rm max}$ & peak wavelength of at the maximum polarization $P_{\rm ext}$\cr
$P_{\rm em}(\lambda)$ & polarization of thermal dust emission\cr

$L$ & bolometric luminosity\cr
$u_{\rm rad}$ & radiation energy density \cr
$u_{\rm ISRF}$ & radiation energy density of the ISRF \cr
$U=u_{\rm rad}/u_{\rm ISRF}$ & radiation strength\cr
$\Gamma_{\rm RAT}$ & Radiative Torque (RAT) \cr
$Q_{\Gamma}$ & RAT efficiency \cr
$\omega$ & grain angular velocity \cr
$\omega_{T}$ &grain thermal angular velocity \cr
$\omega_{\rm RAT}$ & maximum grain angular momentum spun-up by RATs \cr
$\omega_{\rm disr}$ & critical angular velocity for rotational disruption\cr
$T_{\rm sub}$ & sublimation temperature \cr
$R_{\rm sub}$ & sublimation radius\cr
$S_{\rm max}$ & maximum tensile strength of grain material\cr
$a_{\rm disr}$ & grain disruption size by RATD\cr
$a_{\rm disr, max}$ & maximum grain disruption size by RATD\cr
$a_{\rm desr}$ & grain desorption size of ice mantles\cr
$a_{\rm desr, max}$ & maximum grain desorption size of mantles\cr
$v_{\rm drift}$ & drift velocity of grains through the gas\cr
$v_{d}$ & dust grain velocity relative to gas\cr
$v_{\rm th}=(2kT_{gas}/m_{\H})^{1/2}$ & gas thermal velocity\cr
$s_{d}=v_{d}/v_{th}$ & dimensionless drift parameter\cr
$v_{gg}$ & relative grain velocity \cr
$F_{\rm drag}$ & gas drag force \cr
$Y_{\rm sp}$ & sputtering yield\cr
$\bar{A}_{sp}$ & mean atomic mass of sputtered atoms\cr
$R_{\odot}$ & radius of the Sun \cr
$R$ & heliocentric distance from the Sun \cr
\bottomrule
\end{tabular}
\end{center}
\end{table}
\unskip

\begin{table}[H]
\begin{center}
\caption{Glossary of Notations and Meaning (Continued)}\label{tab:notations2}
\begin{tabular}{l l } \toprule
{\textbf{Notation}} & {{\textbf{Meaning}}}\cr
\midrule
$E_{b}$ & binding energy of molecules to the grain surface \cr
$\nu_{0}$ & characteristic vibration frequency of molecules on the icy grain mantle\cr
$B$ & magnetic field strength\cr
$v_{s}$ & shock speed \cr
$v_{n},v_{i}$ & velocity of neutrals and ions in shocks\cr
$\beta$ & electric dipole moment per structure\cr
$\mu $ & electric dipole moment of a grain\cr
$\tau_{ed}$ & electric dipole damping time \cr
$T_{\rm rot}$ & grain rotational temperature\cr
$dn/da $ & grain size distribution\cr
$a_{\rm min}$ & minimum grain size or lower cutoff of the grain size distribution \cr
$a_{\rm max}$ & maximum grain size or upper cutoff of the grain size distribution\cr
$\nu$ & frequency of radiation \cr
$P(\omega,\mu)$ & emission power by a nanoparticle spinning at $\omega$\cr
$j_{\nu}^{a}$ & rotational emissivity from a spinning nanoparticle of size $a$\cr
$j_{\nu}$ & rotational emissivity from all nanoparticles \cr
\bottomrule
\end{tabular}
\end{center}
\end{table}


\reftitle{References}

\end{document}